\documentclass[preprint]{aastex}

\usepackage{hyperref}

\begin{document}
%VB: I suggest a slightly modified title (I don t think, ApJ would allow us
%    to use too many acronyms in the title...)
\title{Reconstructing emission from pre-reionization sources with cosmic infrared background fluctuation measurements by the JWST}

%\author{Team: A. Kashlinsky, J. Mather, R. Arendt, K. Helgason, V. Bromm, H. Moseley, 
\author{A. Kashlinsky\altaffilmark{1,2,3},
 J. C. Mather\altaffilmark{1,4},
K. Helgason\altaffilmark{5},
 R. G. Arendt\altaffilmark{1,6},
V. Bromm\altaffilmark{7},
 S.~H.~Moseley\altaffilmark{1,4}
 }
\altaffiltext{1}{Code 665, Observational Cosmology Lab, NASA Goddard Space Flight Center, 
Greenbelt, MD 20771}
\altaffiltext{2}{SSAI, Lanham, MD 20770}
\altaffiltext{3}{email: Alexander.Kashlinsky@nasa.gov} 
\altaffiltext{4}{ NASA}
\altaffiltext{5}{MPA, Karl-Schwarzschild-Str. 1, 85748 Garching, Germany}
\altaffiltext{6}{CRESST/UMBC}
\altaffiltext{7}{Department of Astronomy, University of Texas, Austin, TX 78712}

%\date{}
%\oddsidemargin .3in
%\evensidemargin .1in
%\footheight 12pt \footskip 30pt
%\topmargin 0.cm
%\textheight 21.5cm
%\textwidth 16.cm
%\renewcommand \baselinestretch{1.5}

% list of standard definitions of symbols, figure commands

\def\plotone#1{\centering \leavevmode
\epsfxsize=\columnwidth \epsfbox{#1}}

\def\wisk#1{\ifmmode{#1}\else{$#1$}\fi}

\def\wm2sr {Wm$^{-2}$sr$^{-1}$ }		%new definition
\def\nw2m4sr2 {nW$^2$m$^{-4}$sr$^{-2}$\ }		% new definition
\def\nwm2sr {nWm$^{-2}$sr$^{-1}$\ }		% new definition
\def\nw2m4sr {nW$^2$m$^{-4}$sr$^{-1}$\ }
\def\Ncut {$N_{\rm cut}$\ }
\def\lt     {\wisk{<}}
\def\gt     {\wisk{>}}
\def\le     {\wisk{_<\atop^=}}
\def\ge     {\wisk{_>\atop^=}}
\def\lsim   {\wisk{_<\atop^{\sim}}}
\def\gsim   {\wisk{_>\atop^{\sim}}}
\def\kms    {\wisk{{\rm ~km~s^{-1}}}}
\def\Lsun   {\wisk{{\rm L_\odot}}}
\def\Msun   {\wisk{{\rm M_\odot}}}
\def\um     { $\mu$m\ }
\def\sig    {\wisk{\sigma}}
\def\etal   {{\sl et~al.\ }}
\def\eg	    {{\it e.g.\ }}
\def\ie     {{\it i.e.\ }}
\def\bsl    {\wisk{\backslash}}
\def\by     {\wisk{\times}}
\def\cosec {\wisk{\rm cosec}}
\def\mic {\wisk{ \mu{\rm m }}}

\def\amin   {\wisk{^\prime\ }}
\def\asec   {\wisk{^{\prime\prime}\ }}
\def\cc     {\wisk{{\rm cm^{-3}\ }}}
\def\deg     {\wisk{^\circ}}
\def\ddeg   {\wisk{{\rlap.}^\circ}}
\def\damin  {\wisk{{\rlap.}^\prime}}
\def\dasec  {\wisk{{\rlap.}^{\prime\prime}}}
\def\approxeq{$\sim \over =$}
\def\abouteq{$\sim \over -$}
\def\percm{cm$^{-1}$}
\def\percmsq{cm$^{-2}$}
\def\percmcub{cm$^{-3}$}
\def\perhz{Hz$^{-1}$}
\def\perpc{$\rm pc^{-1}$}
\def\persec{s$^{-1}$}
\def\peryr{yr$^{-1}$}
\def\te{$\rm T_e$}
\def\tenup#1{10$^{#1}$}
\def\to{\wisk{\rightarrow}}
\def\thin{\thinspace}
\def\uk{$\rm \mu K$}
\def\p{\vskip 13pt}
%\maketitle

\begin{abstract}
%\vspace{1cm}
We present new methodology to use cosmic infrared background (CIB) fluctuations to probe 
%from early sources with source-subtracted CIB fluctuations at NIRCam of the {\it JWST}. We
%start with discussion of the current CIB fluctuation measurements and interpretation, where 
%significant mutually consistent source-subtracted cosmic infrared background (CIB) fluctuations have been identified in the {\it Spitzer} and {\it AKARI} data between $\sim 2$ and $\sim 5 \mu$m, but demonstrate that the situation appears greatly inconsistent at shorter wavelengths in the claims recently 
%advanced with CIBER. We identify an experimental
%configuration with which NIRCam at {\it JWST} can determine
 sources at $10\lsim z \lsim 30$ from a {\it JWST}/NIRCam configuration that will isolate known galaxies to 28 AB mag at 0.5--5\micron. 
 At present significant mutually consistent source-subtracted CIB fluctuations have been identified in the {\it Spitzer} and {\it AKARI} data at $\sim 2-5 \mu$m, but  we demonstrate internal inconsistencies at shorter wavelengths in the recent CIBER data.
We evaluate CIB contributions from remaining
galaxies and show that the bulk of the high-$z$ sources will be in the confusion noise of the NIRCam beam, requiring CIB studies. The accurate measurement of the angular spectrum of the fluctuations and probing the dependence of its
clustering component on the remaining shot noise power would discriminate between the various currently proposed models for their origin and probe the
flux distribution of its sources. We show that the contribution to CIB fluctuations from remaining galaxies is large  at visible wavelengths for the current instruments 
precluding probing the putative Lyman-break of the CIB fluctuations. We demonstrate that with the proposed {\it JWST} configuration  such measurements will enable probing the Lyman break. We develop a Lyman-break tomography method to use the NIRCam wavelength coverage to identify
or constrain, via
the adjacent two-band subtraction, the history of emissions over $10\lsim z\lsim30$ as the Universe comes out of the ``Dark Ages''. We apply the proposed 
tomography to the current {\it Spitzer}/IRAC measurements at 3.6 and 4.5 \micron, to find that it already leads to interestingly low upper limit on emissions
at $z \gsim 30$. 
%We discuss the effects of various astronomical foregrounds and potential systematics from the instrumentation for this measurement.
\end{abstract}

%\newpage
%\tableofcontents
%\clearpage

\section{Introduction}
%VB: slight edits in the first paragraph, including references to recent review
%    papers on Pop III formation theory.
%VB: slight edits (mostly typo corrections) in the other paragraphs of the INTRO

%Formation rates are difficult
%to predict, but upper limits can be placed rather robustly: ${\mathbf{\dot{\rho}}}_{\rm BH}\la
%10^{-3} M_{\odot}$\,Mpc$^{-3}$\,yr$^{-1}$ (e.g. Jeon et al. 2012). 

As the universe emerges out of its Dark Ages, the first structures begin to collapse at redshifts as high as $z\sim 20-30$, hosting the first, so-called Population~III (Pop~III) stars,
supernovae (SNe), and black holes (BHs). The {\it James Webb Space Telescope (JWST)} is expected
to detect a fraction of the bright end of the luminosity distribution of the early universe objects at $z\sim12-15$, but the fainter sources will remain hidden and the predicted number densities of detectable sources are relatively small per ultra-deep field. It is therefore both 
imperative and timely to develop methods that may uncover, or constrain, the history of the overall light production in the 
universe even prior to these epochs.

The Cosmic Infrared Background (CIB) is the collective radiation emitted throughout cosmic history, including from sources
inaccessible to current telescopic studies (see review by
Kashlinsky 2005a). The near-IR part of the CIB at wavelengths $\sim (1-10)\mu$m probes predominantly 
the redshifted stellar, or accreting BH, radiation, and offers an
alternative way to probe the emission from sources at the earliest times. 
Historically, observations of the CIB have taken two complementary approaches. One approach is to
measure the absolute integrated intensity of the CIB at multiple wavelengths,
averaged over large areas. These measurements were the objective of the
{\it COBE}/DIRBE (Hauser et al. 1998), and the {\it IRTS}/NIRS instruments (Matsumoto et al. 2005). The primary difficulty
with the interpretation of these measurements is the large uncertainty associated
with the subtraction of foregrounds, Galactic components (Arendt et al. 1998) and the zodiacal light (Kelsall et al. 1998). Madau \& Silk (2005) pointed out,
using the J-band (1.2 \micron) as an example, that if one were to explain the high {\it mean} CIB levels suggested by studies involving DIRBE (Dwek \& Arendt 1998; Gorjian et al. 2000; Cambresy et al. 2001) and {\it IRTS} (Matsumoto et al. 2005), the energy requirement for such interpretation implied a conversion of a few percent of all baryons into Pop~III stars (see also Kashlinsky 2005b), which does not align with recent theory.
It was proposed that incomplete subtraction of zodiacal light (ZL) may
have caused the estimated mean CIB intensity to be too high (Arendt \& Dwek 2003, Dwek et al. 2005; Thompson et al. 2007; but see Tsumura et al. 2013).

The other approach is to measure the anisotropies or spatial fluctuations of the
CIB, pioneered in DIRBE studies by Kashlinsky et al (1996a,b) and Kashlinsky \& Odenwald (2000). 
This approach is beneficial since at some wavelengths 
and angular scales CIB fluctuations are easier to disentangle from the bright, but relatively smooth foregrounds.

The Universe has been fully ionized with negligible fractions of intergalactic HI remaining by $z\lsim 6-7$ as measurements of the lack of Gunn-Peterson absorption
at those epochs show \citep[see e.g.][and references therein]{gp}. At higher $z$ the efficacy of (re)ionization
is constrained observationally by the CMB polarization measurements of the Thomson optical depth from {\it WMAP} and {\it Planck} CMB data due to both 
homogeneously distributed and clumped ionized gas.
The pre-reionization sources also had to contain the first stars and BHs at very early times. There are strong intuitive reasons to expect measurable CIB 
anisotropies from those early populations 
regardless of whether they were dominated by massive stars or accreting BHs (Kashlinsky et al. 2004; Cooray et al. 2004): 
%\begin{itemize}
%\item 
1) the first stars are predicted to have been massive, with luminosity per unit mass larger than present-day stellar populations by a factor $\sim 10^4$; a similar factor applies to accreting Eddington-limited black BHs;
%\item 
2) their relative CIB fluctuations would be larger because they span a
relatively short time-span in the evolution of the universe; and
%\item 
3) these sources formed at the peaks of the underlying
density field, amplifying their clustering properties. 
%\end{itemize}

Significant progress in the field has been made in the last decade with discovery and measurements of source-subtracted
CIB fluctuations using data from {\it Spitzer} (Kashlinsky et al. 2005, 2007a,b,c - hereafter KAMM1,2,3,4; Arendt et al. 2010 - hereafter AKMM; Kashlinsky et al. 2012) and {\it AKARI} (Matsumoto et al. 2011). Both these
satellites have sufficient resolution and sensitivity to exclude resolved
galaxies to deep limits, so the CIB fluctuation measurements are
dominated by sources that are not otherwise observable with present instruments. It now appears firmly established that the discovered source-subtracted CIB fluctuations cannot be explained by known galaxy populations (KAMM1; Helgason et al. 2012). They
also exhibit an intriguingly strong coherence with the unresolved cosmic X-ray background (CXB) at soft X-rays ([0.5-2]\,keV), while there are no detectable cross-correlations at higher energies (Cappelluti et al. 2013); this is indicative of black hole populations
among the sources of the CIB which is in much higher proportion than in the known galaxy populations. The situation at shorter IR bands is currently 
conflicted and is discussed later.

NIRCam on {\it JWST} will be able to 
identify individual sources to much fainter fluxes than either {\it Spitzer} or {\it AKARI} and with it there is a potential 
to measure the cumulative emissions (i.e. the CIB) produced at still earlier epochs ($z\gtrsim 12-15$). Additionally, the NIRCam wavelength coverage 
will have a built-in 
capability to directly probe the Lyman break of the unresolved populations, provided the instrument noise, astronomical foregrounds and 
foreground galaxy populations can be isolated. Such measurement would provide a fundamental constraint on the 
otherwise inaccessible range of epochs (and fluxes), where as we show the sources are expected to lie in the confusion noise of the {\it JWST} beam.
%, and in this paper we undertake a study, using the latest results and new methodologies, in order to 
%probe the feasibility - and parameters - of this potential experiment with {\it JWST}. 
% This is just a really awkward restatement of what's said above and below. Delete.
%Here we concentrate on
%the near-IR CIB to study the prospects of and the strategies for isolating, with
%{\it JWST}'s NIRCam instrument, the contribution to it from the earliest sources which would remain mostly undetectable. 
With strategies developed here, {\it JWST} will be able to provide critical insight into the origin of the source-subtracted CIB fluctuations detected in {\it Spitzer} and {\it AKARI} measurements, identify the epochs where the fluctuations arise, probe the fluxes of the sources producing them and reconstruct/constrain the history of the emissions via the adjacent two-band Lyman tomography proposed in this paper.
With the particular experimental setup we can address these important, but hitherto unanswered, questions pertaining to the details and the nature of the populations that led the universe out of the ``Dark Ages''. 

This paper is structured as follows:  the CIB fluctuation situation and its implications are discussed in Sec. \ref{sec:cib_status}, where we present 
the general constraints on the sources expected at high-$z$, if they are to explain the observed signal. We identify in Sec. \ref{sec:jwst_pars} an optimal JWST configuration (area, aspect ratio, location, integration time, wavelengths) for probing 
the source-subtracted CIB excess fluctuations after evaluating the levels of the remaining known galaxy populations. Then in Sec. \ref{sec:confusion_floor} we explore the confusion noise properties of the populations expected at $z\gtrsim12-15$ and show that they 
would be within the confusion noise of the {\it JWST}/NIRCam instrument requiring the CIB as a tool to probe them. 
%After discussing the floor from the Galactic cirrus component and its uncertainties in Sec. \ref{sec:cirrus}, w
We quantify determination of the power spectrum of the source-subtracted CIB with this configuration in Sec. \ref{subsec:clustering} and show that the clustering component can be
determined with required fidelity to distinguish these high-$z$ sources from those at later times. The dependence of the clustering CIB component on 
the shot-noise can then be probed to reveal the typical fluxes of these populations as discussed in Sec. \ref{subsec:shotnoise}. Sec.  \ref{sec:lyman} 
shows the 
%the difficulty presented by the CIB fluctuation floor from the remaining known galaxies in the Lyman-break probing with the currently conducted experiments, followed by the numerical evaluation of this floor for the proposed  {\it JWST} configuration. 
difficulty in probing the Ly break of the CIB because of the fluctuations from the remaining known galaxies in current experiments. 
We then evaluate the level of fluctuations from the remaining galaxies in future {\it JWST} experiments. 
We show that the proposed configuration for {\it JWST} would enable direct probing of the epochs associated with sources producing these fluctuations. We then propose a two-band tomography method with NIRCam to reconstruct the history of emissions directly by using the expected Lyman-break shifted differentially into the adjacent NIRCam filters from each range of redshifts (Sec. \ref{sec:tomography}). We apply the new method to the current IRAC-based 
measurements to show that it already leads to useful interesting limits (Sec. \ref{sec:spitzer_tomography}). Finally we discuss the foregrounds and systematics in the measurement (Sec. \ref{sec:foregrounds}). Appendix provides the general background on the CIB fluctuations.

%TBD: rephrase: Specifically:
%\begin{enumerate}
%\item We work out an optimal JWST strategy (area, aspect ratio, location, integration time), wavelengths (Sec. \ref{sec:jwst_pars}) to measure 
%the parameters of the first era emissions,
%
%\item after determining the floors on each from ordinary galaxies  (Sec. \ref{sec:og_floor}),
%
%\item and predicting the confusion noise properties of the remaining sources at high $z$ (Sec. \ref{sec:confusion_floor}).
%
%\item This observational strategy is optimized so we can determine the power spectrum of the source-subtracted CIB below the fixed $m_{\rm lim}$ most accurately (Sec. \ref{subsec:clustering}) and 
%
%\item measure the dependence of clustering on shot-noise  (Sec. \ref{subsec:shotnoise}), then 
%
%\item measure the Lyman-break  (Sec. \ref{subsec:lyman}) of the populations generating the source subtracting CIB.
%
%\item We then propose a method using the 2-band tomography with NIRCam to reconstruct the history of emissions directly by using the expected Lyman-break shifted into the adjacent NIRCam filters from each range of redshifts (Sec. \ref{subsec:tomography}).
%
%\item We finally address and evaluate the levels produced by foregrounds and systematics in the measurement (Sec. \ref{sec:foregrounds}).
%\end{enumerate}
%For completeness, the CIB fluctuation situation is briefly reviewed in Sec. \ref{sec:cib_status} after providing the general background on the CIB fluctuations in Sec. \ref{sec:cib_defs}.

Herebelow, we adopt a spatially flat universe with standard cosmological parameters for matter density, baryon density parameters and the Hubble constant: $\Omega_{\rm m}=0.28$, $\Omega_{\rm bar}=0.045$, $H_0=70$\,km/s/Mpc. Wherever necessary, we also adopt $\sigma_8=0.9$ and denote $h=H_0/(100$\,km/s/Mpc). For reference, with these parameters the coordinate distance to $z=10,15,20$ becomes 9.7, 10.5 and 11 Gpc, while the luminosity distance takes the values of 106.5, 168.3, 231.7 Gpc.

\section{The current status of CIB fluctuations}
\label{sec:cib_status}
%VB: I ve fixed a few typos throughout Section 2.

The general formalism 
of source-subtracted CIB fluctuations, its components and relation to the underlying populations is reviewed in \cite{review}.
We refer to the Appendix for definitions and brief description of the quantities used to characterize and describe source-subtracted CIB fluctuations in this
paper: 1) shot-noise 
power ($P_{\rm SN}$), which dominates small angular scales, 2) $P_{\rm clus}$ the power spectrum of the clustering component which appears dominant at angles 
much larger than the beam and is defined by the clustering of the emitting sources and their brightness evolution, 3) the cross-power $P_{12}$ between two 
different wavelengths $\lambda_1,\lambda_2$, 4) the coherence ${\cal C}_{12}\equiv \frac{P_{12}^2}{P_1P_2}$, and 5) other key quantities. The mean square fluctuation 
on angular scale $2\pi/q$ is defined as $\frac{q^2P}{2\pi}$ and the relation between the cyclical wavenumber $q$ and multipole $\ell$ is 
$\ell\simeq q$(in radian$^{-1}$). It is important to reiterate that coherence {\it must be always below unity}, 
${\cal C}\leq 1$: namely, in any two wavelength bands, populations cannot be more coherent with each other than they are with themselves.

\subsection{Current measurements}

Here we summarize the current status of the CIB fluctuation measurements and their theoretical implications. For reasons that will become apparent from the discussion below, we divide the current CIB measurements into two categories: 1) at wavelengths 2--5 \micron\ using {\it Spitzer} and {\it AKARI} results, and 2) 1--1.6 
micron from 2MASS, NICMOS and CIBER.

\subsubsection{{\it Spitzer} and {\it AKARI}:  CIB fluctuations over 2--5 \micron}

Using deep integration data from {\it Spitzer}/IRAC this team has developed the methodology to robustly identify source-subtracted CIB fluctuations at 3.6 
and 4.5 \micron. The methodology is discussed in detail in \cite[][hereafter AKMM]{akmm} and utilizes the self-calibration technique of \cite{fixsen} to produce the images with the fidelity
required to probe the faint CIB signals expected from the first stars era. The assembled maps are clipped of resolved sources whose extended outer parts 
are then removed using a variant of the CLEAN \citep{clean} procedure down to the specified level of shot-noise with the noise power spectrum evaluated
directly from the time-differenced $A-B$ maps. This allowed to identify and measure the large-scale
fluctuation from clustering at the various shot-noise levels. In a later and independent analysis using {\it AKARI} satellite \cite{akari} have confirmed these results and importantly extended the measurement to $\sim 2$\micron. 

\begin{figure}[ht!]
%\plotone{fig_cib.eps}
\includegraphics[width=6.5in]{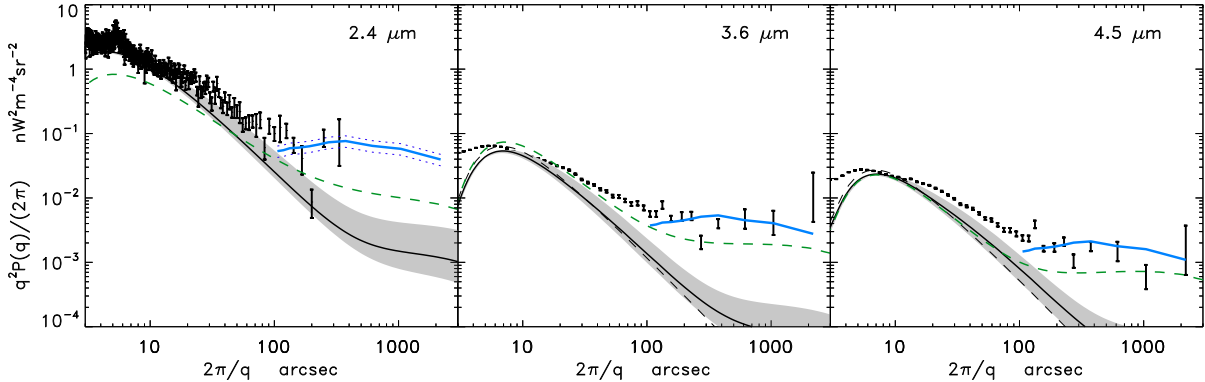}
\caption{\small Mean squared source-subtracted CIB spatial fluctuations vs angular scale at 2.4, 3.6, 4.5 \micron. Black dashes show the shot-noise component remaining in the IRAC maps. Black solid line shows the ``default'' reconstruction of the contribution from remaining known galaxy populations with uncertainty shown by shaded area from \cite{kari}. Blue solid line shows the template of the high-$z$ $\Lambda$CDM model from \cite{seds} which used the analytical fit from \cite{sugiyama}; it is extrapolated to the 2.4 \um\ data from the IRAC channels using the $\lambda^{-3}$ energy spectrum as proposed in \cite{akari} with the uncertainty marked with blue dots. Green line shows a proposal by Cooray et al. (2012) of the intrahalo light model for the fluctuations. {\bf Left}: AKARI results from \cite{akari}. {\bf Middle and right} panels show the RAC-based measurements from \cite{seds}. 
}
\label{fig:cib}
\end{figure}

Fig. \ref{fig:cib} sums up the current measurements of the source-subtracted CIB fluctuations from {\it AKARI} and {\it Spitzer} that cover the range of 2 to 5 \micron. 
These results, and others, on the source-subtracted CIB fluctuations now appear to say that:
%\begin{enumerate}
%\item There are significant source-subtracted CIB fluctuations with only low levels of the shot noise, but 
%substantial clustering component; these diffuse maps do not correlate with the mask or with the removed sources,
%\item These 
%fluctuations do not correlate at any detectable level with the optical ACS sources out to AB magnitude $>28$,  
%\item AKARI-based analysis extended the 
%measurement to 2.4 mic and suggests a Rayleigh-Jeans type SED of these sources, 
%\item Our SEDS-based measurements extended the 
%CIB angular spectrum to ~1 deg, 
%\item Cross-correlation analysis with X-ray data 
%suggests that unresolved CXB and CIB are coherent at a remarkably high level; this is true for soft X-ray band (0.5-2 keV), with no 
%coherent signal appearing at harder X-ray bands, 
%\item These fluctuations cannot be accounted for by known sources (and they have the 
%spatial template consistent with origin in high-z LCDM distributed populations). Any interpretative effort must account for all of these.
%\end{enumerate}
\begin{enumerate}

\item The residual CIB fluctuations at 3.6 and
4.5 \um\ have two components: i) on scales
$>30''$ the fluctuations are dominated by the
clustering of the sources,
ii) smaller scales appear presently dominated by the
shot noise from the remaining (unresolved) sources.

\item  {\it AKARI}-based analysis extended the CIB fluctuation measurement to 2.4 $\mu$m and suggests an approximately Rayleigh-Jeans type spectral energy 
distribution ($\nu I_\nu \propto\lambda^{-3}$) of the sources producing them \citep{akari}.

\item The CIB fluctuation spectrum has now been measured out to $\sim 1^\circ$; it appears consistent with the high-$z$ 
$\Lambda$CDM clustering power spectrum and is the same (within the uncertainties) in different directions on the sky as
required by its cosmological origin \citep[hereafter K12]{seds}.

\item The diffuse maps, from which the fluctuations have been measured, do not correlate with the mask or with the extended parts of 
the removed sources 
% \citep[hereafter KAMM1, AKMM]{kamm1,akmm}
(KAMM1, AKMM).

\item The clustering component of the fluctuations does not correlate at any detectable level with the optical ACS sources out to $m_{\rm AB} \!>$28 and $<$0.9$ \mu$m
% \citep[hereafter KAMM4]{kamm4}. 
(KAMM4). 
In other words, the Lyman break wavelength is red-shifted
beyond the longest ACS wavelength (0.9\,\um) with the detected CIB fluctuations arising within the first Gyr of the universe, unless the CIB anisotropies
come from more local but extremely faint ($L<2\times 10^7L_\odot$) and so far unobserved galaxies. 

\item The source-subtracted CIB fluctuations have been shown to be highly coherent with the soft [0.5--2] keV unresolved X-ray 
cosmic background (CXB) with no detectable cross-correlation appearing at harder X-ray energies \citep[hereafter C13]{c13}. 
Such a population is expected to contribute to the CXB at $z>$7, with first black holes 
growing rapidly to form the observed massive QSOs at $z\sim 5-6$.  

\item It is now firmly established that the clustering component of the source-subtracted 
CIB fluctuations strongly exceed what can be produced by known galaxy populations after extrapolating to lower luminosities \citep[][see Fig. \ref{fig:cib}
in this paper]
{kari} and the same applies to the discovered CXB-CIB coherence \citep{kari2}.

\item The  source-subtracted CIB fluctuations appear with low shot noise, while exhibiting a substantial clustering component 
\citep[hereafter KAMM3]{kamm3}, which indicates the origin of the clustering component in very faint populations (currently $\lesssim 20$ nJy).

\item The clustering component does yet appear to start decreasing as the shot noise is lowered from 7.8 hr/pix to $>21$ hr/pix exposures 
(see below in Sec. \ref{subsec:shotnoise}).
\end{enumerate}

\begin{figure}[ht!]
%\plotone{fig_cib.eps}
\includegraphics[width=4in]{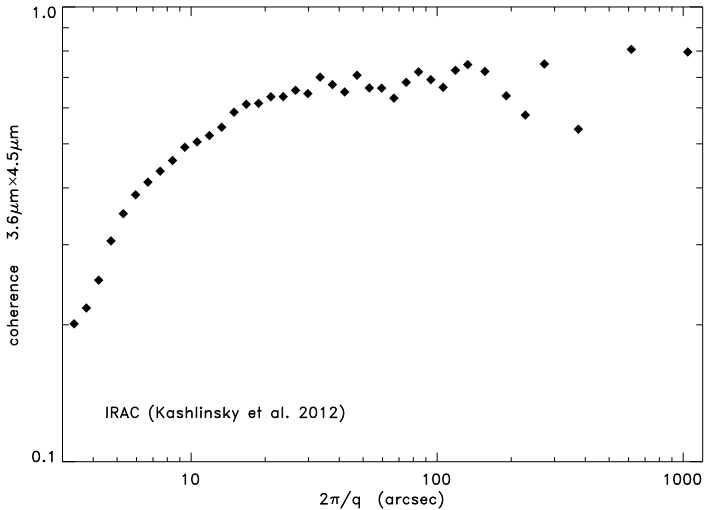}
\caption{\small Coherence between the 3.6 and 4.5 \micron\ CIB fluctuation data from \cite{seds}. As required mathematically the coherence is always below unity, independently of errors. At small scales the CIB fluctuations are dominated by shot-noise and non-linear clustering terms from remaining galaxies, which are removed independently at each band. Hence the coherence at small angular scales is low rising gradually to its value close to unity at larger angular scales, which are dominated by the new populations not removed from the data.
}
\label{fig:coherence_seds}
\end{figure}

Fig. \ref{fig:coherence_seds} shows the coherence between the source-subtracted CIB fluctuations from {\it Spitzer} at 3.6 and 4.5 \micron\ \citep{seds}. The figure shows
a consistent picture of the CIB measurements obtained with IRAC: 1) the coherence is always bounded from above by unity, 2) with small scales dominated by the remaining 
known galaxy populations, which are independently removed at the two bands and so are less coherent than 3) the large scales, where new populations dominate, which  cannot be resolved with {\it Spitzer} and, hence, were not yet removed.

If at high $z$, the CIB excess of $\sim 1$ nW m$^{-2}$ sr$^{-1}$ associated with these new populations is within the current constraints from $\gamma$-ray absorption,
which dictate an  upper CIB limit of 7 and 4.6 nW m$^{-2}$ sr$^{-1}$ at 3.6 and 4.5 \micron\ \citep{hess}.

\subsubsection{2MASS, {\it HST} and CIBER: 1--1.6 \micron}

\begin{figure}[ht!]
%\plotone{fig_cib.eps}
\hspace{-2.5cm}
\includegraphics[width=8in]{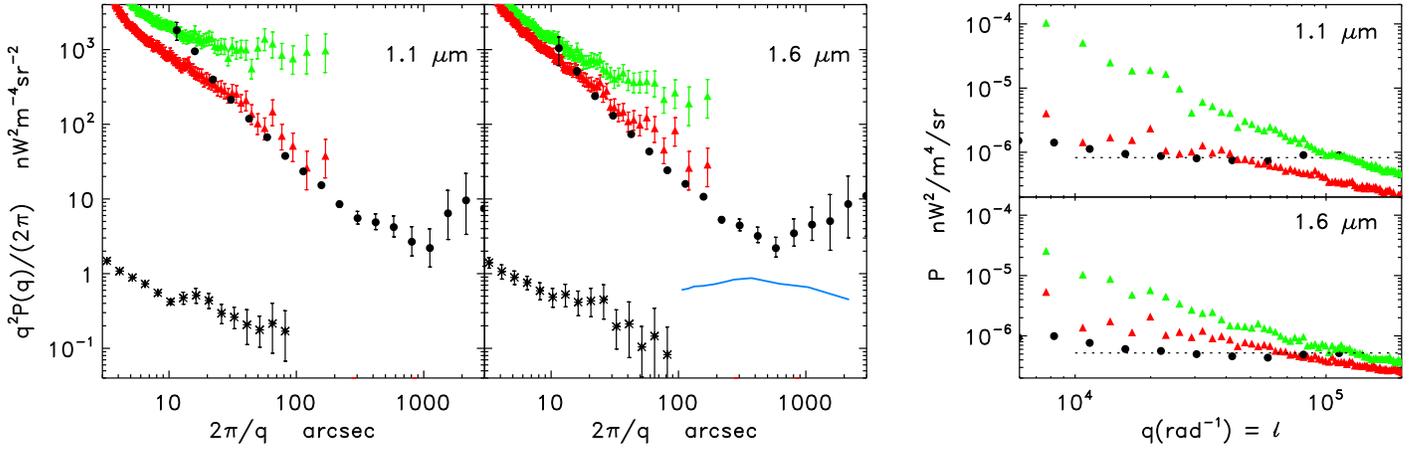}
\caption{\small Red and green triangles show CIB fluctuations from 2MASS analysis using the deepest and least deep removal threshold in the seven patches from \cite{2mass}. Asterisks are NICMOS-based CIB fluctuation results from \cite{thompson,thompson2}. Blue line is the $\Lambda$CDM clustering normalized to the {\it Spitzer} measurements and then extrapolated as $\lambda^{-3}$ to 1.6\micron.  Filled circles show the fluctuations from CIBER \citep{ciber}. Right panels show the power spectra from 2MASS \citep{2mass,odenwald} and CIBER \citep{ciber} zooming in on small angular scales to demonstrate the apparent inconsitency between 2MASS and CIBER: whereas in the 2MASS study, which removed deeper sources, non-linear galaxy clustering is clearly detected and still dominates the shot noise, that component somehow disappeared in the shallower CIBER study where only shot noise appears to remain. To ease on clarity the error bars, shown in the left panels, are not plotted on the right. For reference, the 2MASS analysis removed sources down to Vega magnitudes of 19.2 (green symbols) to 20 (red) at 1.1 \micron, and 18.7 (green) to 19.2 (red) at 1.6 \micron, which is fainter than the CIBER removal
at the same wavelengths. The best fit white noise for CIBER numbers is shown with horizontal dotted lines.
}
\label{fig:nicmos}
\end{figure}

At wavelengths shortward of 2\micron\ source-subtracted fluctuations have been measured in 2MASS \citep{2mass,odenwald} and HST/NICMOS \citep{thompson,thompson2} surveys. These were recently supplemented with the results from CIBER \citep{ciber}. While the 2MASS and NICMOS are 
mutually consistent, they both disagree with the CIBER published results, which also appear to be internally inconsistent as discussed below.
Fig. \ref{fig:nicmos} summarizes the current measurements of the source-subtracted CIB fluctuations at 1.1 and 1.6 \micron. 

We now discuss each of the measurements in chronological order:

$\bullet$ {\bf 2MASS} standard star survey was used by \cite{2mass,odenwald} to evaluate source-subtracted CIB fluctuations in seven square patches of 
$512''\times 512''$ out to angular scales $2\pi/q\sim 200''$ after galaxies have
been removed down to Vega magnitude of $\sim 18.7-20$ (AB magnitudes $\sim 20-21$) in the J, H, K$_s$ photometric bands. The fraction of sky
removed with the resolved sources was less than  10\% in the analysis. They detect CIB 
fluctuations with the clearly non-white-noise spatial spectrum produced by (evolving) non-linear clustering from remaining galaxies with $P\propto q^{-n}$ and
the slope varying between $n=1.4$ for the shallowest removal and $n=0.6$ for the deepest. Figure 2 of \cite{2mass} shows the evolution of the slope (and amplitude) of the non-linear 
clustering component with increasing depth and demonstrates that {\it at these thresholds and bands the non-linear clustering still clearly dominates the shot-noise 
term from remaining galaxies at all scales exceeding $\sim 1''$}. Red triangles in Fig. \ref{fig:nicmos} show the source-subtracted CIB fluctuations from \cite{2mass} 
at the deepest thresholds probed with that data and green triangles correspond to the least deep of the seven 2MASS CIB patches with sources
excised $\sim 0.5$ mag brighter.

$\bullet$ {\bf NICMOS}-based CIB fluctuations  at 1.1 and 1.6\micron\ were studied by \citet{thompson,thompson2} after progressively eliminating galaxies down to much fainter fluxes than in 2MASS using data from the NUDF field. After removing identified sources down to  AB magnitude of $\sim 27.7$
\citep[$5\sigma$,][]{thompson2005}, 93\% of the map
remained for robustly direct power spectrum evaluation. The remaining CIB fluctuations are plotted with black asterisks in Fig. \ref{fig:nicmos}. At the magnitude
limits corresponding to the depth reached in the 2MASS study of \cite{2mass,odenwald}, the NICMOS results agree very well with that study and both show the
non-linear clustering from remaining known galaxies, which dominates the shot-noise component.

\begin{figure}[ht!]
%\plotone{fig_cib.eps}
\includegraphics[width=4in]{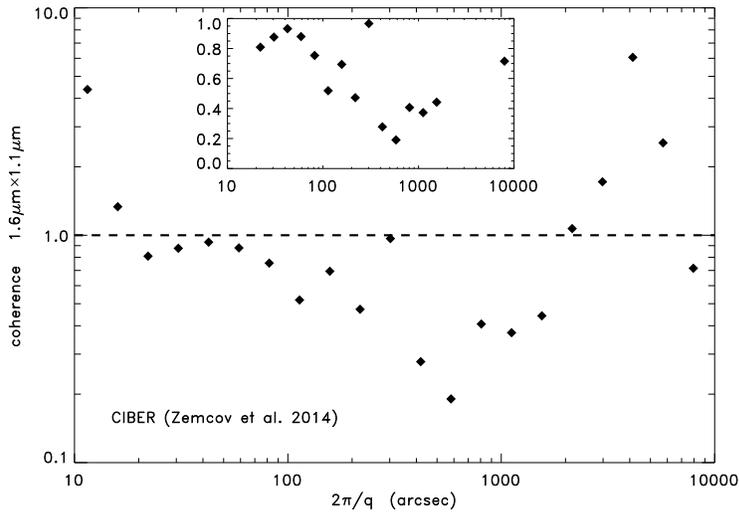}
\caption{\small The coherence from data in Fig. 1 of  the \cite{ciber} CIBER results at 1.1 and 1.6 \micron. Much of the data is above the upper bound of unity, marked with horizontal dashed line. At face value this 
would imply the impossible situation of CIB sources at two different bands being more coherent than they are with themselves. The inset shows the range of 
$0< {\cal C} <1$, where no systematic trend is present in contrast to the 
{\it Spitzer} measurements displayed in Fig. \ref{fig:coherence_seds}. At large scales \cite{ciber} claim their measurements are dominated by Galactic
cirrus which should lead to coherence close to, but smaller than, unity contrary to what is seen in their data shown in the inset.
}
\label{fig:coherence_ciber}
\end{figure}
$\bullet$ {\bf CIBER} experiment has recently suggested CIB fluctuations at 1.1 and 1.6 \micron\ shown in Fig.  \ref{fig:nicmos}  with filled circles \citep{ciber}.  After removing galaxies to Vega magnitude of 17.5 at 3.6\micron\ only $\sim 30\%$ of sky was left for Fourier 
analysis. 
It is claimed in the paper that the data strongly favor the intrahalo light model and, because they do not fit the epoch-of-reionization (EoR) modeling 
of \citet[][a]{cooray-eor}, it is further claimed that it rules out high-$z$ origin of the fluctuations. No accounting is made in the interpretation of 
the measured high coherence of  source-subtracted CIB with unresolved soft-band CXB, which \cite{ciber} peculiarly termed ``partial correlation'', or the lack of correlation with extended wings of subtracted galaxies. As shown in Fig.  \ref{fig:nicmos}  there appears to be a conflict between the CIBER data and the earlier and mutually consistent analysis of CIB fluctuations at the same
wavelengths from 2MASS \citep{2mass,odenwald} and NICMOS \citep[]{thompson}. 2MASS data clearly show that non-linear clustering, with a slope distinctly different from shot-noise, still dominates out to at least fainter magnitudes than probed by \cite{ciber}. However, no such component appears in 
the data from the CIBER analysis, which on scales below $\lsim 100''$ shows no deviations from shot-noise with constant $P(q)$. Although the discrepancy is obvious, we quantify it as follows: we evaluate the best fit amplitude of the shot-noise power, $P_{\rm SN,CIBER}$, implied by the CIBER data (shown with dotted lines in the right panel of Fig. \ref{fig:nicmos}), and then evaluate the value of the reduced $\chi^2_{dof}(2MASS|P_{\rm SN,CIBER})$ for the $N_{dof}=63$ data points of 2MASS for each of the wavelengths and removal thresholds (green and red in the figure). The resultant $\chi^2_{dof}(2MASS|P_{\rm SN,CIBER})$ is between 4.4 and 130.6 in all cases and implies zero probability that CIBER results implying such white-noise component are consistent with the 2MASS CIB data (for reference the 99.9\% confidence level for $N_{dof}=63$ is bounded by $\chi^2_{dof}<1.64$). 
It is unclear how, in the presence of such
significant fog produced by this diffuse CIB, \citet{thompson,thompson2} could remove individual sources down to AB mag of $\sim 28-29$ to get to the 
data shown with asterisks in the figure. Additional questions appear after inspecting the coherence reconstructed with the CIBER analysis measurements 
which is shown in  Fig. \ref{fig:coherence_ciber}.  As the figure shows the coherence is %- remarkably -
greater than unity over much of the data. This appears paradoxical. 
A possibility arises that this occurred because of correcting
for the noise bias, which \cite{ciber} evaluated indirectly from a model. However, this possibility appears to be in conflict with the magnitude of the upward 
fluctuation and should apply to high multipoles. The excess coherence at low multipoles should be much less affected unless the noise
dominates the large scales as well leaving the entire signal at most  at a modest signal/noise level. Even where the 
coherence  is below unity it does not display any systematic behavior in stark contrast to the {\it Spitzer}-based results in Fig. \ref{fig:coherence_seds}. 
This demonstrates that a possibility must be considered  that these 
CIBER results cannot be of cosmological/astrophysical origin\footnote{A possible flaw may arise from the application of the MASTER formalism to the 
Fourier-based analysis of such heavily masked sky. This 
could have been checked with the correlation function analysis as in \cite{k2007}, which was not shown in \cite{ciber}.}.

Constraints from the observed $\gamma$-ray absorption \cite{hess} give upper limits of 17 and 14 nW m$^{-2}$ sr$^{-1}$ at 1.1 and 1.6 \micron. This is to be 
compared with the resolved CIB from faint galaxy counts estimated by \cite{keenan} at $11.7^{+5.6}_{-2.6}, 11.5^{+4.5}_{-1.5}$ nW m$^{-2}$ sr$^{-1}$ at these bands. Thus no
more than $\sim 4-8$ nW m$^{-2}$ sr$^{-1}$ currently appears feasible in CIB excess at these wavelengths.

\subsection{Remaining ``ordinary'' populations}
\label{subsec:og}

HRK12 provide a robust heuristic way of reconstructing CIB fluctuations from galaxy populations spanning wavelengths from UV to mid-IR out to $z\sim 6$. The assembled database now covers over 340 luminosity function (LF) surveys \citep{hk12,kari2}, and the HRK12 methodology allows to fill in the redshift cone with known galaxies across the required wavelengths. The reconstructed populations are on average described by the ``default'' model of HRK12 and are bracketed by the high- and low-faint-end (HFE and LFE) extremes of the LF extrapolation to very faint luminosities still consistent with the LF surveys. The accuracy of the reconstruction is verified by the remarkably good fits to the newly measured, and much deeper than before, IRAC counts \citep{ashby1,ashby2}.  As Fig. \ref{fig:cib} shows the galaxies remaining in the {\it Spitzer} and {\it AKARI} data account well for the shot-noise term, but
appear to produce too little CIB to explain the large-scale clustering component, which then must arise in new populations. 

This reconstruction is adopted in our discussion below. 
% with the HFE/LFE modes bracketing the systematic uncertainties in the CIB produced by the known galaxy populations.

%Discuss per HRK12. Fig. \ref{fig:dndm} shows the counts reconstructed per HRK12 and the CIB from known populations fainter than a given magnitude. 

\subsection{Energy requirements of the CIB fluctuations and new populations}

\subsubsection{General considerations}

The bolometric flux produced by populations containing a fraction $f$ of the baryons in the Universe after they have converted their mass-energy into radiation with efficiency $\epsilon$ is given by:
\begin{equation}
F_{\rm tot} \simeq \frac{\epsilon f}{z_{\rm eff}}\; \frac{c}{4\pi}\rho_{\rm bar} c^2 \simeq 
9.1\times 10^5 \frac{\epsilon f}{z_{\rm eff}} \;\frac{\Omega_{\rm bar}h^2}{0.0227} \;\; {\rm nW\ m}^{-2}\ {\rm sr}^{-1}
\label{eq:cib_excess_th}
\end{equation}
where $z_{\rm eff}\equiv 1/\langle(1+z)^{-1}\rangle$ is a suitably averaged effective redshift. Here the $\rho_{\rm bar} = \Omega_{\rm baryon} \frac{3H_0^2}{8\pi G}$ is the comoving baryonic density and the redshift factor accounts for the radiation energy density decreasing with expansion as $\propto (1+z)^{-4}$ vs. the matter density $\propto (1+z)^{-3}$.

Power spectrum of CIB fluctuations from the new populations can be characterized with an amplitude at some fiducial scale and a template. The CIB fluctuation at, say $\sim 5'$ which was used for such normalization in K12, as measured with {\it Spitzer} and {\it AKARI} can be integrated to give the net integrated CIB flux fluctuations over the wavelengths of the detections leading to:
\begin{equation}
%\delta F(2-5\mic) = \int_{\lambda_{2,IRAC}}^{\lambda_{1,AKARI}}\; 
\delta F_{2-5\mic}(5') = \int^{IRAC}_{AKARI}\; 
\left(\frac{q^2P_\lambda}{2\pi}\right)^{1/2}\; \frac{d\lambda}{\lambda} = \delta F_{4.5\mic}(5')\;\left(\frac{(4.5/2.4)^\alpha -1}{\alpha}\right) \simeq 0.09\ {\rm nW\ m}^{-2}\ {\rm sr}^{-1}
\label{eq:cibfluc_bol}
\end{equation}
where $\nu \delta I_\nu \equiv [\frac{q^2P_\lambda}{2\pi}]^{1/2}$ is the CIB flux fluctuation in nW m$^{-2}$ sr$^{-1}$ and we assume per Fig. \ref{fig:cib} that it 
scales with wavelength as $\nu \delta I_\nu \propto \lambda^{-\alpha}$ with $\alpha \simeq 3$; for $\alpha=-2$ the above expression gives $\delta 
F(2-5\mic) \simeq 0.065$ nW m$^{-2}$ sr$^{-1}$. In the above expression we have taken the {\it AKARI} and {\it Spitzer}/IRAC filters to have the integrated range of 
2--5 \um\ as per Fig. \ref{fig:NIRCam} and the ``nominal'' central values of the filters were plugged into the middle expression above. Populations at high $z
$, which are strongly biased and span a short period of cosmic time, are expected to produce $\Delta_{5'}\equiv \delta F/F(2-5\mic)\sim 10\%$ relative CIB fluctuations on 5 arc minute scales. 
Such populations would then require producing about $F_{\rm CIB} \sim 1$ nW m$^{-2}$ sr$^{-1}$ in the integrated flux at near-IR wavelengths $2.4-4.5\mic$ 
(KAMM3). If populations at lower redshifts and spanning longer cosmic periods with less biasing were to explain the measurement, they would require 
production of much larger CIB, which would be comparable to the net CIB flux at 3.6 and 4.5 \mic\ from all the known galaxies out to $m_{\rm AB} \gtrsim 
26$ \citep{fazio,review,ashby1}. If its $\lambda^{-3}$ SED extends to 1.6 \um, the integrated CIB fluctuation excess from the new populations would be higher 
at $\delta F(5')\sim 0.3$ nW m$^{-2}$ sr$^{-1}$ over the 1.6--5 \um\ range leading to $F_{\rm CIB}(1\!-\!5\micron) \lsim 3$ nW m$^{-2}$ sr$^{-1}$ still within the errors of the current 
conservative CIB measurements of \cite{thompson}.

The shot-noise power can also be written
as $P_{\rm SN}\simeq S_\nu(\bar{m}) F_{\rm tot}(> m_{\rm lim})$,
where $F_{\rm tot}(> m_{\rm lim})$ is the net CIB flux from the
remaining sources. Hence we express the shot noise power in units of nJy$\cdot$nW m$^{-2}$ sr$^{-1}$ such that a population producing a mean CIB level of 1 nW m$^{-2}$ sr$^{-1}$ has typical flux of $S=P_{\rm SN}$ nJy. 
The measured levels of the shot-noise do not currently reach the regime of 
 attenuation of the large-scale fluctuation from clustering; the point where this happens would then probe the flux of the typical sources responsible for this CIB component. The deepest current limits reached are $P_{\rm SN}=(26,14)$ nJy$\cdot$nW m$^{-2}$ sr$^{-1}$ at (3.6,4.5)\um. Since $P_{\rm SN}\sim S F_{\rm CIB}$, these limits coupled with the above, imply the upper limits on the typical fluxes of the sources producing them \citep{kamm3}
 \begin{equation}
 S_{(3.6,4.5)\micron}\; \lesssim \; (26,14)\; \frac{F_{\rm CIB}}{\rm nW/m^2/sr} \;\; {\rm nJy}
 \label{eq:sfromsn}
 \end{equation}
 Such objects would have $m_{\rm AB} \gtrsim 28-29$ and may have fluxes well below what can be probed individually with the {\it JWST}. At wavelengths approaching 1\micron, where observations of $\gamma$-ray absorption restrict any CIB excess $F\simeq Sn_2$ over that from known galaxies to be at most a few nW m$^{-2}$ sr$^{-1}$, the same argument implies the new sources to appear at fluxes 
 \begin{equation}
 S_{\lambda\sim 1\micron} < 10 {\rm nJy}\; \left(\frac{n_2}{10^{11}{\rm sr}^{-1}}\right)^{-1}\left(\frac{F_{\rm CIB}(\lambda\sim 1\micron)}{3\;{\rm nW/m^2/sr}}\right)\left(\frac{\lambda}{1\micron}\right)^{-1} 
 \label{eq:s_1micron}
 \end{equation}
This corresponds to AB magnitudes fainter than $m_{\rm AB}\gsim 29-30$ 
 around 1\micron\ being in the range currently not accessible to galaxy counts surveys.
 
 A {\it lower} limit on the projected surface density, $n_2$, of these new sources can be estimated in a similar manner by writing the shot-noise power from these sources as $P_{\rm SN}^X \sim F_{\rm CIB}^2/n_2$ (KAMM3). Since the measured shot-noise at $P_{\rm SN} \sim 10^{-11}$nW$^2$/m$^4$/sr (KAMM2) represents an upper limit on the shot-noise from the new populations, their number per {\it JWST} beam of area $\omega_{NIRCam}$, ${\cal N}_2$, can be bounded from {\it below} as:
 \begin{equation}
{\cal N}_2 \gtrsim 0.1\; \left(\frac{F_{\rm CIB}}{1\;{\rm nW/m^2/sr}}\right)^2\;\left(\frac{P_{\rm SN}^X}{10^{-11}{\rm nW^2/m^4/sr}}\right)^{-1} \; \frac{\omega_{NIRCam}}{10^{-12}\;{\rm sr}}
\label{eq:n2_kamm3}
\end{equation}
Confusion intervenes when there are less than 50 beams/source \citep{condon}, so the above shows that the bulk, perhaps all, of the new populations would be well within the confusion noise of the NIRCam beam.
 
\subsubsection{High-$z$: emissions from first star epochs}
\label{subsec:cib-hiz}
%VB: Sasha, here I ve included some more detailed edits. I ve also included
%    a few questions for you.

The expectation is that at any given time the first collapsing halos will contain a mixture of stellar populations: Pop~III stars, formed out
of still pristine gas, and characterized by an initial mass function
(IMF) that is still very uncertain, but is thought to be biased towards high masses (reviewed in Bromm 2013); and Pop~II stars, formed
out of already metal-enriched material, described by a normal,
Salpeter-like IMF 
%Theoretically the total star formation rate density (SFRD)
%at these redshifts ($10<z<20$) is thought to be $\sim 10^{-2}M_{\odot}$\,Mpc$^{-3}$\,yr$^{-1}$,
%whereas the, again highly uncertain, Pop~III contribution has been estimated to peak at
%$\sim 10^{-4} - 10^{-3} M_{\odot}$\,Mpc$^{-3}$\,yr$^{-1}$
(e.g., Bromm \& Loeb 2006; Campisi et al. 2011).
In addition to these stellar sources, the production of H and He ionizing
photons from accreting BHs could play a major
role in shaping the high-$z$ IGM (e.g., Mirabel et al. 2011; Jeon et al. 2012, 2014).
Such BH activity could originate in either the relics of massive Pop~III or
Pop~II stars, or in the direct collapse of primordial gas clouds to
$\sim 10^5 M_{\odot}$ massive BHs (Bromm \& Loeb 2003; Volonteri \& Bellovary 2012). Accreting BH populations would also contribute to the cosmic X-ray background (CXB), due to their high-energy, non-thermal spectral component (e.g. Ricotti et al. 2005, Yue et al. 2013b). Cooray et al. (2012a) and Yue et al. (2013a) argue, that extrapolating the $z\sim 8$ UV LF to much higher redshifts would lead to fairly low levels of the CIB from the purely stellar component.  Yue et al. (2013b) ascribe the CIB fluctuations and their coherence with the soft X-ray CXB (C13) to direct collapse black holes, which with the right absorption properties ($N_{\rm H}$) at $z\gtrsim 15$ can explain  both the CIB fluctuations and the CIB-CXB coherence.

We now provide a general constraint that the sources at high $z$ have to satisfy if they are to explain the CIB fluctuation measurement. If the faint contributors to the source-subtracted CIB fluctuations discussed above lie at the very early epochs associated with the first stars, galaxies and BHs,
%(reviewed in Bromm \& Yoshida 2011), 
then they have to contribute as follows:
\begin{itemize}
\item {\bf Massive stars}, such as hypothesized to dominate the first stars era, are radiation-pressure dominated, and emit nearly at the Eddington limit. In addition, they are close to fully convective with the entire stellar mass taking part in the hydrogen burning \citep{bromm01,schaerer}. This leads to the  high overall efficiency of $\epsilon_*=0.007$. Hence, a fraction of
\begin{equation}
f_{P3}\sim 1.4\times 10^{-3}\;\; \left(\frac{z_3}{10}\right)\; \left(\frac{\Delta_{5'}}{0.1}\right)^{-1},
\label{eq:f_star}
\end{equation}
or about 0.1\% of the entire baryons, would have had to go through very massive stars at high redshifts in order to explain the flux level of $\sim 1$ nW m$^{-2}$ sr$^{-1}$ implied by Equ.\,\ref{eq:cib_excess_th},\ref{eq:cibfluc_bol}, if the CIB fluctuations were produced by these stars at high $z$. Note that this argument is approximately independent of metallicity in as much as the convection encompasses the entire star. If the excess CIB extends to 1.6 \um with the $\lambda^{-3}$ SED, the fraction in eq. \ref{eq:f_star} would rise to $\sim 0.6\%$, which is still an order of magnitude below what \cite{madau-silk} deemed problematic.
\item For {\bf normal Pop~II stars}, described by a Salpeter IMF, the effective efficiency is an order of magnitude lower since only a small core burns hydrogen. This leads to the overall efficiency being at least an order of magnitude lower than that of massive stars
%the hydrogen burning, 
requiring that if such populations were to explain the flux level  of $\sim 1$ nW m$^{-2}$ sr$^{-1}$ in Equ.\,\ref{eq:cib_excess_th},\ref{eq:cibfluc_bol}, they had to contribute 
\begin{equation}
f_{P2}\gtrsim 0.01 \left(\frac{\epsilon}{7\times10^{-4}}\right)\left(\frac{z_3}{10}\right)\; \left(\frac{\Delta_{5'}}{0.1}\right)^{-1}. 
\end{equation}
If the excess CIB extends to 1.6 \um with the $\lambda^{-3}$ SED, the fraction of normal stars required to explain such CIB fluctuations would rise to $\gtrsim 6\%$.
\item A qualitatively different contribution to the CIB arises from the
death in {\bf energetic SNe} of massive Pop~III and Pop~II stars at high $z$.
Here, the kinetic energy of the explosion, and that from the radioactive decay 
of the newly synthesized Ni will be converted into radiation (e.g., Ritter et al. 2012).
However, the
overall efficiency for this conversion is extremely small. As a representative
example, if we assume a kinetic explosion energy of $E_{\rm SN}\sim 10^{51}$\,erg,
and compare this with the rest energy of a $M_{\rm SN}=10 M_{\odot}$ progenitor,
typical for standard core-collapse SNe, we find an overall efficiency of
$\epsilon_{\rm SN}\simeq E_{\rm SN}/(M_{\rm SN} c^2)\lesssim 10^{-4}$.
It is thus clear that SN feedback will not be an important source of ionizing radiation, compared
with massive stars and BHs (see below), and we are thus justified in neglecting their
contribution to the CIB.
\item In case of {\bf accreting black holes}, $\epsilon_{\rm BH}$ can be as high as 0.4 for maximally-rotating Kerr-holes, and in any case is much greater than that of H-burning. Thus, BHs can contribute significantly even with much smaller fraction than stars. Namely
\begin{equation}
f_{\rm BH} \sim 5\times 10^{-5}\;\; \left(\frac{z_3}{10}\right)\; \left(\frac{\Delta_{5'}}{0.1}\right)^{-1} \left(\frac{\epsilon}{0.2}\right)^{-1}
\label{eq:f_bh}
\end{equation}
\item {\bf Dense stellar systems} (DSS), where direct stellar collisions may occur as a result of stellar dynamical evolution in the early Universe \citep{kr83} releasing large amounts of energy would provide an additional component to the net CIB balance. This contribution is hard to quantify a priori in a robustly model-independent way, but may contribute significantly to the overall CIB budget. The point, however, is that DSS's  ultimately lead to formation of a single very massive object and BH formation and can achieve radiative efficiency comparable to that of BH accretion \citep{begelman-rees}, so even a small fraction of early population systems, if they evolve to reach the DSS stage, would require a low fraction, $f_{\rm DSS}$, comparable to $f_{\rm BH}$ if they are to explain at least part of the CIB excess.
\end{itemize}
This discussion suggests the proportions of the baryons in the various different sources required to explain the CIB fluctuation measurements.  In reality any baryonic populations will be described by the high-$z$ $\Lambda$CDM template (only weakly dependent on redshift at $z\gsim 10$)  shown in blue in Fig. \ref{fig:cib}. The normalization will then be fixed by a mixture of the various contributors with degenerate proportions. Once they are normalized to the given overall CIB, their spatial fluctuation will resemble the blue line in Fig. \ref{fig:cib} and later in the paper. We will thus refer to the "high-$z$ $\Lambda$CDM" as a general and wide class of high-$z$ models of sources of the various nature and note that the current CIB measurements alone cannot break the degeneracy in their fractions. However,  measurements of the CXB-CIB coherence provide important insight into this questions and the current results of C13 imply that accreting BHs are responsible for a significant fraction, perhaps most, of the {\it Spitzer} CIB fluctuation signal (KAMM1-4, AKMM and K12).

\subsubsection{Intermediate-$z$: IHL}
\label{subsec:ihl}

An alternative to the high-$z$ explanation has been recently proposed by \cite{cooray12}, 
where the CIB signal originates at $z\sim$ 2--4 (green dashes in Fig. \ref{fig:cib} and Fig. \ref{fig:clustering_sn_irac} below) in stars tidally stripped off parental galaxies, 
the so-called intrahalo light (IHL); the intrinsic faintness of
these features keeps them out of reach of direct telescopic searches.  

The difficulties that this scenario faces when 
confronted with the entirety of the CIB measurements, elucidated in \citet[][see Sec. 4.3 there]{kari2}, were not addressed in the later publication of the
same team \citep{ciber}. 
%An alternative origin was recently proposed by \cite{cooray12} in the form of starlight scattered between galaxies as the result of mergers and tidal interactions. 
%In this model, significant amount of light can avoid being masked along with resolved galaxies and produce large scale anisotropies. This model is also the favoured explanation of the CIBER measurement \citep{ciber} where the intrahalo light model was updated. 
In that most recent form, IHL arises mostly at $z<0.5$. However, there remain a number of observational and theoretical challenges which make the IHL 
interpretation problematic. First and foremost, all tests that have been conducted so far have failed to reveal any spatial correlation between the fluctuation signal 
and extended emission from detected galaxies. If the IHL were to arise from stars originally formed within galaxies, the unresolved fluctuations should produce a 
measurable spatial correlation with the spatial distribution of resolved galaxies. The apparent absence of such correlations with i) the subtracted outer parts of 
galaxies, ii) artificial halos placed around galaxies, and iii) the insensitivity to the increased area of source masking, all present challenges for the IHL model. These 
observational tests are described in detail in \citet[][see also KAMM1]{akmm}. Second, the IHL cannot account for the measured correlation with the soft X-ray 
background \citep[see ][]
{kari2}. Third, the IHL does not seem to reproduce the observed blue color of the CIB fluctuations. Finally, the light-to-mass ratio of the IHL is calibrated based on 
intracluster light (at $2.7 \times 10^{14}M_\odot$), and extrapolated as a power-law down to much lower mass scales. The bulk of the IHL is therefore associated with 
low-mass systems so that it requires low-mass systems to host IHL exceeding their own stellar light. This results in IHL comparable to the integrated energy produced 
by the entire galaxy populations. We further show below in Sec. \ref{subsec:shotnoise} that the currently presented IHL models, in addition, violate 
the CIB fluctuation measurements at progressively lower shot-noise.

%IHL is by nature an elusive component. 
If IHL exists in required quantities, the superb sensitivity of JWST should 
permit its
%in general make it more manageable for 
detection and possible 
subtraction. The NIRCam resolution of 0.03--0.06$^{\prime\prime}$/pixel will also allow robust tests of behavior of the signal as the masked area is increased/
decreased. In addition, the interconnected nature of galaxies and the hypothetical IHL provides a clear prediction for JWST: the large scale fluctuation signal should 
steadily decrease as galaxies, and their wings, are subtracted to deeper levels and larger radii. As noted in Section 2.1.1, current instruments do not show this 
behavior and  JWST will be capable of placing stronger constraints on IHL. Furthermore, the low-$z$ IHL component should drop out in a cross-correlation tomography  
analysis developed in Section \ref{sec:tomography} which by design isolates a high-$z$ emission.

\section{JWST and NIRCam: requirements and experimental parameters}
\label{sec:jwst_pars}

Our goal in this section is to arrive at a {\it JWST}/NIRCam experimental configuration to optimize the determination of the following: 1) epochs of the CIB sources
producing the source-subtracted fluctuations, 2) constrain/determine the range of individual brightness of the new sources, 3) probe the energy spectrum of CIB fluctuations, 4) measure the power spectrum from clustering at every wavelength to reasonable accuracy, and 5) develop methodology to isolate the history of emission production from the early sources. After presenting for completeness the NIRCam filters to be used in this experiment, followed by discussion of the residual CIB from known galaxy populations, we zero in on an optimal experimental setup for arriving at this measurement in a reasonable number of the {\it JWST} observing hours.

\subsection{NIRCam filters}

%\subsection{JWST NIRCam prospects}

\begin{figure}[ht!]
\includegraphics[width=4in]{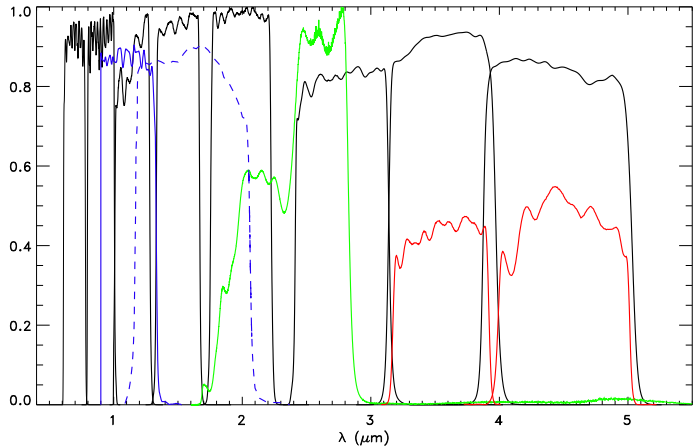}
\caption{\small Transmission curves of the NIRCam W filters are shown in black. Green shows the AKARI $2.4\mu$m filter; red corresponds
to IRAC 3.6 and 4.5 $\mu$m bands. CIBER filters at 1.1 and 1.6 $\mu$m are shown in blue adopted from \cite{bock}. The 2MASS (and NICMOS) filters are 
%standard and 
not plotted in this already crowded figure; they are shown in e.g. Fig. 6 of \cite{review}.}
\label{fig:NIRCam}
\end{figure}

We will select all eight of the NIRCam wide filters for the proposed study. They cover the required range of wavelengths, have sufficient sensitivity and are optimal for reducing the net integration time due to their wide bandwidth. Fig. \ref{fig:NIRCam} compares the NIRCam W-filters with the IRAC, AKARI 
and CIBER ones.

Beam size determines where confusion noise becomes important. For the following discussion the net beam areas were calculated by integrating simulated PSFs generated by using WebbPSF\footnote{\url{http://www.stsci.edu/~mperrin/software/psf_library/}}. The integrated PSFs for the eight 
NIRCam bands
considered here are $\omega_{\rm NIRCam} = \\(0.15, 0.11, 0.11, 0.14, 0.19, 0.33, 0.51, 0.75)\times 10^{-12}$ sr in order of increasing NIRCam wavelength.

\subsection{Remaining foreground galaxies}
\label{sec:og_floor}

Any measurement of CIB fluctuations requires good understanding of extragalactic galaxy populations. HRK12 showed that the small scale power in current measurements is well understood in terms of shot noise from unresolved galaxy populations, $m_{\rm AB}\gtrsim 25$. The same populations however, are unable to account for the large-scale clustering signal. Assuming that the faint-end slope of currently measured luminosity functions continues out to still fainter levels, HRK12 make a testable prediction of the contribution of faint galaxies to the CIB fluctuations measured with JWST. 
%Figure 10 shows the CIB fluctuations from galaxies fainter than $m_{\rm AB}=28$ in the NIRCam bands.

\begin{figure}[ht!]
\includegraphics[width=6.5in]{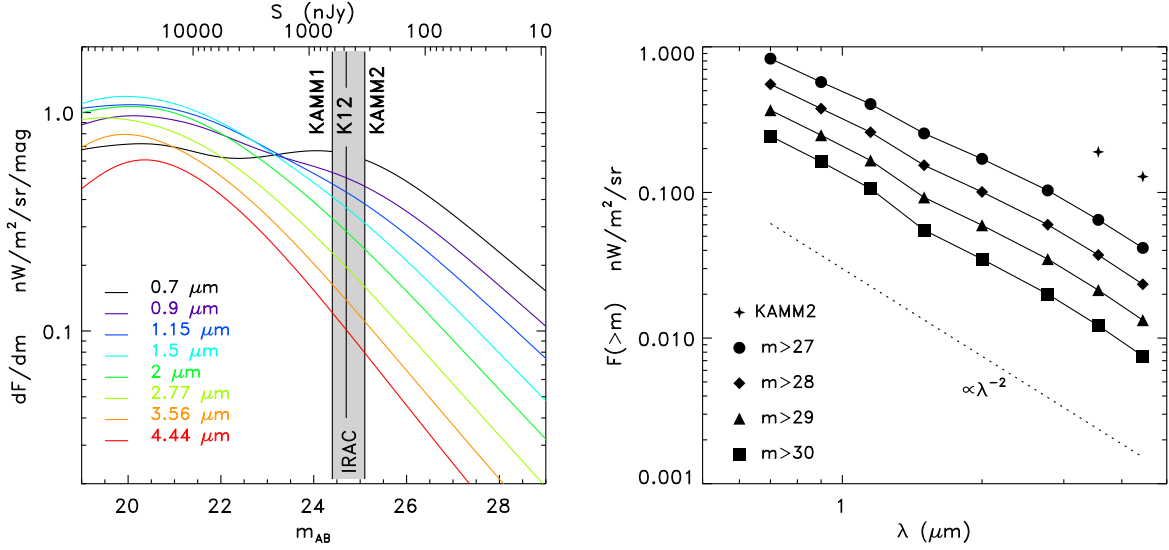}
\caption{\small  {\bf Left}: Differential flux contribution per $dm$ reconstructed via HRK12 methodology for counts at all NIRCam bands from 0.7 to 4.44 \micron. {\bf Right}: CIB flux from galaxies fainter than a given $m$ is plotted vs wavelength. Two asterisks show the {\it Spitzer} magnitude removal threshold of \citep{seds}. The known populations give CIB which scales as $\propto \lambda^{-2}$ distinctly different from the $\propto \lambda^{-3}$
measured by \cite{akari}.
}
\label{fig:dndm}
\end{figure}

With its superior resolution and sensitivity, {\it JWST} is expected to subtract galaxies down to $m_{\rm AB}\sim 29-30$, reducing shot noise considerably. The slope of galaxy counts, extended to these magnitudes, is such that the net reduction in shot noise is progressively less towards shorter NIR wavelengths (see Fig. \ref{fig:dndm} and Sec. \ref{sec:lyman}). The shot noise at these levels is always dominated by galaxy populations at intermediate redshifts i.e. $z \sim 1-4$ at 1--5\mic. The shot noise contribution of currently observed $z\gtrsim 8$ populations, characterized by a steep faint-end Schechter LF slope of $\alpha \simeq -2$, is subdominant at all relevant magnitudes. This contribution may however be detectable in the shot noise in a NIRCam cross-correlation analysis of the CIB fluctuations (see Section \ref{sec:tomography}).

Faint galaxies are expected to exhibit the same low clustering--shot noise ratio in {\it JWST} maps as in current observations. This lack of power at large scales led HRK12 to the conclusion that faint galaxies are unable to account for the signal measured by {\it Spitzer} and {\it AKARI}. The observed behavior of the amplitude of the large-scale CIB signal in the process of subtracting $m_{\rm AB} \gtrsim$26 sources in NIRCam maps will be revealing of its nature. 
In the discussion below we adopt the HRK12 heuristic way of reconstructing CIB fluctuations from galaxy populations spanning wavelengths from UV to mid-IR out to $z\sim 6$ using the database of $>340$ luminosity function (LF) surveys \citep{kari2}. With it we robustly fill in the redshift cone with known galaxies across the required wavelengths after extrapolating the LF to faint luminosities. The reconstructed populations are on average described well by the ``default'' model of HRK12 and the accuracy of the reconstruction is verified by the remarkably good fits to the newly measured and much deeper than before IRAC counts \cite{ashby1,ashby2}.  Given the good fits of the default model to the counts data this reconstruction is presented in this section for our estimates of the CIB produced by the known galaxy populations.

\begin{figure}[ht!]
%\plotone{fig_shotnoise.eps}
\includegraphics[width=6.5in]{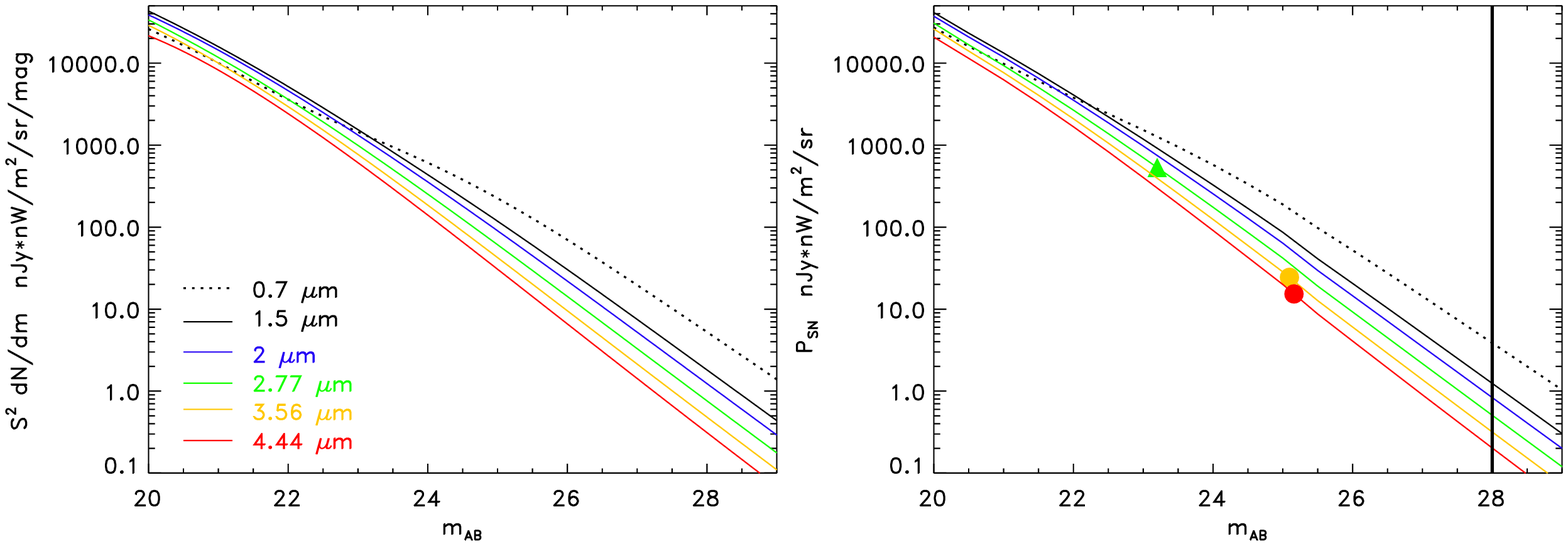}
\caption{\small {\bf Left}: Differential shot noise power, $S^2dN/dm$, vs the removal magnitude threshold for NIRCam filters  reconstructed with the HRK12
methodology.  {\bf Right}: Integrated shot noise vs the removal magnitude threshold reconstructed via HRK12.  Triangle corresponds to the limit reached in AKARI and circles to that in KAMM2. The limit of the 400hr observing configuration is shown with a vertical line.
}
\label{fig:shotnoise}
\end{figure}
Fig. \ref{fig:dndm} shows the reconstructed CIB-related properties for known galaxy populations.
Fig. \ref{fig:shotnoise} shows the reconstructed shot-noise power at NIRCam bands for the fainter exposures compared with what was reached in the {\it Spitzer} and {\it AKARI} measurements. The shot-noise levels would be low enough to enable the Lyman-break determination of the CIB and this component also sets limits on the measurement of the CIB from new populations. Note the very different behavior of the shot-noise at the 0.7 $\mu$m band, which is driven by the different slope of source counts there compared to those at near-IR; this will have implications in what follows. 

Unless otherwise noted, when discussing the relevant numbers for the {\it JWST} configuration below we will  conservatively show
the HFE reconstruction, which provides an {\it upper} bound on the CIB contributions from remaining galaxies.

\subsection{Experimental setup}
\label{sec:setup}

 Here the goal is to optimize
the potential {\it JWST} use for measuring CIB from populations which are still fainter, and potentially at still earlier times than in 
the current studies. 
%Minimize the contributions from foregrounds, measure w high accuracy 1) the spectrum of the 
%fluctuations, 2) the Lyman break of these populations, and 3) their SED.
%To first order the necessary requirements to enable the CIB science are 1) we should reach low enough CIB from remaining known galaxies at all NIRCam bands to reliably probe any excess CIB components, 2) we should reach low enough shot-noise levels to reach the shot-noise expected from the new populations and to probe the clustering component over both small and large scales, and 3) we need to  minimize/optimize the ratio of dead-to-integration time while  given 1) and 2). 
The parameters of observation we need to optimize for the CIB science are: 1) area of the contiguous field, $A$, 2) net integration time, $t_{\rm total}$, 3) geometry/aspect-ratio of the field, given 4) the wavelengths to use as well as 5) we need to  minimize/optimize the ratio of dead-to-integration time.

\begin{figure}[ht!]
\includegraphics[width=3.4in]{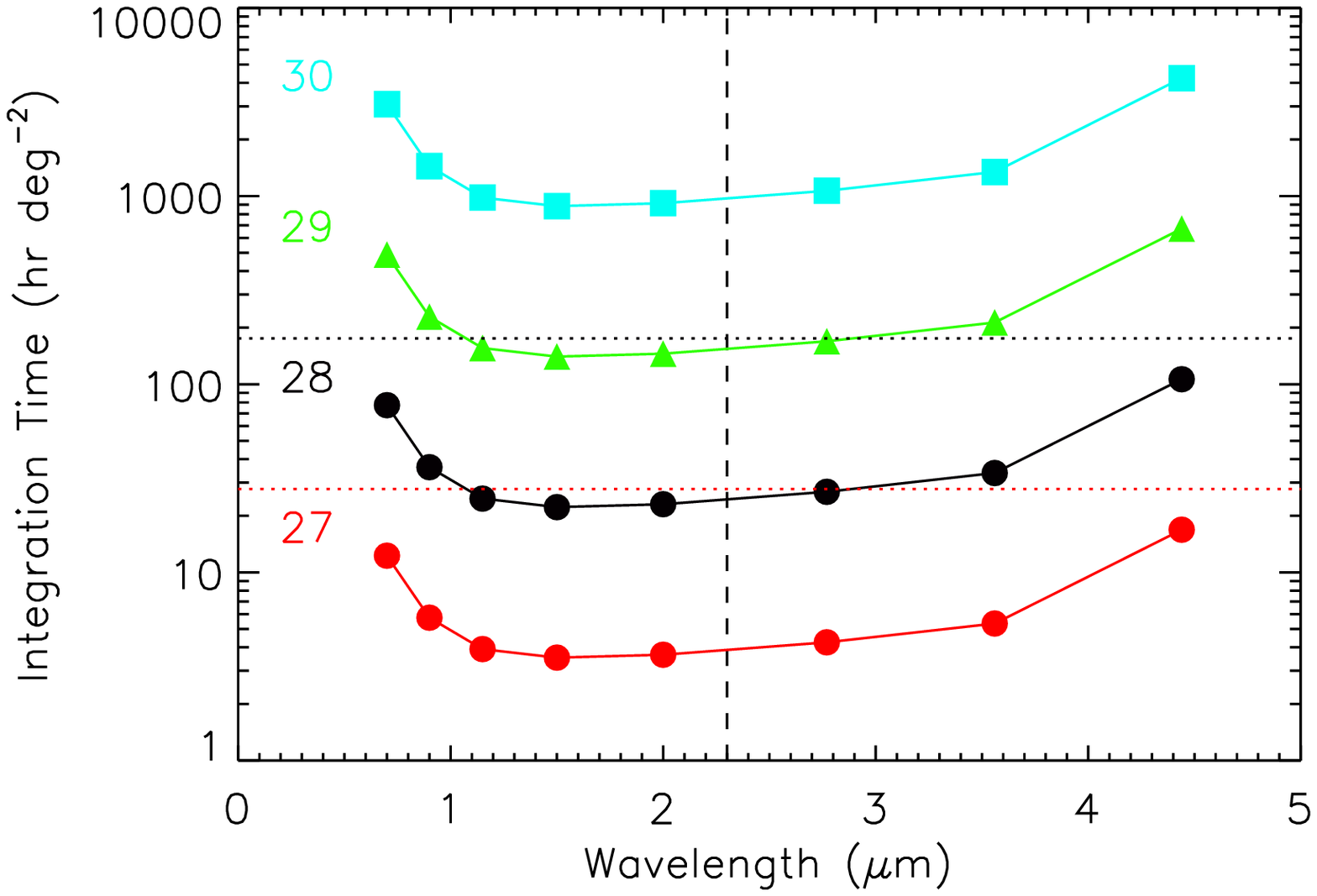}
   \includegraphics[width=3.4in]{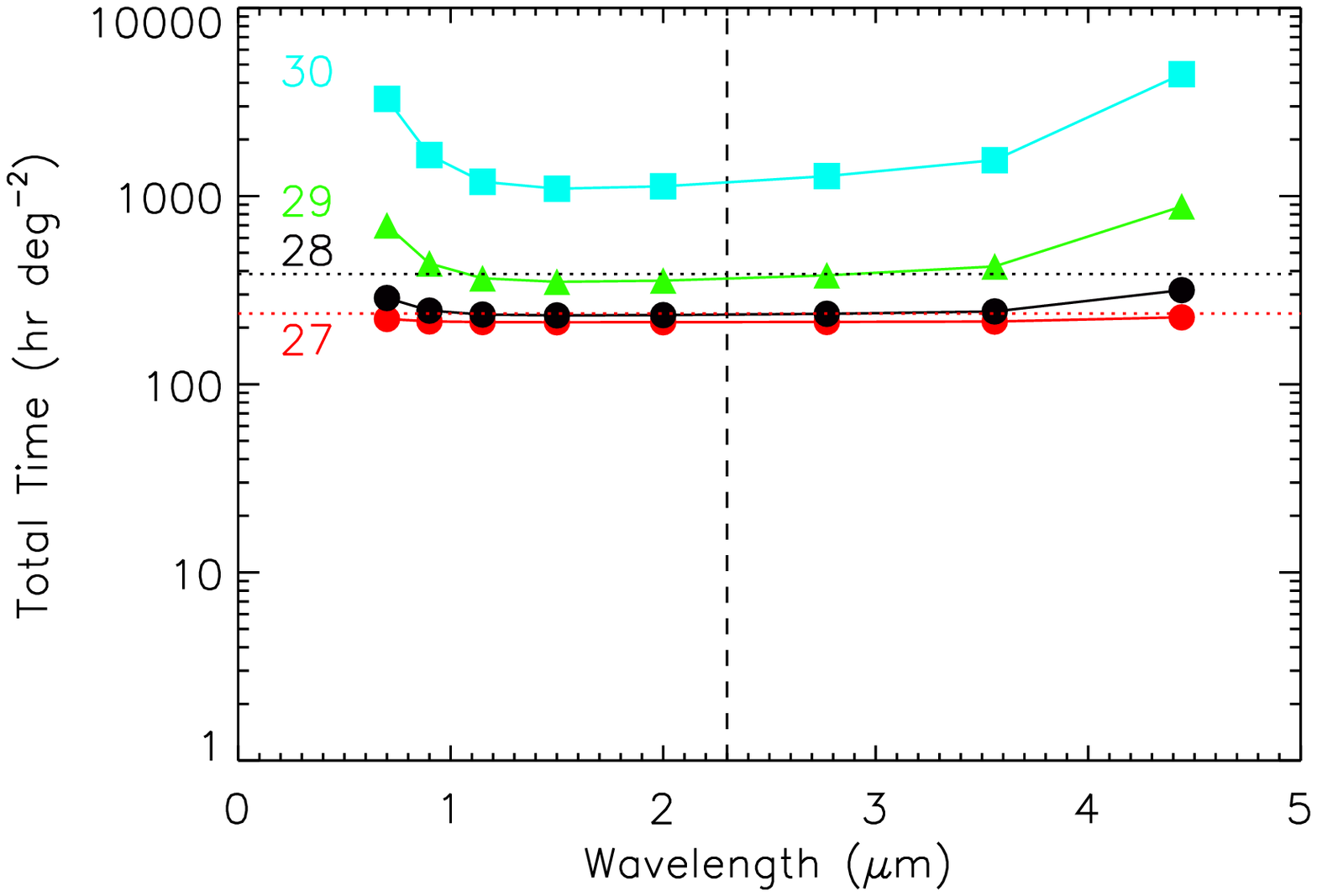} 
\caption{\small {\bf Left}: The estimated integration time per square degree needed to survey to 
depths of $m_{AB} =$ 27, 28, 29, and 30 (at $3\sigma$) in each
of the NIRCam wide bandpass filters. 
The total times needed to survey in all 8 filters are indicated
by dotted lines ($m_{AB} =$ 27 and 28 only).
An additional $\sim210$ hr deg$^{-2}$ will be required
for telescope overheads. {\bf Right}: Total time (including overheads) per square degree
   needed to survey to depths of $m_{AB} = $27, 28, 29 , and 30 (at $3\sigma$)
   in each of the wide NIRCam bands. The cumulative time to obtain observations
   in all 8 bands is indicated as the dotted line ($m_{AB} = $27 and 28 only).
   The vertical dashes separate the NIRCam short wavelength channels from the
   long wavelength channels; the two channels will be observed concurrently with the {\it JWST}/NIRCam.}
\label{fig:integration}
\end{figure}

When performing a wide survey, the total time required to map 1 deg$^2$
to a depth of $S$ $(3\sigma)$ in one filter with NIRCAM is roughly 
\begin{equation} 
t_{\rm total} = [93(S/S_0)^{-2} + 210] \;\;\; \frac{\rm hr}{\rm deg^2}
\end{equation}
where the first term on the right is the net integration time
and $S_0$ is the quoted NIRSPEC 
sensitivity\footnote{\url{http://www.stsci.edu/jwst/instruments/nircam/sensitivity/table}}
in $10^4$ s at $10\sigma$, and the second term is the overhead time 
(estmated using the Astronomer's Proposal Tool (APT v22.2), and similar
to programs in the Science Operations Design Reference Mission, Revision C). 
Short and long wavelength channels can be used simultaneously, but there 
will be a small increase in the overhead if filter changes are added at each pointing. 
The $3\sigma$ limit is used throughout because we study the background that 
remains after the individual sources have been removed (at $3\sigma$). 
Fig. \ref{fig:integration} gives the integration times to reach chosen magnitude
limits.

CIB science dictates that we settle on a configuration requiring $\simeq 400$ hrs of NIRCam mapping for all wide filters of a region of 1 deg$^2$ out $m_{\rm AB} =28$. A shallower exposure of a region of  6.3 deg$^2$ out to $m_{\rm AB} =27$, which leaves more known galaxies but allows better measurement of the spatial spectrum, would require 6 times more integration time because of the overhead and is not realistic. The considered possibility allows deeper removal of contributions from known galaxy populations, and also has sufficient accuracy when probing the spatial spectrum of the source-subtracted CIB component. We also consider square vs. rectangular geometries, which determine the accuracy of the angular spectrum measurement and the range of scales it covers. It appears that probing $\sim 1$deg$^2$ to AB mag of 28 at 3-sigma would provide answers to the above questions and
enable the science developed below.

%TBD: Add plot of noise power $P_{\rm noise}$ vs. $\lambda$ for this configuration.

\section{Probing the power spectrum of the CIB from high-$z$ populations with JWST}

We start this section with showing that the high-$z$ populations very generally lie in the confusion noise of the {\it JWST}, which requires CIB for studying these early sources. Then we address the measurability of the spatial spectrum of source-subtracted CIB in the presence of remaining populations of known galaxies.  We end this section with discussion of the dependence of the clustering component on the shot-noise to discriminate between the models for the origin of the fluctuations. 

\subsection{Confusion noise limits on identifying the new populations}
\label{sec:confusion_floor}

The emergence of the first sources of light at the end of the cosmic dark
ages is largely governed by the ability of primordial gas to cool (e.g., Bromm 2013).
In the absence of any metal coolants, prior to the dispersal of the first
heavy elements from Pop~III supernovae, there are two principal cooling channels
in the early universe. At temperatures in excess of $\sim 10^4$\,K, line radiation
from atomic hydrogen, predominantly concentrated in the Lyman-$\alpha$ transition,
provides very strong cooling. However, in bottom-up, hierarchical structure
formation, the first DM halos are characterised by shallow gravitational potential
wells, with correspondingly low virial temperatures, $T_{\rm vir}$.
%\simeq \frac{G M m_{\rm H}}{k_{\rm B} R_{\rm vir}} $.

Halos with $T_{\rm vir}\lesssim 10^4$\,K will thus not be able to activate
atomic hydrogen cooling. In such low-$T_{\rm vir}$ systems, the so-called
minihalos, cooling has to rely on molecular hydrogen. The H$_2$ formation
chemistry in the absence of dust grains is catalyzed by free electrons
left over from the epoch of recombination, with a rate that relies on
the gas temperature. For sufficient H$_2$ production, temperatures
of $\sim 10^3$\,K are required. This effect selects DM halos with
$T_{\rm vir}\sim 10^3$\,K, minihalos, as the formation sites for the
first (Pop~III) stars. Molecular hydrogen, however, is fragile, and can
easily be destroyed by soft-UV photons in the Lyman-Werner (LW) bands.
Such a pervasive LW background is expected to rapidly emerge in the
aftermath of the initial Pop~III star formation (e.g., Johnson et al. 2008).
It has therefore been argued that the first galaxies, defined as systems
that can sustain self-regulated star formation, will be hosted by more massive
DM halos (Bromm \& Yoshida 2011). Indeed, ``atomic cooling halos'' with
$T_{\rm vir}\gtrsim 10^4$\,K are considered promising candidates for first-galaxy
hosts, as they would not have to rely on H$_2$ as a coolant, and could instead
tap into the much more efficient, and resilient, atomic hydrogen channel.
Thus, in summary, there are two characteristic scales for DM host halos, expressed
in terms of $T_{\rm vir}\sim 10^3$\,K and $\sim 10^4$\,K, where the former is
predicted to host the first stars, and the latter the first galaxies.

With the adopted cosmological parameters the comoving radius containing total (dark matter + baryons) mass $M\equiv M_6 10^6M_\odot$ is  $r_M=13 M_6^{1/3} h^{-1}$kpc. At turn-around $v_{\rm virial} \sim Hr$, so the virial temperature of the haloes is $T_{\rm vir} \sim 120 M_6^{2/3} (1+z)$K. Below we will normalize the  density field underlying the first halos to the $\Lambda$CDM concordance 3-dimensional power spectrum, $P(k)$, which determines the rms density fluctuation, $\sigma_M \equiv [\langle \delta M/M\rangle]^{\frac{1}{2}}$ over the volume containing mass $M$. At the present epoch the latter is  normalized to $\sigma_8$ as: 
\begin{equation}
\sigma_M^2 = \sigma_8^2 \frac{\int P(k) W_{\rm TH}(kr_M)k^2dk}{\int P(k) W_{\rm TH}(kr_8)k^2dk}
\label{eq:sigma_mass}
\end{equation}
with $W_{\rm TH}$ being the top-hat window function and $\sigma_8=0.9$. In linear regime, the above expression can simply be multiplied by the linear growth factor from redshift $z$ to give the dispersion of the density field at $z$.

\begin{figure}[ht!]
\includegraphics[width=4in]{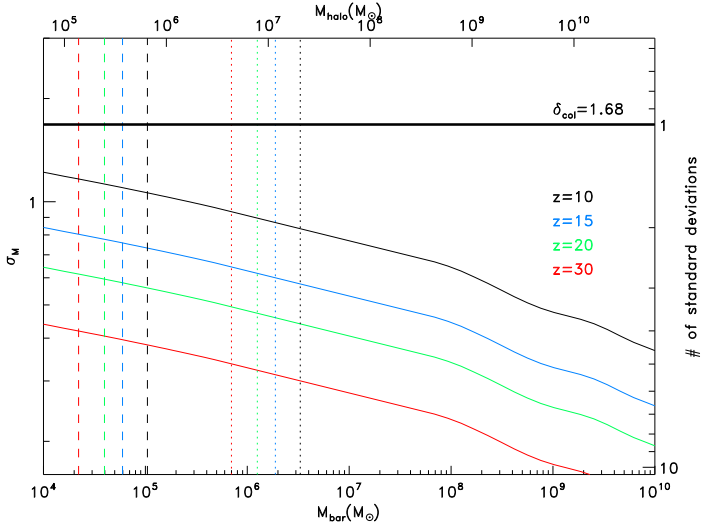}
\caption{\small The rms density fluctuation of density field over mass contained in each halo from eq. \ref{eq:sigma_mass}; at these epochs its
amplitude scales $\propto (1+z)^{-1}$. Thick horizontal line corresponds to the collapse threshold $\delta_{\rm col}=1.68$. Lower horizontal axis 
shows the mass in baryons, the upper shows the total halo mass. The right vertical axis shows the number of standard deviations at
each epoch, $\eta \equiv \delta_{\rm col}/\sigma_M(z)$, corresponding to the given mass. Black, blues, green, red colors correspond to $z=10, 15, 20, 30$. Vertical dashed lines delineate the region of masses with $T_{\rm vir}>10^3$K, the dotted vertical lines correspond to $T_{\rm vir}>10^4$K.}
\label{fig:sigma_m}
\end{figure}

In order to evaluate the abundance of collapsed haloes at these epochs we adopted the matter power spectrum from CMBFAST computed for $1/k>0.1 h^{-1}$Mpc. At smaller scales, $1/k$, the density field power is not well evaluated by CMBFAST due to the complex physics there, although in inflationary models the asymptotic power spectrum should reach $P(k)\propto k^{n_s}$ with $n_s\rightarrow-3$ as $k\rightarrow \infty$. We then extrapolated the CMBFAST-based power to smaller $1/k$ using the lowest range of the linear scales computable. In this way the spectral index eventually reaches $n_s=-2.6$ at $1/k\simeq 1h^{-1}$kpc and the resultant density field is elevated over what it would have been with $n_s=-3$. The density field at various epochs of relevance here is shown in Fig. \ref{fig:sigma_m}. The vertical dashed (dotted) lines in the figure delineate 
haloes with $T_{\rm vir}\geq 10^3 $K ($10^4$K).

%TBD:  discuss the range of minimal $T_{\rm vir}$ from the point of H2 cooling and atomic cooling in presence of first sources.

We now proceed to estimate the projected angular number density of the first haloes. Briefly, we adopt the flat Universe with the metric $ds^2=cdt^2-(1+z)^{-2}[dx^2+x^2d\omega]$ with $x(z)\equiv cH_0^{-1}{\cal D}_z$ being the comoving coordinate distance with ${\cal D}_z=\int_0^z dz/\sqrt{\Omega_{\rm m}(1+z)^3+1-\Omega_{\rm m}}$. The comoving volume is given by $dV_c=4\pi x^2dx(z)$ and the projected angular density from sources per steradian at redshifts greater than $z$, $n_2(>z)$, is related to the evolving 3-D comoving density $n_3$ via
\begin{equation}
n_2(>z)=R_H^3 \int_z^\infty n_3 {\cal D}_z^2\frac{dz}{\sqrt{\Omega_{\rm m}(1+z)^3+1-\Omega_{\rm m}}}
\label{eq:n3gen}
\end{equation}

The 3-D density of collapsed haloes is typically approximated by the general Press-Schechter (1974) formalism: the probability of halo of total mass $M_h$ 
to collapse at $z$ is $P_M={\rm erfc}(\eta/\sqrt{2})$ where $\eta\equiv \delta_{\rm col}/\sigma_M(z)$ is shown in Fig. \ref{fig:sigma_m}. The 3-D 
comoving density for haloes per mass interval $dM$ is then $2(\rho_{\rm m}/M_h) dP_M/dM_h$ with $\rho_{\rm m}=\Omega_{\rm m}\frac{3H_0^2}
{8\pi G}$ being the comoving matter mass density. The evolving comoving density of haloes of mass greater than $M_h$ is now
$n_3 =\int_{M_h}^\infty 2(\rho_{\rm m}/M_h) dP_M/dM_h dM_h$, which in the limit of large $\eta\gg1$ can be approximated as $n_3(>M_h,z) \simeq 
2\rho_{\rm m}M_h^{-1} P_M$ \citep{k93}\footnote{This approximation in any case represents a {\it lower} limit on $n_3(>M_h)$ since $\int_{M_h} \frac{dP_M}
{dM}\frac{dM}{M}> \frac{P_M}{M_h}$ for positive $P_M$.}. Finally this reduces to
\begin{equation}
n_2(>M_h,>z) = \frac{3\Omega_{\rm m}}{2\pi}  \int_z^\infty \frac{R_H}{r_g(M_H)} \times P_M(z) {\cal D}_z^2\frac{dz}{\sqrt{\Omega_{\rm m}
(1+z)^3+1-\Omega_{\rm m}}}
\label{eq:n3}
\end{equation}
which is driven by the ratio of the Hubble to the halo Schwarszchild radii, $R_H/r_g(M_h) = 4.5 \times 10^{16} (M_h/10^6M_\odot)^{-1}$.
Over the range of redshifts corresponding to first halo collapse, $10<z<40$, the value of $x$ remains fairly constant spanning $2.3\leq{\cal D}_z\leq 
2.8$ over $10\leq z \leq 50$, while $P_M$ rises very rapidly toward the lowest $z$ associated with these sources. Hence, the integral above is dominated by the lowest $z$ and the projected angular 
number density of halos is high at $n_2 \sim 10^{16} (M_h/10^6M_\odot)^{-1} P_M(1+z)^{-3/2}$sr$^{-1}$, which shows the difficulty of attempting to study these 
objects  individually due to confusion noise of an instrument with the $\sim 10^{-12}$sr beam. At the highest redshifts, the value of $P_M$ may 
be low enough to overcome the confusion, but there the (faint) sources would be well below the {\it JWST} detection. This already shows that 
CIB and its fluctuations would be critical in studying this epoch with the NIRCam instrument.

\begin{figure}[ht!]
%\plotone{fig_shotnoise.eps}
\includegraphics[width=4in]{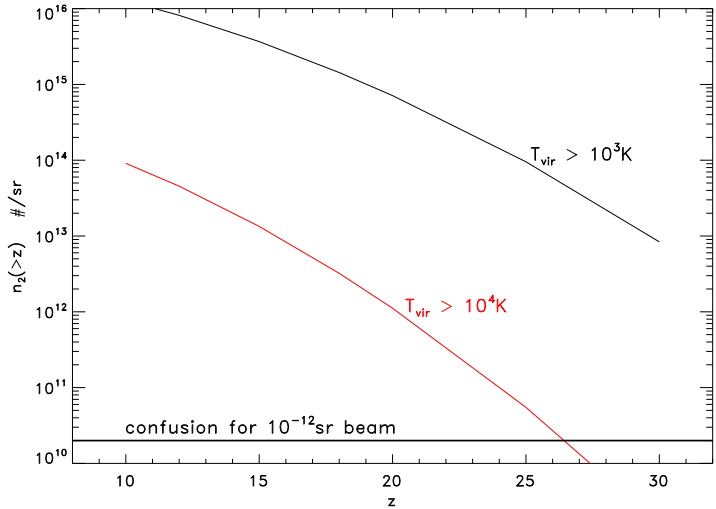}
\caption{\small  The projected angular density of early luminous haloes at redshifts greater than $z$ assuming stars and accreting BHs form when $T_{\rm vir} \geq 10^3$K (black) and $10^4$K (red). Horizontal thick solid line shows the confusion limit for a beam of $10^{-12}$ sr in area.
}
\label{fig:confusion}
\end{figure}

\begin{figure}[ht!]
%\plotone{fig_shotnoise.eps}
 \includegraphics[width=7in]{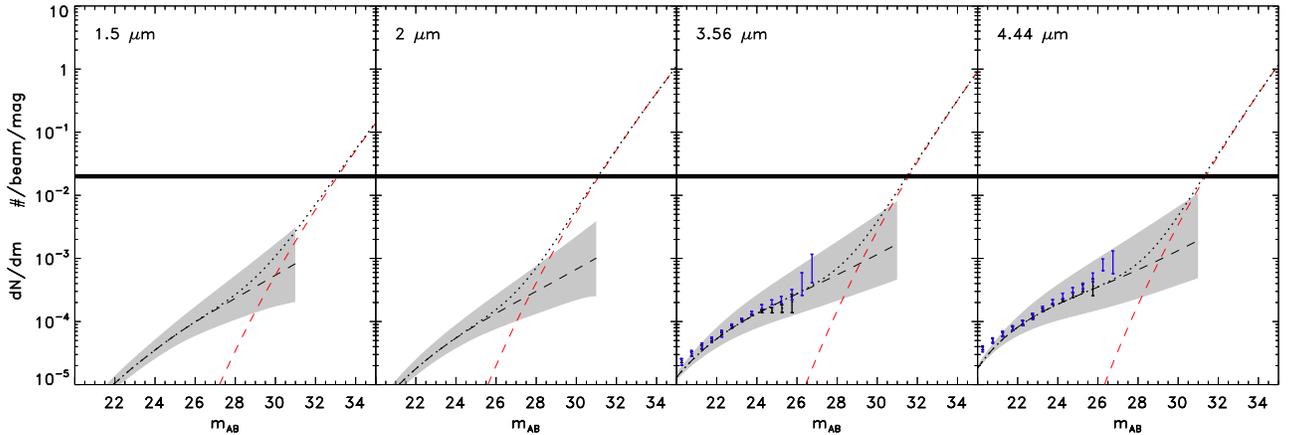}
\caption{\small  
Reconstructed counts due to known galaxies per HRK12 are shown with shaded region spanning the HFE to LFE limits; the default model of HRK12 is
shown with black dashes. High-$z$ sources from a model that reproduces CIB fluctuations from Helgason et al (2015) are shown with red dashes. The sum of the default reconstruction with the high-$z$ component is shown with dots. Thick horizontal line correspond to confusion noise limit of 50 beams per source. No new counts data at 1--2\micron\ have appeared since HRK12, where 
Figure 5 of that paper compared the existing counts to the reconstruction. 
%the existing counts are compared to the reconstruction in Figure 5. 
At 3.6 and 4.5 \micron\ the new counts data from IRAC are shown 
from \cite{ashby1} with black error bars and \cite{ashby2} in blue.
}
\label{fig:confusion_counts}
\end{figure}

Using this formalism, we have computed the projected angular density of collapsed luminous haloes assuming they form stars and accreting black holes when their virial temperature exceeds $T_{\rm vir} =10^3, 10^4$K. The results are shown in Fig. \ref{fig:confusion} and are generally well above the confusion limit for a $10^{-12}$sr beam. These sources would then be well within the confusion of the {\it JWST}, or at redshifts that put them below its detection limit, which demonstrates the need of CIB studies to probe the bulk of the new high-$z$ populations. Fig. \ref{fig:confusion_counts} shows the counts per each NIRCam beam due to known populations reconstructed per HRK12 method compared with an example of a model of star formation at high $z$ from Helgason et al. (2015, in preparation). The model assumes stars form with a somewhat heavy IMF, with typical masses of $\sim 10M_\odot$, forming continuously until $z=12$ in this particular example. Halos with virial temperatures $>10^3$K are able to form stars such that the fraction of baryons in these stars satisfy constraints in Sec. \ref{subsec:cib-hiz}. In this particular model, confusion intervenes around $m_{\rm AB}\sim 30-32$, but it is shown to illustrate
that the bulk, likely most, of the early systems would be within the confusion noise of the NIRCam beam requiring CIB to study their era.
  
We mention toward the end of this section a possible uncertainty due to extrapolation of the initial power spectrum to the small scales relevant for this analysis, an extrapolation due to the inadequacy of software packages such as CMBFAST for probing the very small scales. 
Any small deviation from the $n=-3$ power spectrum asymptote at these
scales may result in a larger halo abundance that what is adopted here. This in turn would lead to greater CIB at given individual source brightness.
Additionally, modification of the particle physics may be relevant as well as possible (small) non-Gaussianity coupled with the simplifications 
assumed in the Press-Schechter formalism at very high peaks of the underlying density field.

\subsection{Clustering component of source-subtracted CIB}
\label{subsec:clustering}

We now turn to estimating how well  one would measure source-subtracted CIB fluctuations from new populations with the configuration worked out
in Sec. \ref{sec:jwst_pars} in the presence of
remaining known galaxy populations.

\begin{figure}[ht!]
%\plotone{fig_shotnoise.eps}
\includegraphics[width=4in]{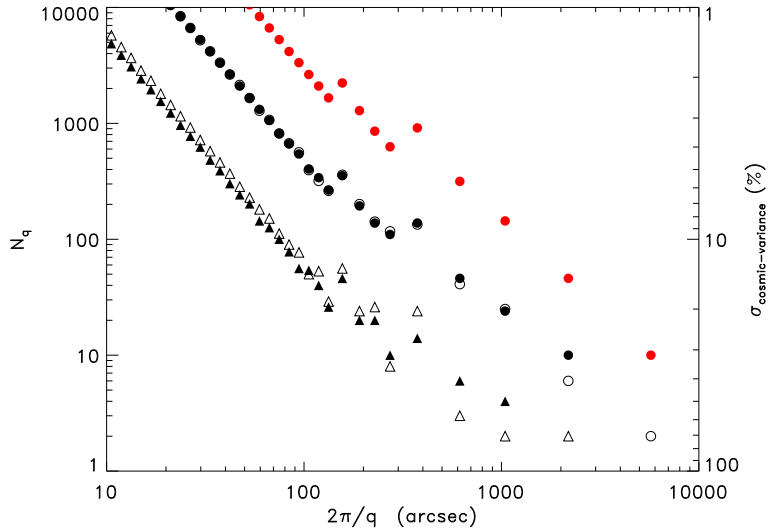}
\caption{\small  Number of independent  Fourier elements at each scale and cosmic variance for the  proposed configurations: 1 deg$^2$ area is shown with black circles for square (filled) and rectangle of 1:4 aspect ratio (open). Triangles show the number of Fourier elements for the current CIB fluctuation results from \cite{seds} for two regions: UDS of $21'\times21'$ (filled) and EGS of $8'\times 62'$ (open). For the same aspect ratio, the number of Fourier elements increases in proportion to the area, so e.g. using $\simeq 6.3$ deg$^2$ region will decrease the errors by a factor of $\sqrt{6}\simeq 2.5$ as shown with red circles. The relative uncertainty from cosmic/sample variance in determining the power is shown on the right vertical axis.
Bumps in the slope of $N_q$ as a function of $2\pi/q$ are caused by discrete 
changes in the relative widths (or spacing) of the chosen 
bins at each angular scale.
}
\label{fig:cv}
\end{figure}

The relative cosmic (sampling) variance when determining the power at scale 
$2\pi/q$ is $\sigma_P/P = N_q^{-1/2}$, where $N_q$ is the number of independent
Fourier elements averaged in the bin around angular frequency $q$. 
At small angular scales (high frequencies) we choose logarithmically spaced bins such that 
$\Delta q \sim q$ and thus the number of Fourier elements within the ring of 
radius $q$ in the Fourier domain is
$N_q \sim 2\pi q \Delta q = 2\pi q^2$. 
However at the largest scales (low frequencies), the discrete nature of the data 
requires shifting to larger relative bin widths.
%Given the size and the geometry of the selected region, the relative cosmic 
%(sampling) variance when determining the power at scale $2\pi/q$ by 
%averaging over $N_q$ independent Fourier elements is $\sigma_P/P = N_q^{-1/2}$. 
Fig. \ref{fig:cv} shows the relation between the observing area size and shape, and the accuracy for measuring the shape and the amplitude of the power spectrum of CIB fluctuations.
% at each of the NIRCam bands for the proposed configuration of 400 hours of NIRCam observations. 
If the source-subtracted CIB originates at high $z$,  the angular spectrum of its fluctuations should peak around $10'-12'$ corresponding to the horizon at matter-radiation equality projected to the epoch of the sources.
On these scales the CIB power spectrum can be probed with better than about 10-20\% statistical accuracy, provided contribution from  remaining known galaxies is small. 

\begin{figure}[ht!]
\includegraphics[height=3.5in,width=7in]{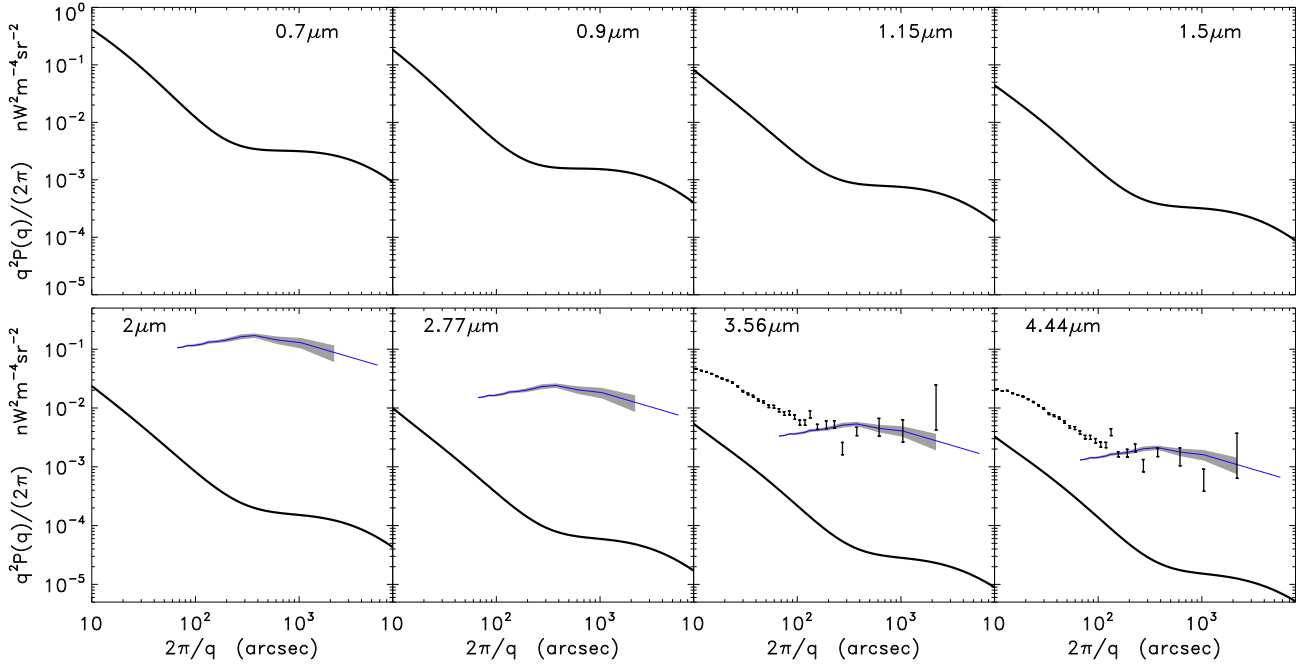}
\caption{\small  Mean squared CIB fluctuation from 0.7 to 4.44 \micron. Reconstructed contribution from known galaxies fainter than $m_{\rm AB}= 28$ is shown for the 
high-faint-end case of the HRK12 model representing the {\it upper} reconstruction limit.  
Error bars show the source-subtracted signal from Spitzer measurements out to degree scale \citep{seds}; at shorter wavelengths, {\it AKARI} results indicate that the CIB fluctuations increase as $\propto \lambda^{-3}$ \citep{akari} equivalent to the displayed square fluctuations going as $\propto \lambda^{-6}$. 
%Orange vertical bars show the range of cirrus fluctuations according to Fig. \ref{fig:ism}. TBD: red dashed-triple-dotted line shows cirrus for the LH according to ... 
}
\label{fig:cib_fluc}
\end{figure}

Fig. \ref{fig:cib_fluc} shows the upper bound from the HFE reconstruction of the source-subtracted CIB fluctuations for the parameters envisaged here. Its levels at all scales $\gsim 0.5'$ are comfortably  below the {\it Spitzer} and {\it AKARI} levels on the measured CIB fluctuation at all NIRCam wavelengths longward of 2\micron\ and 
also at shorter bands, provided the source-subtracted CIB there is at least at the levels shown in the lower panels. Of course, if the CIB fluctuations
originate in sources predominantly at high $z$, the CIB fluctuations should drop significantly (or be absent) at sufficiently short wavelengths because of the Lyman break in the energy spectra of these sources. This is discussed in some detail in Sec. \ref{sec:lyman}

In real measurement the power spectrum is measured from the cut sky, where Fourier harmonics are not strictly orthogonal. Correction must be made 
then for masking if the mask is reasonably modest \citep{seds}, or the correlation function computed instead if the masking fraction of available pixels is high \citep{k2007}. Fig. \ref{fig:confusion_counts} shows that out to $m_{\rm AB} = 28$, known galaxies would occupy $<0.01$
%less than 1\% 
sources per beam and with the
sky removal around the $\sim3\sigma$ clipping threshold the total number of pixels (noise plus sources) lost to clipping should be modest and
smaller than the fraction of $\sim 25\%$ in the {\it Spitzer}-based analysis, where it was already shown that correcting for masking leads to
small corrections in power of less than a few percent \citep{seds}. In any event, for the expected fraction of pixels lost to clipping, one can use the procedure 
outlined in Appendix of K12, whereby initially assumed power spectra templates are iteratively processed through the mask to evaluate the best fit
spectrum and its systematic and statistical uncertainties. Given the expected clipping fraction levels, the systematic correction to the power from masking
will very likely be within the statistical uncertainties in Fig. \ref{fig:cv}.

\subsection{Source-subtracted CIB fluctuations vs. shot-noise}
\label{subsec:shotnoise}

%Add per Helgason, Bromm, Kashlinsky (2014) modeling - TBD. Discuss DCBH shot-noise.

Assuming the source-subtracted CIB 
must ultimately originate in discrete sources, the measured excess fluctuation implies that there must be fainter sources numerous enough to account for the measurements. When populations responsible for the large-scale CIB fluctuation (from clustering) also dominate the shot noise, the two components become coupled with $P_{\rm SN} \propto S\cdot \delta F$. 
Eventually reaching to sufficiently low levels of shot noise with deep exposures should result in attenuation of the large-scale fluctuation from clustering; the point where this happens would then probe the flux of the typical sources responsible for this CIB component.  Determining observationally where this occurs will provide important clues to the nature of this population and confront several theoretical models for the origin of the CIB fluctuations with hard data, which differ measurably in the dependence of the clustering component on the underlying shot/1halo-noise term. There are two distinct possibilities in this context: 1) the amplitude of the clustering signal will remain constant (reducing at most by a fraction $<10$\% corresponding to the contribution of known galaxies) implying that the underlying population lies still beyond the detection threshold of JWST; 2) the signal will steadily diminish in amplitude as the underlying sources become resolved and removed.

\begin{figure}[h!]
%\plotone{fig_shotnoise.eps}
\includegraphics[width=7in]{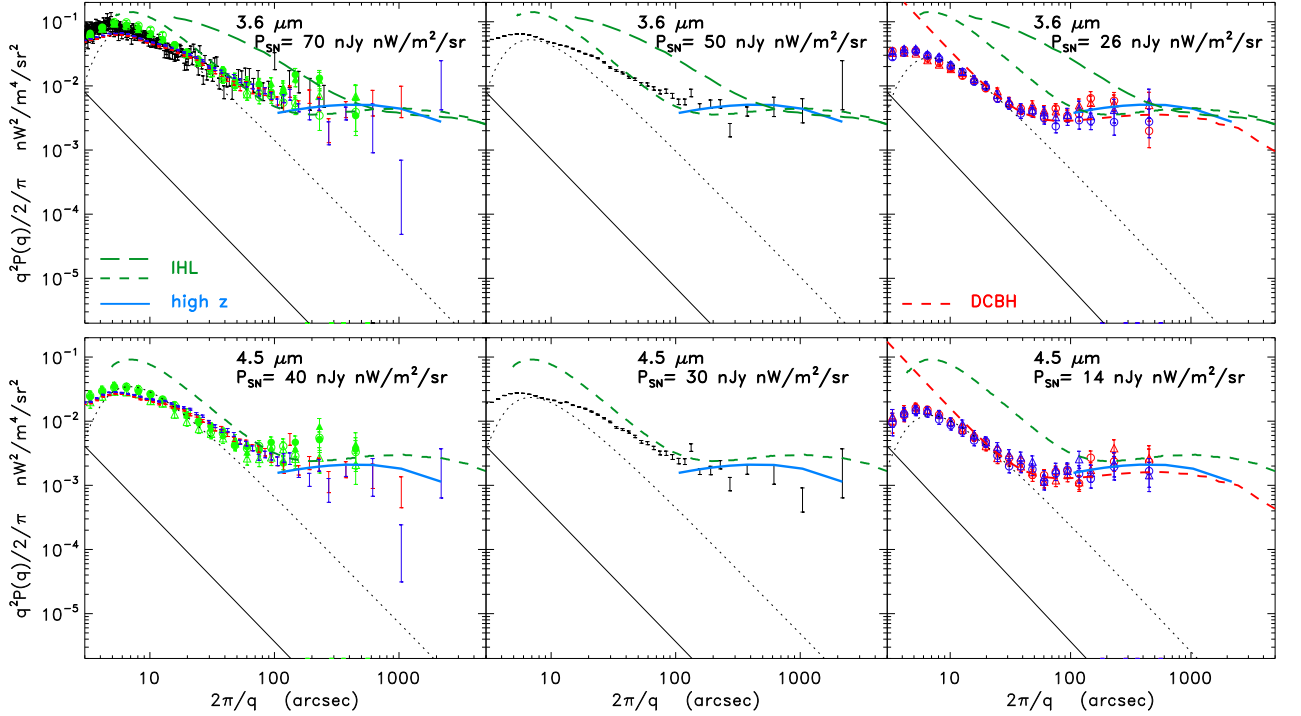}
\caption{\small  Current {\it Spitzer}/IRAC-based measurements at different shot-noise levels from KAMM1, K12 and KAMM3 measurements. Upper panels correspond to 3.6\um, lower to 4.5 \um. Dotted lines show the shot noise remaining in the {\it Spitzer}/IRAC data, which is convolved with the IRAC beam. {\it No decrease of the large-scale clustering component is yet apparent at these shot-nose levels}. Black solid lines show the shot noise component from remaining known galaxies at $m_{\rm AB}\geq 28$ at the two longest NIRCam wavelengths that will reached the 
configuration proposed in Sec. \ref{sec:jwst_pars}. IHL models, shown in green from \cite[][short dashes]{cooray12} and \cite[][long dashes]{ciber}, 
 appear inconsistent with the data already available in 2007. Template of high-$z$ $\Lambda$CDM model normalized to the IRAC data is shown in blue. The DCBH model of \cite{yue}, which accounts for both the CIB fluctuations and the CXB-CIB coherence, is plotted with red dashes only at the lower current shot-noise levels (for clarity) without convolving with the IRAC beam. Its shot noise is below the currently reached levels, but well above what this experiment setup will reach.
}
\label{fig:clustering_sn_irac}
\end{figure}
Fig. \ref{fig:clustering_sn_irac} shows  the current measurements binned by the progressively lower shot-noise  from KAMM1, K12 and KAMM2. The figure demonstrates that the
current IRAC-based deep integrations have not yet reached the regime where the large-scale CIB component from clustering starts decreasing
with the shot-noise, implying that the shot-noise arises in different populations than the clustering component. The figure shows that the IHL modeling of \cite{cooray12,ciber} fails to account for the CIB fluctuations measured at lower shot-noise levels.
%, the data which although available were ignored in the modeling by those authors. 
The DCBH model of \cite{yue} is consistent with the data, but comes tantalizingly within the reach of being testable with the
NIRCam configuration here. 

 \begin{figure}[h!]
%\plotone{fig_shotnoise.eps}
\includegraphics[width=6.5in]{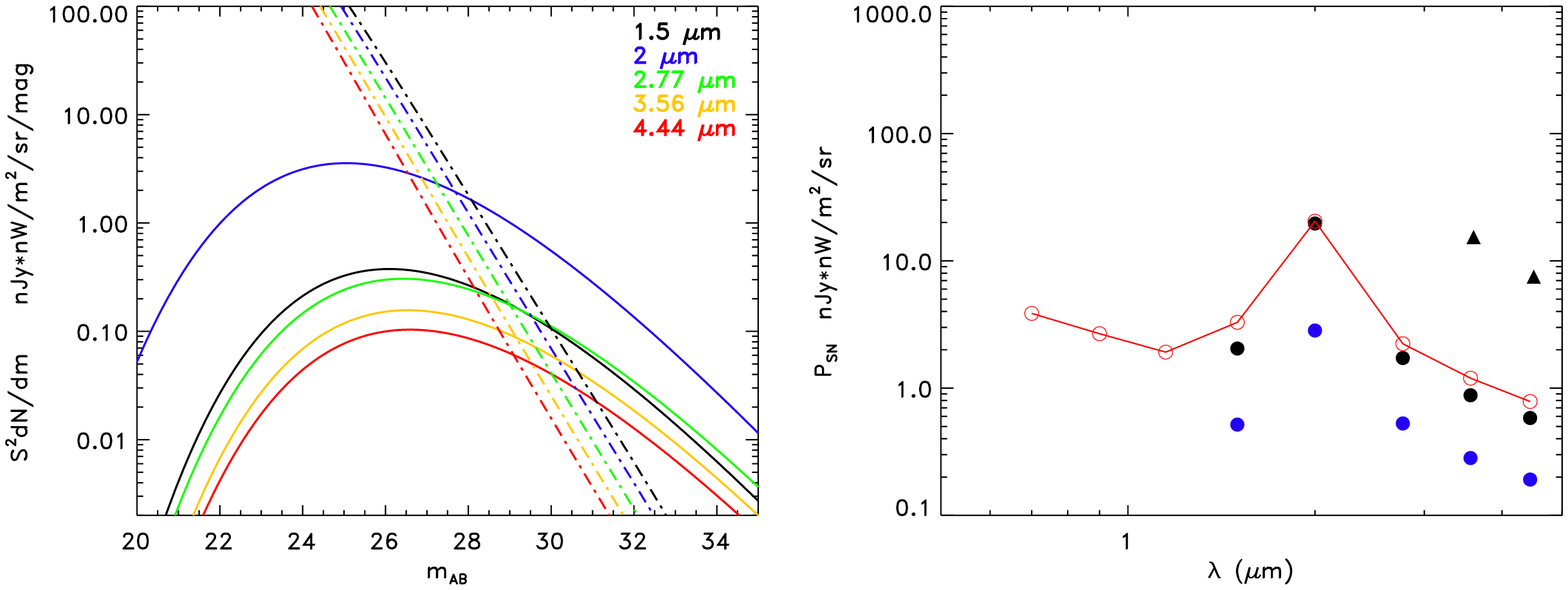}
\caption{\small  {\bf Left}: Differential shot noise contribution $S^2 dN/dm$. Color notation is shown in the upper right corner. Dashed-dotted lines are known galaxy populations 
reconstructed in HRK12; thick solid lines are from high-$z$ star modeling of Helgason et al (2015) as described in the main text. {\bf Right}: Triangles show the shot noise
from the DCBH model of Yue et al (2013). Shot noise from the Helgason et al (2015) modeling model is shown as an example at NIRCam bands with 
filled black circles; blue circles show only the contribution from $m_{AB} > 28$. Red  open circles correspond to the sum total of the shot noise from known galaxy
populations at $m_{\rm AB} \ge 28$. The hump between 1 and 2 $\micron$\ corresponds to the Lyman emission toward the end of the "first stars era" for that particular illustrative model. 
}
\label{fig:sn_jwst}
\end{figure}
Fig. \ref{fig:sn_jwst} shows the build-up of the shot noise from known galaxy populations (dashed-dotted lines) compared with the advanced 
models at high $z$: in the right panel the shot-noise of the DCBH model of \cite{yue} is marked with triangles, and filled black circles are from the set of models of high-$z$ star formation (Helgason et al. 2015, in prep.) discussed in Sec. \ref{sec:confusion_floor}. The build-up of the shot noise by the
known and new high-$z$ sources at the NIRCam wavelengths is illustrated in the left panel.

To sum up this discussion, the current measurements indicate that the current data on the shot noise and clustering components indicate that 1) there is no coupling yet between the shot-noise and clustering CIB levels in the {\it Spitzer} measurements, 2) the required CIB source flux is $\lesssim 20$ nJy, 
and 3) the IHL model, as presented, is in conflict with the data. The expected fluxes where the clustering component will be expected to couple to
the progressively decreasing shot noise can be reached with the NIRCam exposures here. This will provide important new information on the 
nature of sources producing the CIB fluctuations and their epochs. 
\section{Measuring the Lyman-break of the source-subtracted CIB fluctuations}
\label{sec:lyman}

At present, there is no direct measurement of the redshifts associated with the sources that produced the source-subtracted CIB fluctuations. However, unlike other cosmic backgrounds where direct measurement of the epochs (or redshifts) is unobtainable, for this CIB component one can, in principle, measure the redshifts that its sources inhabited since their individual energy spectra would have no emission below the Lyman-break wavelengths at those epochs
\citep{santos,sf03,kagmm,cooray,akmm,seds,cooray12}.  Physically, such cutoff would appear in energy above 1) the Lyman-$\alpha$ 
energy (10.2 eV) if the emission is fully absorbed in the gas cocoons surrounding the sources, or 2) the Lyman-continuum 
energy (13.6 eV) if the emitted photons freely escape \citep[][]{santos}. Therefore, the component of the CIB fluctuations from populations at redshifts greater than $z_{\rm s}$, should {\it not} correlate with the diffuse background below
 $\lambda_{\rm Ly-cutoff}\simeq \frac{1+z_{\rm s}}{10}\mu{\rm m}$. This potentially offers a prospect of measuring the redshifts where the background is produced with several current CIB programs now gearing toward this measurement using a variety of space- and stratosphere-borne instruments.

%Unlike other cosmic backgrounds where direct measurement of the epochs (or redshifts) is unobtainable, for this CIB component one can, in principle, measure the redshifts where its sources lived via the existence of the Lyman-break in their individual emissions around optical wavelengths 
%\cite{santos,sf06,kagmm,cooray,akmm,seds,cooray12}.  Physically, such cutoff would appear in energy above 1) the Lyman-$\alpha$ 
%energy (10.2 eV) if the emission is fully absorbed in the gas cocoons surrounding the sources, or 2) the Lyman-continuum 
%energy (13.6 eV) if the emitted photons freely escape \citep[][]{santos}. Therefore, the component of the CIB fluctuations from redshifts greater than $z_{\rm s}$, they should {\it not} correlate with the diffuse background below
% $\lambda_{\rm Ly-cutoff}\simeq \frac{1+z_{\rm s}}{10}\mu{\rm m}$. Elaborate: TBD different modes of evolution.
 
Here we at first provide robust empirical estimates of the floor for such measurement with the currently operating instruments which arises from known galaxy populations remaining in the various {\it currently planned}
experiments before discussing the {\it JWST} prospects in this regard. It turns out that, given the parameters of the current 
experiments, this floor with its current systematic uncertainties provides a highly non-negligible component which likely affects a robust determination of the epochs associated with the sources producing 
the CIB fluctuations using these configurations. 
%Presenting the quantitative estimates for this floor and its systematic uncertainties for use
%in the current experiments is the purpose of this section. 
Foreground contributions will increase this floor further. 

Diffuse CIB maps at wavelength $\lambda$ can be decomposed into independent (i.e. additive in quadrature) contributions as ``ordinary galaxies'' ($g$), the unknown/new population ($X$) and foregrounds ($f$) at pixel $\vec{x}$:
\begin{equation}
\delta_\lambda (\vec{x}) = g_\lambda(\vec{x}) + X_\lambda(\vec{x})+f_\lambda(\vec{x})
\label{eq:del_cib}
\end{equation}
%In the absence of masking the Fourier transformation decomposes $\delta$ into {\it statistically independent} harmonics with amplitudes $\Delta(\vec{q}) = \int \delta(\vec{x}) \exp(-i \vec{q}\cdot \vec{x}) d^2\vec{x}$. 
%CIB fluctuations $\delta_\lambda (\vec{x})$ are characterized by the  angular power spectrum as a function of the angular scale 
%$2\pi/q$ defined as $P(q)=\langle |\Delta(\vec{q})|^2\rangle$,
%where the average is performed over all phases of the given $q$. The cross-power describing the correlations between fluctuations at different wavelengths (1,2) is
%$P_{\rm 1\times2} (q) = \langle \Delta_{1}(q) \Delta^*_{2}(q)\rangle = {\cal R}_{1}(q) {\cal R}_{2}(q) + {\cal I}_{1}(q) {\cal I}_{2}(q)$ with ${\cal R, I}$
%standing for the real, imaginary parts of $\Delta(\vec{q})$.  The cross-power is a real quantity which can assume positive or negative values. 
%If masking eliminates a substantial fraction of pixels, these harmonics are no longer statistically independent and the correlation function should be used instead \citep[e.g.][]{k2007}. 
The different components in eq. \ref{eq:del_cib} are independent and add in quadrature in the auto-power without contributing to the cross-power. Additionally, the hypothesis of the Lyman-break in the $X$-component implies that the cross-power of $X$ vanishes between $\lambda_1 <\lambda_{\rm Ly-cutoff}\equiv\lambda_{\rm Ly}(1+z_s)$ and $\lambda_2>\lambda_{\rm Ly-cutoff}$. With this assumption the autopower of eq. \ref{eq:del_cib} is:
\begin{equation}
P_{\lambda}(q)=\left\{\begin{array}{cc} P^g_{\lambda}(q) + P^X_{\lambda}(q) + P^f_{\lambda}(q) & \lambda\geq \lambda_{\rm Ly}(1+z_s) \\  P^g_{\lambda}(q) + P^f_{\lambda}(q) & \lambda< \lambda_{\rm Ly}(1+z_s)
\end{array}\right.
\label{eq:autopower}
\end{equation}
while the cross-power  of eq. \ref{eq:del_cib} becomes for wavelengths shortward of the Lyman break:
%\begin{equation}
%P_{\lambda_1 \times\lambda_2}(q)=\left\{\begin{array}{cc} P^g_{\lambda_1\times\lambda_2}(q) + P^X_{\lambda_1\times\lambda_2}(q) + P^f_{\lambda_1\times\lambda_2}(q) & \lambda_2\geq \lambda_{\rm Ly-cutoff} \\  P^g_{\lambda_1\times\lambda_2}(q) + P^f_{\lambda_1\times\lambda_2}(q) & \lambda_2< \lambda_{\rm Ly-cutoff}
%\end{array}\right.
%\label{eq:crosspower}
%\end{equation}
\begin{equation}
P_{\lambda_1 \times\lambda_2}(q)= P^g_{\lambda_1\times\lambda_2}(q) + P^f_{\lambda_1\times\lambda_2}(q) \; ; \;\lambda_2< \lambda_{\rm Ly}(1+z_s)
\label{eq:crosspower}
\end{equation}
Thus quantities of interest here are $\Delta P=P_{\lambda_1}-P_{\lambda_2}$ and $P_{\lambda_1 \times\lambda_2}$ which ideally should be (much) smaller than $P_{\lambda_2}$ under the Lyman-break assumption at $\lambda_1<\lambda_{\rm Ly}(1+z_s)$, so constraining the epochs $z_s$ here is set by the floor from the remaining known galaxy populations and foregrounds. The two quantities require somewhat different interpretation: $\Delta P$ would probe whether the power at
fiducial wavelength, $\lambda_1$ assumed to be below the Lyman-break of sources at $1+z_s>\lambda_1/\lambda_{\rm Ly}$, is much smaller than that at $\lambda_2$ where the CIB excess is measured. At the same time, $P_{\lambda_1 \times\lambda_2}$ probes 
the absence of sources at $\lambda_1$ which dominate the clustering component of the CIB fluctuations at $\lambda_2$. Testing for each of these propositions is limited by the levels of the floor determined by remaining galaxies and foreground emissions in each experimental setup; this level is the subject of this discussion. 
%Because that floor turns out very substantial in the current experiments, we do not discuss here how big a drop would constitute a detection of the Lyman-break, beyond noting that in Lyman-break galaxy searches one is usually satisfied with a flux drop of TBD (power drop of TBD$^2$) to determine a Lyman break (ref TBD).

%After subtraction of the instrument noise from e.g. time-differenced data, the measured power-spectrum of source-subtracted CIB is made up of two components: 1) a white-noise component due to the shot-noise from remaining populations (convolved with the instrument beam), which may also include a (approximately white noise) ``1-halo'' term (white-noise convolved with the typical halo size), and 2) the component arising from clustering of the unresolved CIB sources. 

%The shot-noise term is straightforwardly measured from the diffuse masked maps and its amplitude fixes the magnitude threshold of the unresolved populations remaining in the maps (KAMM3). This in turn determines the level of the known galaxy ($g$) contributions to the cross-power and its statistical and systematic uncertainties. 

For coeval sources or foregrounds (such as cirrus), the coherence ${\cal C}_{12}\equiv \frac{P_{12}^2}{P_1P_2}\simeq 1$, and the 
cross-power can be approximated as $P_{\lambda_1 \times\lambda_2}(q) \simeq [P_{\lambda_1}P_{\lambda_2}]^{1/2}$. Thus the level of the cross-power between visible and IR bands from remaining known galaxies is to good accuracy $P^g_{\rm vis\times IR}\simeq [P^g_{\rm vis}P^g_{\rm IR}]^{1/2}$. Hence, if one is left with a significant and systematically uncertain power from remaining known galaxies at visible bands, this may hinder probing of the Lyman-break of the source-subtracted CIB fluctuations. 

\subsection{Limitations for the Lyman-break probing with current instruments}
Because of atmospheric fluctuations probing the Lyman break of the source-subtracted CIB fluctuations, observed at the levels of $\lsim 0.1$ nW m$^{-2}$ sr$^{-1}$ at around 2--4\micron, are best done   with space- or stratosphere-borne instruments.  E.g. the ground-level atmospheric fluctuation at $1''$ around 2\um\ is  $\sim 6,000 t^{-\frac{1}{2}}_{\rm int}(\rm sec)$ nW m$^{-2}$ sr$^{-1}$ for 1m diameter mirror \citep{odenwald} requiring impossibly long
integrations even with the largest telescope mirrors. In space, currently, only datasets obtained with ACS/WFC3 instruments onboard {\it HST} provide the necessary wavelengths for such imaging to cross-correlate with the CIB data from
current measurements at 1--5\micron.

The ACS instrument onboard {\it HST} currently offers the best prospect for probing cross-correlations with the source-subtracted CIB in {\it Spitzer/AKARI} measurements \citep{kamm4}. The net area covered with {\it Spitzer} measurements of K12 based on the SEDS project \citep{ashby1}, corresponding to the deepest source
removal threshold at $m_{\rm AB}\sim 25$ for sub-degree scales, is a little less than 1,000 arcmin$^2$, of which a much smaller area is covered with {\it HST}/ACS observations (\url{http://candels.ucolick.org}).   {\it AKARI} IRC
2.4 $\mu$m CIB measurements, after removing sources down to $m_{AB}\simeq 23.2$, resulted in the fluctuation measured to $\sim 5^\prime$ over a field of $\simeq 80$ arcmin$^2$.
CIBER Wide-Field Imager
\citep{bock} % (see \url{http://ciber.caltech.edu}) 
has bands at 
1.1 and 1.6\ $\mu$m with $7''$ resolution and achieves removal down to only $m_{\rm AB}\simeq 18.4$ at 3$\sigma$  from the net area of $2^\circ\times2^\circ$ per one field-of-view. In the near future, the CIBER-2 instrument will be equipped with 6 bands spanning 0.5-2 $\mu$m \citep{ciber2}. 

\begin{figure}[ht!]
%\plotone{fig_lbreak_current.eps}
\includegraphics[width=6.5in]{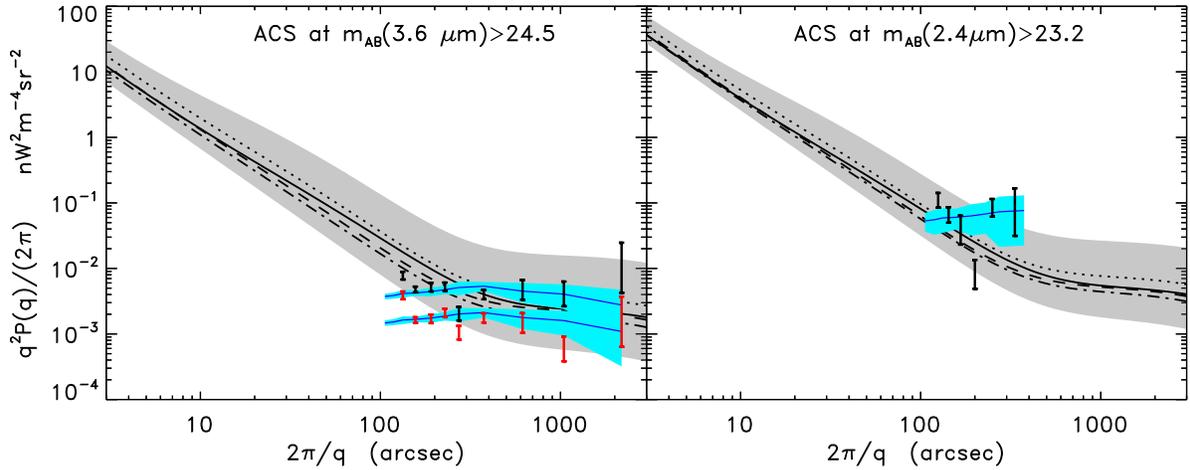}
\caption{\small   {\bf Left and right}: Auto-power at the four ACS bands  from galaxies remaining in the {\it Spitzer} and {\it AKARI} CIB configurations computed using the HRK12 reconstruction. Solid, dotted, dashed and dash-dotted lines correspond to the ``default'' reconstruction in HRK12 for the (B, V, i, z) bands respectively. Black/red error bars in the left panel show the K12 measurements at 3.6/4.5 \mic\ at scales $\geq 2^\prime$; error bars in the right panel show the AKARI measurements at the same range of scales for 2.4 \mic. Azure region represent the cosmic variance range of the measured power for the $\Lambda$CDM template in Fig. \ref{fig:cib}.  
%{\bf Right}: Dotted, solid and dashed lines correspond to the default HRK12 reconstruction at the 0.9, 1.1 and 1.6 \mic\ for CIBER. In each panel light shaded areas show the spread defined by the HFE and LFE limits. Blue/red solid line show the high-$z$ $\Lambda$CDM template fit to the power at {\it AKARI} levels and assuming the $\lambda^{-3}$ SED with the azure shade showing the cosmic (sampling) variance in a square of 2$^\circ$ on the side.
}
\label{fig:lbreak_auto_current}
\end{figure}
Fig. \ref{fig:lbreak_auto_current} compares the contributions from the remaining galaxies in each configuration with the signal measured in {\it Spitzer} and {\it AKARI} %or expected at CIBER 
bands. The floor from the galaxies respectively remaining in these configurations appears high and highly systematically uncertain to preclude
a direct probe of $\Delta P$ form the auto-power measurements there. Once the measurement at two wavelengths $\lambda_1$ and $\lambda_2<\lambda_1$ is made, the auto-power from the X-component can be  determined as $P_{\lambda_1}^X=P_{\lambda_1}^{\rm measured}-P_{\lambda_1}^g$ at $\lambda_1$ which is probed to lie below the putative Lyman break wavelength of the X-population. The error on it  is $\sigma_{\lambda_1}^X=\sigma_P + \sigma_{\lambda_1}^g$. Here $\sigma_P$ is the error on the measurement
of the total power, which is at best that from the cosmic variance, and the second term is the {\it systematic} uncertainty of constraining remaining galaxy contribution, which is added linearly. The latter is in practice approximated by the HFE line. 
Thus we appear to be in the regime of $\sigma_{\lambda_1}^X/P_{\lambda_{IRAC}}^X> 1$.

\begin{figure}[ht!]
%\plotone{fig_lbreak_current.eps}
\includegraphics[width=6.5in]{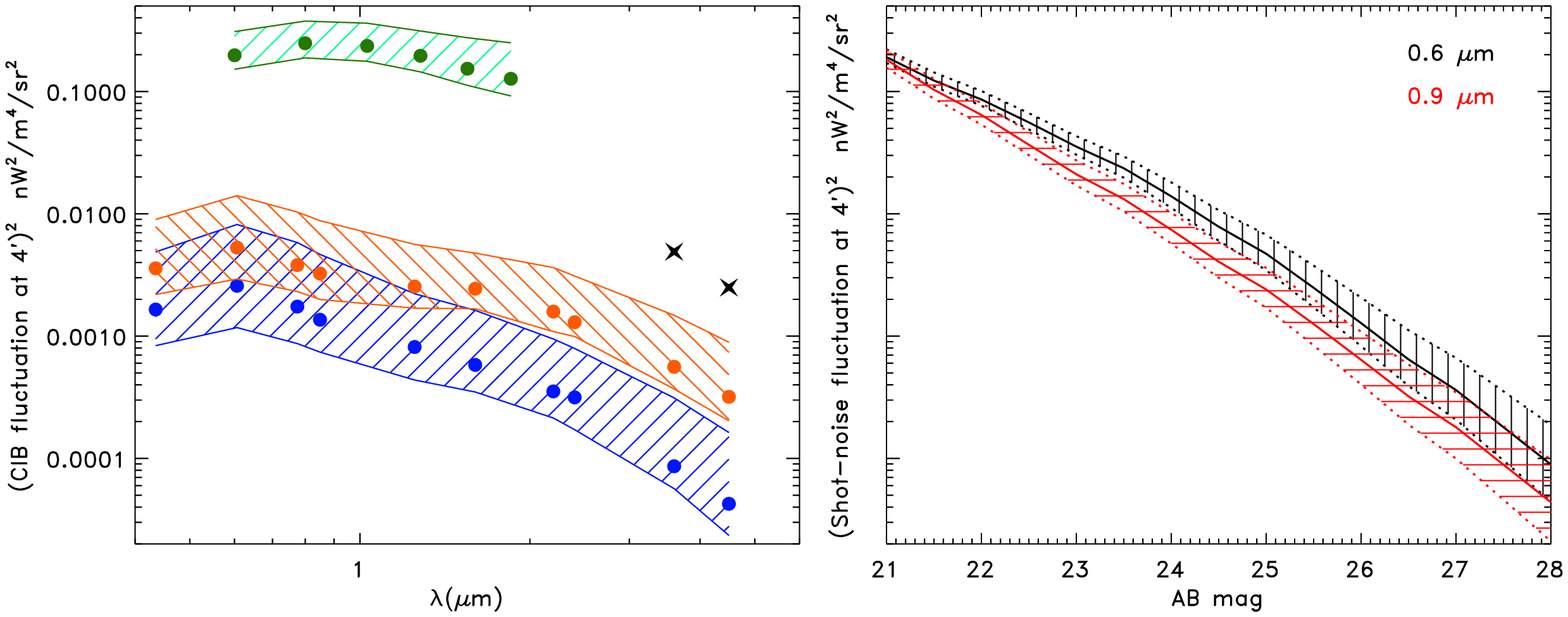}
\caption{\small {\bf Left}: Net CIB fluctuation from clustering and shot-noise at $4'$ due to galaxies remaining in the various current configurations whose
limitations are defined by either the measured shot-noise levels for AKARI (orange) and {\it Spitzer} (blue) or projected magnitude removal thresholds for CIBER2 (green) \citep{ciber2}.  Filled 
circles correspond to HRK12 default reconstruction with the dashed areas marking the limits due to the HFE and LFE limit 
extrapolations. The black 4-pointed stars show the CIB fluctuation levels in the {\it Spitzer} measurements of K12.  The fluctuations from the remaining galaxies in the CIBER2 configuration appear particularly large compared to the CIB signal to preclude its Lyman-break probe in that experiment. 
The fiducial scale of $4'$ is shown because the fluctuations at larger scales remain approximately constant (Fig. \ref{fig:lbreak_auto_current}). {\bf Right}: 
Contribution to the CIB fluctuations at $4'$ from the shot-noise component by known galaxies fainter than the horizontal axis at ACS bands of 0.6
and 0.9 \um.  The vertical axis limits are the same as in the left panel. The filled regions correspond to the HFE to LFE limits. The panels show that at visible bands the diffuse fluctuations from
remaining galaxies are at levels comparable to the CIB fluctuations representing an obstacle to a robust Lyman-break probe for the IRAC-based CIB fluctuations with these instruments.
}
\label{fig:lbreak_cross_current}
\end{figure}

The net known galaxy contribution to the CIB power spectrum is bounded from {\it below} by the shot-noise fluctuation. In fact the dominant term for the fluctuation at visible bands appears to come from galaxy shot noise, which is fixed by the measured galaxy counts.  The galaxy counts, $dN/dm$, at visible wavelengths appear to be such (e.g. Fig.5 of HRK12) that the shot-noise power decreases with increasing $m$ significantly slower than at the IRAC channels. Consequently one needs to go to much deeper integrations in order to reach larger reduction factors in the background fluctuations from the remaining galaxies at visible bands, where the putative Lyman-break is expected to affect energy spectra of high-$z$ sources assumed to be responsible for the source-subtracted CIB fluctuations discovered with {\it Spitzer} and {\it AKARI}. 
%Much of that significant level arises in the harder-to-integrate-down shot-noise term, which is observationally constrained by the measured counts in the visible bands. 
Fig. \ref{fig:lbreak_cross_current} demonstrates explicitly the importance of the shot-noise term (left
panel) and the resultant highly non-negligible level of the source-subtracted fluctuation in the diffuse light at visible wavelengths. It appears that one must eliminate
remaining sources down to magnitudes much fainter than feasible for the current experiments with {\it Spitzer} and {\it AKARI} %and CIBER 
in order to probe the 
possible Lyman-type break of the CIB signal at the levels shown in Fig. \ref{fig:cib}.

%With current experimental facilities, 
%even if the {\it systematic} uncertainty of the reconstructed CIB from known remaining galaxies were reduced to negligible, the contribution driven by the substantial CIB levels from the remaining galaxies at $\lambda_1$ due to cosmic variance will require very large fields to be covered. Indeed, the relative error on $P_{\lambda_1}$ from the cosmic uncertainty variance at, say, $\theta_0$, when measured over the field of area $A$ would be $\simeq \theta_0/\sqrt{A}$ and so to achieve a statistically significant (say at the $\alpha$-sigma level) measurement of negligible contribution at $\lambda_1$ from sources contributing to the near-IR measurements one would require $A\gsim 0.1 (\theta_0/4')^2 (\alpha/5)^2 (P_{\lambda_1}^g/P_{\lambda_2}^X)$ deg$^2$, which appears considerable.

%TBD- remove 0.9 \micron from \ref{ig:results_crosspower}!!

\begin{figure}[ht!]
%\plotone{fig_cross_current.eps}
\includegraphics[width=6.5in]{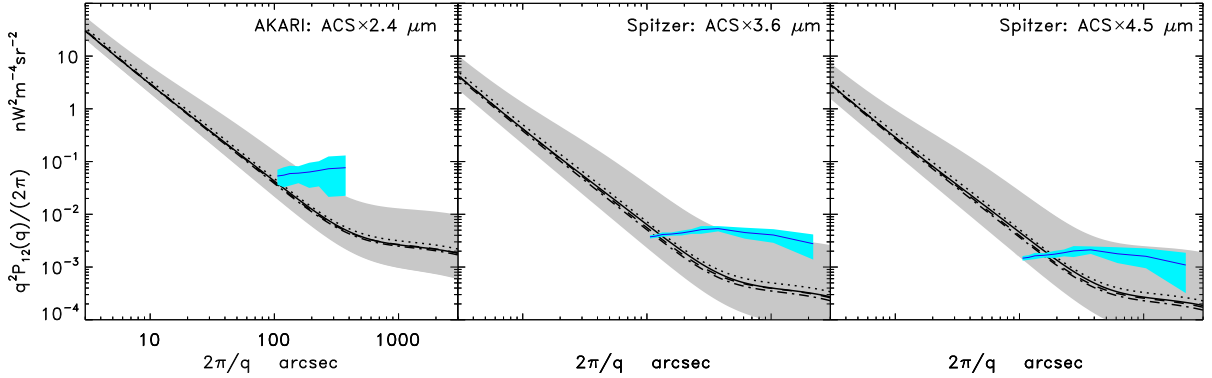}
\caption{\small   Cross-power due to remaining galaxies at ACS bands vs. AKARI and Spitzer measurements of the excess CIB at
the limiting magnitudes shown in Fig. \ref{fig:lbreak_auto_current}.  Same notation as in Fig. \ref{fig:lbreak_auto_current}. The panel shows that the shot-noise component by itself represent a formidable obstacle to the Lyman-break probe for the IRAC-based CIB fluctuations unless populations significantly fainter than AB mag $>27$ are removed.
}
\label{fig:results_crosspower}
\end{figure}

Fig. \ref{fig:results_crosspower} compares the substantial levels of the cross-power with the visible bands from the
remaining known galaxy populations with the CIB signal measured or expected in the various configurations.  Similar argumentation 
to that of the auto-power applies to the cross-power determination, where the {\it systematic} uncertainty from the HRK12 galaxy reconstruction appears at least comparable to the auto-power amplitude detected in the IRAC measurements; being systematic it does not integrate down. 

%{\bf {\it Spitzer} IRAC}
%The CIB fluctuations results obtained by KAMM1-4, AKMM and K12 used IRAC deep exposures at 3.6 and 4.5 \mic, reaching the shot-noise levels of $P_{\rm SN} \sim 30$ nJy$\cdot$nW m$^{-2}$ sr$^{-1}$ . This shot-noise corresponds to sources eliminated from maps out to AB mag of 24.5-25 (KAMM1,KAMM3, HRK12, K12). The fluctuation angular spectrum has been measured by K12 to $\sim1^\circ$ from two large fields obtained with the SEDS project \cite{ashby1} and demonstrated to be consistent with the other five smaller fields used in KAMM1,KAMM2 . 

Even if the {\it systematic} uncertainty of the reconstructed CIB from known remaining galaxies were reduced to negligible, the contribution driven by the substantial CIB levels from the remaining galaxies at $\lambda_1$ due to cosmic variance will require very large fields to be covered. Indeed, the relative error on $P_{\lambda_1}$ from the cosmic variance uncertainty at, say, $\theta_0$, when measured over the field of area $A$ would be $\simeq \theta_0/\sqrt{A}$ and so to achieve a statistically significant (say at the $\kappa$-sigma level) measurement of negligible contribution at $\lambda_1$ from sources contributing to the near-IR measurements one would require $A\gtrsim 0.1 (\theta_0/4')^2 (\kappa/5)^2 (P_{\lambda_1}^g/P_{\lambda_2}^X)$ deg$^2$, which appears considerable.
It appears that at these depths the current experiments will not be able to probe the Lyman-break because of the substantial levels,
and their systematic uncertainties,  of the remaining foreground galaxies. At the depths and wavelengths of IRAC/Spitzer data used in KAMM, the situation is the least pessimistic, but probing the Lyman break is still a tall order. 

Green symbols in Fig. \ref{fig:lbreak_cross_current} show the squared fluctuation at $4'$ in the diffuse light produced by known galaxies projected to remain for the CIBER2 experiment \citep{ciber2}. The systematic uncertainty in this signal  due to HFE vs LFE limits of reconstructions is shown as the green  hatched region. Given the much shallower integrations at the six bands shown (at $m_{\rm AB}\sim 21$ at 3$\sigma$, \citep{ciber2}) these levels are very substantial.   The cross-powers can be estimated from the figure to good accuracy as $P_{\lambda_1\lambda_2}\simeq \sqrt{P_{\lambda_1}P_{\lambda_2}}$ and its systematic uncertainty would be over one order of magnitude above the CIB fluctuation power measured 
at 3.6\mic\ of {\it Spitzer}. This 
suggests that the foreground fluctuations from galaxies remaining at visible bands in the CIBER2 \citep{ciber2} experiment
are large and highly uncertain compared to  the source-subtracted CIB discovered with {\it Spitzer} even in the highly unrealistic case that its levels rise as rapidly as $\lambda^{-3}$ toward $\sim(1-1.5)\mic$. This appears to present highly significant obstacles in probing the Lyman break of the CIB with CIBER2.

%It is important to note that the dominant term for the fluctuation at visible bands appears to come from galaxy shot noise, which is fixed by the counts.  The net known galaxy contribution is bounded from {\it below} by the shot-noise fluctuation. The galaxy counts, $dN/dm$, at visible wavelengths appear to be such (e.g. Fig.5 of HRK12) that the shot-noise power decreases with increasing $m$ significantly slower than at the IRAC channels. 

\subsection{Lyman-break probing with NIRCam}

\begin{figure}[ht!]
\includegraphics[height=4in,width=7in]{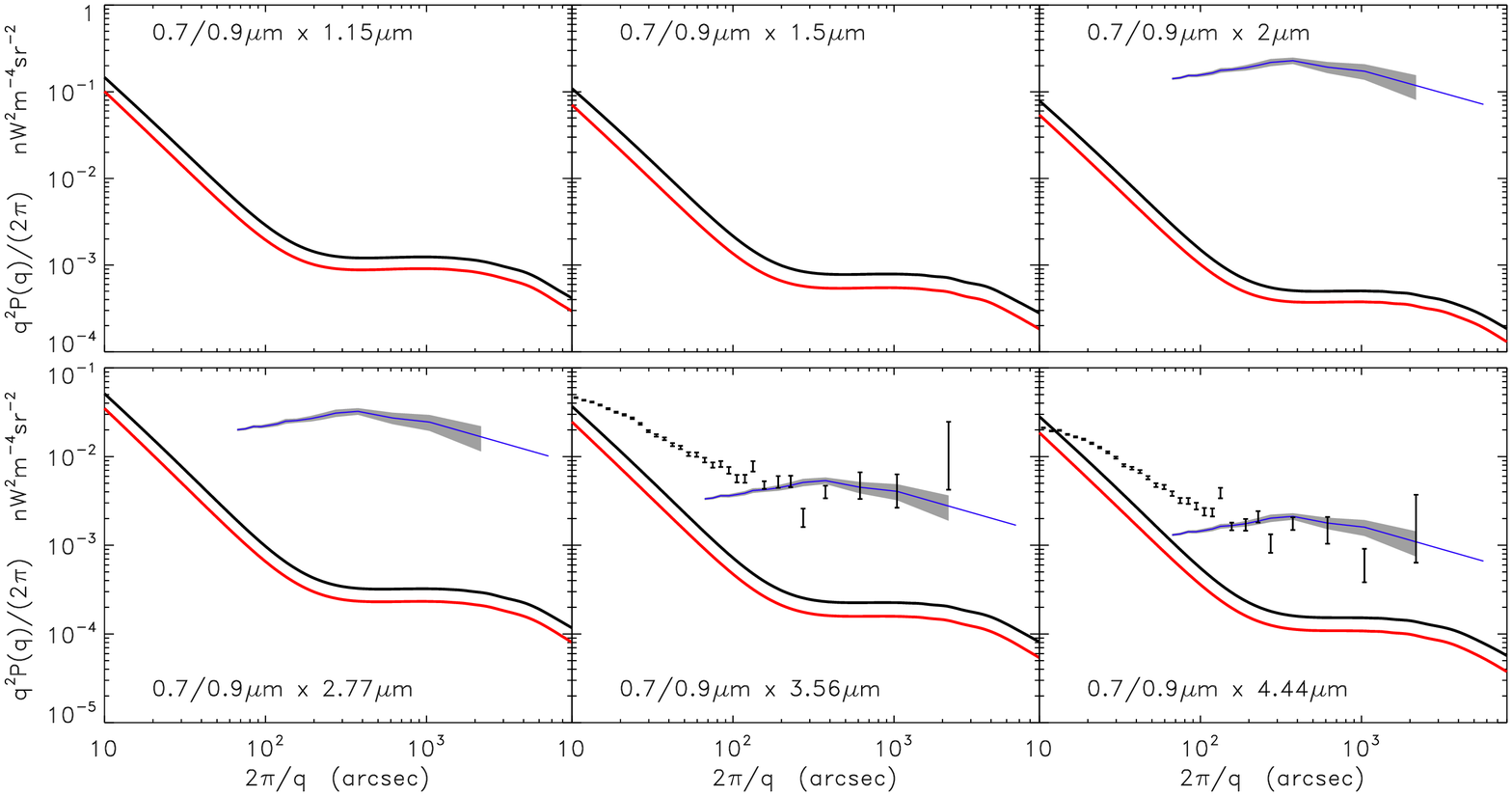}
\caption{\small  Cross-power of the mean squared CIB fluctuation from 1.5 to 4.44 \micron\ form galaxies fainter than $m_{\rm AB}=28$ 
with the 0.7 (black) and 0.9 (red) \micron\ bands. {\it This cross-power determines the floor for measuring the Lyman-break decrease from 
the  CIB sources.}  Error 
bars show the source-subtracted signal from Spitzer measurements out to degree scale from \cite{seds}}
\label{fig:cib_cross_fluc}
\end{figure}
Consequently one needs to go to much deeper integrations in order to reach larger reduction factors in the background fluctuations from them at visible bands, where the putative Lyman-break is expected to affect the high-$z$ sources assumed to be responsible for the source-subtracted CIB fluctuations discovered with {\it Spitzer} and {\it AKARI}.
Fig. \ref{fig:cib_cross_fluc} shows the floor on the Lyman break measurement for CIB fluctuations from the cross-power from
remaining known galaxy populations with the JWST configuration proposed here,  and shows that in this way the Lyman break will be easily probed. Fig. \ref{fig:cib_fluc} and \ref{fig:cib_cross_fluc} demonstrate explicitly that, in contrast to the current experiments, with the proposed
experimental configuration the {\it JWST} will be able to achieve the Lyman-break measurement in the source-subtracted CIB power spectrum.

\section{NIRCam-based tomography: reconstructing the history of emissions directly from two-band cross-power}
\label{sec:tomography}

%discuss the outline.

\subsection{Lyman-break tomography with NIRCam}
\begin{figure}[ht!]
\includegraphics[width=5in]{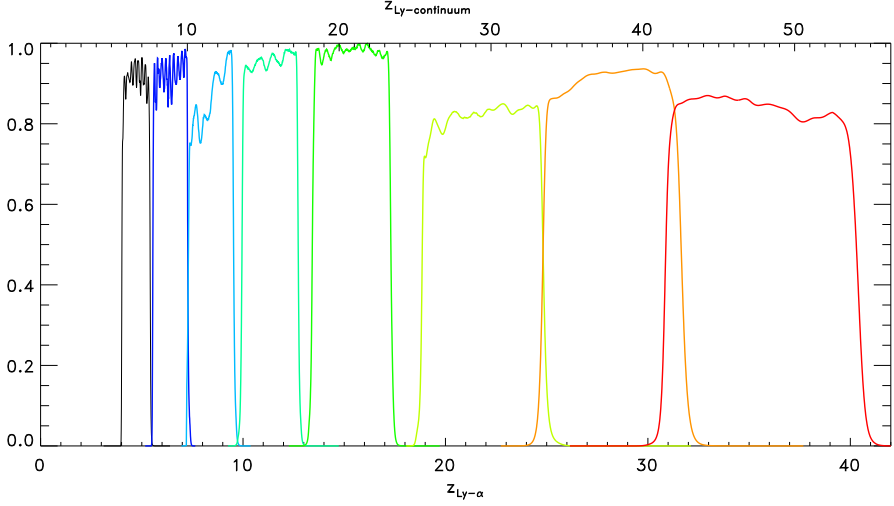}
\caption{Transmission curves of the NIRCam W filters are shown in black. Lower horizontal axis 
shows the Lyman-$\alpha$ redshift for each filter the Lyman-continuum redshift is shown on the top axis. 
NIRCam filters thus probe the entire range of epochs of when the Universe 
would be coming out of the Dark Ages.}
\label{fig:lbreak}
\end{figure}

The positioning of the NIRCam wide filters allows for {\it continuous} probing of the extragalactic diffuse light from below $<1$\um\ to $\sim 5$ $\mu$m. Because one expects no emissions from individual sources below the Lyman wavelength this offers an interesting and unique opportunity to differentiate emissions that enter a longer wavelength filter due to higher $z$ sources from the {\it adjacent} shorter wavelength filter where such sources would not contribute because of the Lyman cutoff in their spectra. The critical wavelength may be as large as that of the Ly$\alpha$ at 0.1216 \micron\ (10.2 ev), but certainly no emissions would be expected below the Lyman continuum wavelength at 0.912\um\ (13.6ev). The sources emitting radiation at epochs filled with even small amounts of neutral hydrogen will likely have their Ly$\alpha$ photons, which escape the parental halos, absorbed by the intergalactic matter (IGM) \citep{loeb-rybicki}. Observations of the Gunn-Peterson absorption trough in quasars indicate the presence of HI in the IGM at $z\gtrsim 6-7$. Since the optical depth to Ly-$\alpha$ photons shortward of the rest $\lambda_{\rm Ly\alpha}\simeq 0.12\mu$m by the uniformly distributed HI with $\Omega_{\rm HI}\sim \Omega_{\rm bar}$ is $\tau\sim 10^5$,
%\gtrsim 10^5$  \citep[e.g.][and references therein]{barkana-loeb}, 
the Lyman-break will occur at the rest Ly$\alpha$ until HI has been exhausted and the Universe reionized to $1-x_e<10^{-5}$. This would argue for the Lyman-break occurring at the rest Ly$\alpha$ wavelength for the sources lying at $z\gtrsim 10$. Fig. \ref{fig:lbreak} shows the positioning of the wide NIRCam filters as function of the redshift of the Lyman-cutoff for the emissions probed by them.

We will assume therefore an absolute cut-off in the energy spectrum of populations so that CIB emissions at $z\geq z_{\rm Ly-break}(\lambda_2)$ are absent at Band 1. Here
\begin{equation}
z_{\rm Ly-break}=\frac{\lambda_{\rm NIRCam}}{\lambda_{\rm Ly-break}}-1
\label{eq:lam_break}
\end{equation}
The uncertainty in $z_{\rm Ly-break}$ is about 20\% reflecting the difference between the cutoff at the Lyman continuum when Lyman photons freely propagate and the Ly$\alpha$ when they are fully absorbed by the gas in the halo and/or nearby IGM (e.g. Santos et al. 2002). The proposed configuration with NIRCam will
provide seven subsequent pairs of the eight adjacent NIRCam wide filters to be used in the differential tomographic measurement here. The subsequent filter will contain the extra populations at  $z>z_{\rm Ly-break}$ compared to the adjacent filter(s) shortward in wavelength.

We first consider a toy model to illustrate the principles involved: denote with $\delta_1$ and $\delta_2$ the diffuse flux fluctuation in bands 1 and 2 and let $\Delta$ be its Fourier transform. 
Then at each band the net flux in each pixel is:
$\delta_1 = f_1 + n_1 \;\; ; \; \; \delta_2 = \alpha_{12} f_1 + n_2 + \Delta F$, where $f$ is the common CIB flux to both bands, 
$\alpha_{12}$ is its spectral energy distribution such that $f_2=\alpha_{12}f_1$,
and $\Delta F$ is the CIB flux in Band 2 which is absent at shorter wavelengths ($\lambda_1$) because of the Lyman-break of
the sources producing it, and $n$ is the instrument noise. The power spectra of the diffuse light after removing resolved sources down 
to some shot-noise level as per IRAC work is measured as 
$P_{11}(q)=\langle \Delta_1(q)\Delta^*_1(q)\rangle \; ; \; P_{22}(q)=\alpha_{12}^2P_{11} + 
P_{\rm \Delta F}(q) \; ; \; P_{12}(q)= \alpha_{12} P_{11}$.
The auto power spectra are determined {\it after} the noise ($A-B$) subtraction and the cross-power $P_{12}$ does not have the noise 
component since the noise at Bands 1 and 2 in uncorrelated. 
So given the data on $P_{11}, P_{22}$ and $P_{12}$ 
we can determine - at each $q$ - the unknowns, $\alpha_{12}$ and 
$P_{\rm \Delta F}$. 
Having measured $P_{11}, P_{22}$ and $P_{12}$, we can solve the above for $P_{\rm \Delta F}$, {\it the power 
spectrum from CIB emissions at $z\geq z_{\rm Ly-break}(\lambda_2)$  to be given by} $P_{\rm \Delta F} = P_{22} - \frac{P_{12}^2}{P_{11}}$.
If $P_{\rm \Delta F}=0$ we recover the full coherence between channels 1,2 with $P_{12}^2=P_{11}P_{22}$.

%\subsection{Lyman tomography with NIRCam}
%The above toy model assumes that the colors between the two bands are constant. 
We now generalize this discussion to
cosmological populations, both known or expected to exist and cluster with the ($\Lambda$CDM) 3-D power spectrum,
$P_3(k, z)$, at redshift $z$. We define with $P_{11}(q,<z), P_{22}(q,<z), P_{12}(q<z)$ to be the auto- and cross-power spectra
at angular scale $2\pi/q$ from populations out to a given redshift. The coherence between the CIB from these is given by:
\begin{equation}
{\cal C}(q, <z) \equiv \frac{[P_{12}(q, <z)]^2}{P_{11}(q, <z)P_{22}(q, <z)}
\label{eq:coherence}
\end{equation}
as per Kashlinsky et al (2012). Note that the coherence is for the same redshift range, $<z$, at both wavelengths.
Populations present via contributions to the CIB 
at Band 2, but not the other wavelengths, would lead to ${\cal C} <1$
{\it provided they are numerous and bright enough to contribute to the source-subtracted CIB fluctuation signal}.

The projected 2-D auto-power spectrum of the CIB produced out to redshift $z$ at band 1
is given by the relativistic Limber equation (see Appendix):
\begin{equation}
P_{11}(q,<z_{\rm Ly-break}(\lambda_1))= \int_0^{z_{\rm Ly-break}(\lambda_1)}
 (\frac{dF_{\lambda_1^\prime}}{dz^\prime})^2 Q(qd_A^{-1}; z^\prime) dz^\prime
\label{eq:limber_1}
\end{equation}
where $d_A$, the comoving angular distance to $z$,  whose integration range 
extends to $z_{\rm Ly-break}(\lambda_1)$. The same equation applies to Band 2, except we write it as:
\begin{equation}
P_{22}  (q,<z_{\rm Ly-break}(\lambda_2))= \int_{z_{\rm Ly-break}(\lambda_1)}^{z_{\rm Ly-break}(\lambda_2)} 
(\frac{dF_{\lambda_2^\prime}}{dz^\prime})^2 Q(qd_A^{-1}; z^\prime) dz^\prime
\;\;+P_{22}(q, <z_{\rm Ly-break}(\lambda_1))
\label{eq:limber_2}
\end{equation}
The first term in the RHS above arises from populations inaccessible to Band 1, so the cross-power between these two bands is:
\begin{equation}
P_{12} = \int_0^{z_{\rm Ly-break}(\lambda_1)}
 \frac{dF_{\lambda_1^\prime}}{dz^\prime} \frac{dF_{\lambda_2^\prime}}{dz^\prime} Q(qd_A^{-1}; z^\prime) dz^\prime
\label{eq:cross-power}
\end{equation}

Now the excess CIB power arising between redshifts $z_{\rm Ly-break}(\lambda_1)$ 
and $z_{\rm Ly-break}(\lambda_2)$, can be expressed as:
%\begin{eqnarray}
%\Delta P  & = & \int_{z_{\rm Ly-break}(\lambda_1)}^{z_{\rm Ly-break}(\lambda_2)} 
%(\frac{dF_{\lambda_2^\prime}}{dz^\prime})^2 Q(qd_A^{-1}; z^\prime) dz^\prime + \nonumber \\
%& & P_{22}  (q,<z_{\rm Ly-break}(\lambda_1)) [ 1- {\cal C}(q, <z_{\rm Ly-break}(\lambda_1))]
%\label{eq:excess_cib}
%\end{eqnarray}
\begin{equation}
P_{\Delta  F} \equiv \int_{z_{\rm Ly-break}(\lambda_1)}^{z_{\rm Ly-break}(\lambda_2)} 
\left(\frac{dF_{\lambda_2^\prime}}{dz^\prime}\right)^2 Q(qd_A^{-1}; z^\prime) dz^\prime  
 = P_{22}|_{\rm DATA} -  \left(\frac{P_{12} ^2}{P_{11}}\right)_{\rm DATA} \; \frac{1}{{\cal C}(q, <z_{\rm Ly-break}(\lambda_1))} 
\label{eq:excess_cib}
\end{equation}

%The meaning of eq. \ref{eq:excess_cib} is clear: $\Delta P$ contains contributions from fluctuations that are present in 
%Band 2, but not in Band 1, and vice versa 
%as well as contributions to $P_{22}$ from populations which are incoherent between the two bands (leading to ${\cal C}
%<1$). The latter would also contain contributions from single channel artifacts and other systematic components which are 
%uncorrelated between the bands. 

Eq. \ref{eq:excess_cib} connects the quantities directly measured from the data, $P_{22},P_{11}, P_{12}$ to the excess CIB power 
from emissions at $z_{\rm Ly-break}(\lambda_1)< z< z_{\rm Ly-break}(\lambda_2)$ and the coherence of the CIB emissions between channels 1 and 2 
from sources out {\it the same redshift}. In other words the measurable quantity, LHS below, is:
\begin{equation}
\left( P_{22} -  \frac{[P_{12} ]^2}{P_{11}}\right)_{\rm DATA} = P_{\Delta  F} + 
\left(\frac{P_{12} ^2}{P_{11}}\right)_{\rm DATA}\; \frac{[1-{\cal C}(q, <z_{\rm Ly-break}(\lambda_1))]}{{\cal C}(q, <z_{\rm Ly-break}(\lambda_1))} 
\label{eq:excess_cib_1}
\end{equation}
We note that the quantity on the left-hand-side of the above equation is {\it always positive} for correctly measured CIB, since the coherence ${\cal C}=\frac{[P_{12} ]^2}{P_{11}P_{22}}\leq 1$.
%; this will be violated for the CIBER-claimed measurements as discussed below.

Provided the coherence term is sufficiently close to 1 (discussed below) one can reconstruct the history of emissions from two-band NIRCam cross-
and auto-power analysis all the way to redshifts given in Fig. \ref{fig:lbreak}. The lack of coherence (${\cal C} \ll1$), if found for CIB sources, would by itself be an important result as well. In any event, since coherence is bounded by 1 from above, eq. \ref{eq:excess_cib_1} imposes an {\it upper limit} on any emissions at $z>z_{\rm Ly-break}(\lambda_1)$.

The auto power spectrum at Band 2 is measured to within the cosmic variance uncertainty of 
$\sigma_{2}\simeq P_{22}/\sqrt{N_q}$, where 
$N_q$ is the number of independent Fourier elements which went into determining $P_{22}$ (see Fig. \ref{fig:cv}). 
Because $N_q$ then depends on the field 
configuration and parameters, with a properly selected observational configuration one can measure $P_{\rm \Delta F} $ down to the 
level of $\sigma _2$. The error on the $P_{\rm \Delta F} $ is 
\begin{equation}
\sigma_{P_{\rm \Delta F}}\simeq P_{22}\left(\frac{6}{N_q}\right)^{1/2} \simeq 0.017 P_{22} \left(\frac{N_q}{2\times10^4}\right)^{-1/2}
\label{eq:sigma_tomography}
\end{equation}
where we have substituted the net $N_q$ at $>30''$ in area of $A=$1 deg$^2$ square region ($N_q\propto A$) evaluated per Fig. \ref{fig:cv}. With {\it Spitzer} data we are in the regime where the large-scale power (say at scales $>100''$) is dominated by the clustering component from new populations. If one assumes a template at those scales, such as $\Lambda$CDM at a specific $z$, the overall error on its amplitude ($A_{5'}$ in the terminology of \cite{seds}) will be given by the total $N_q$ at such scales. E.g. for $2\pi/q>100''$ and a 1 deg$^2$ square field, the total $N_q=2,024$ leading to 5\% sampling uncertainty in determining the band-averaged amplitude for a given template. Moreover, by reducing the shot-noise from remaining known galaxies to much lower levels than in the IRAC measurements with the deeper integrations, the proposed experiment would isolate the CIB fluctuations from the new populations at smaller angular scales leading to a higher $N_q$ overall. Coupled with the expectation that CIB from the new populations will increase toward shorter wavelengths, should allow probing the history of emissions at the redshifts shown in Fig. \ref{fig:lbreak} to levels below $\lesssim 1\%$ of those measured at the individual NIRCam filters.

To see how well this reconstruction of emissions to high $z\lesssim 30-40$ works we need to compute the contamination term - the second term in the RHS of eq. \ref{eq:excess_cib_1}.  Before we do this, we note - with the current {\it Spitzer}-based measurements at least - that there are two ranges of scales 1) whereas the large-scales appear dominated by the clustering CIB component which is much greater than that remaining from known galaxy populations, 2) the small scales are explained by the shot-noise from known remaining galaxies and contain a much smaller shot-noise component from the new populations. Since we do not know at what integration depth the clustering component will start decreasing in sync with the shot-noise, we discuss the prospect of measurement the clustering component of the CIB in eq. \ref{eq:excess_cib_1} at progressively larger redshifts.

%
%TBD: need to compute/evaluate: 1) $C(q)$ vs. scale and limiting magnitude from know populations as in Helgason et al (2012), 
%2) coherence from new high-$z$ populations - choose a high-$z$ model and do semi-analytical analysis, 
%3) artifacts contributions, 4) etc.
%
At large scales where the clustering component dominates over the shot noise, the {\it Spitzer/AKARI} measurements indicate that the new component, ``X'', produces CIB fluctuations with power significantly exceeding that from known galaxy populations, denoted by ``g''. I.e. at the {\it Spitzer}/IRAC band ``1'' (3.6\mic) we have with the
proposed experimental configuration $P_{1}^{\rm g} \equiv \epsilon_1 P_1^{\rm X}$ with $\epsilon\sim 3-5\%$. Thus we can evaluate the ``confusion'' term in eq. \ref{eq:excess_cib_1} to be determined by:
\begin{equation}
\frac{1-{\cal C}}{{\cal C}} \simeq [\frac{1-{\cal C_{\rm X}}}{{\cal C_{\rm X}}}]\;\; +\frac{\epsilon_1+\epsilon_2 -2 \sqrt{{\cal C}_{\rm g}/{\cal C}_{\rm X}} (\epsilon_1\epsilon_2)^\frac{1}{2} + O(\epsilon^2)}{{\cal C}_{\rm X}}
\label{eq:pollution}
\end{equation}
%If ${\cal C}_{\rm g}\simeq {\cal C}_{\rm X}$ the last term above becomes $O(\epsilon^2)$.
In other terms in the regime established from the {\it Spitzer} and {\it AKARI} measurements the confusion is determined by the level of coherence of the {\it new} populations with small/negligible (of order a few percent) contribution from the remaining known galaxy populations.  If the new populations are at early times, it is likely that they are highly coherent with ${\cal C}_{\rm X}$ being very close to 1. If so, with the proposed tomography, {\it JWST} can identify the contribution of progressively higher-$z$ populations in Fig. \ref{fig:lbreak} down to the level of a few to $\lesssim 1$ percent of the power measured at Band 2.  At small scales, where the shot-noise dominates, we may well end up measuring the incoherence of the (dominant) shot-noise due to differential source subtraction at the two bands. We will, however, have determined in the course of this experiment whether the clustering component is correlated with the reduction in the shot-noise down to very low levels of $< 1$ nJy$\cdot$nW m$^{-2}$ sr$^{-1}$ (Fig.  \ref{fig:shotnoise}).

Assuming the measurement has been done, the results will include the following possibilities:
1) Remaining floor from remaining known galaxies, 
2) Term due to incoherence of the new populations, or
3) The power emitted from $z_{\rm Ly-break}(\lambda_1)<z<z_{\rm Ly-break}(\lambda_2)$.
Eq. \ref{eq:pollution} shows that the first possibility is attenuated by a factor of $\epsilon$.
The proposed NIRCam-based tomography will thus result in three possible outcomes: 1) we will probe the CIB from $z>z_{\rm Ly-break}$ down to the floor fixed by $O(\epsilon)/C_X$, 2) probe the lack of coherence of the new sources, or 3) measure the CIB produced at  $z_{\rm Ly-break}(\lambda_1)<z<z_{\rm Ly-break}(\lambda_2)$ for each pair adjacent of bands. In the worst event we will have measured the {\it upper} limit on the CIB at $z_{\rm Ly-break}(\lambda_1)<z<z_{\rm Ly-break}(\lambda_2)$ for each pair of bands to unprecedentedly low levels. With this we will measure excess over eight filters (of eight $\Delta z$) and this gives consistency check vs. LCDM 
distribution at those $z$. Fig. \ref{fig:lbreak} shows how far in redshift the NIRCam filters can probe the CIB emissions with this tomography method, which may be competitive with/complementary to the prospective 21 cm measurements for properly designed configurations.

%TBD: discuss the shot-noise regime, where we may reach also that $P_{\rm SN}^X \gg P_{\rm SN}^g$.

\subsection{Application of Lyman-break tomography to current {\it Spitzer}/IRAC data}
\label{sec:spitzer_tomography}

Fig. \ref{fig:NIRCam} shows the good similarity of the IRAC shortest band filters to the longest two filters of NIRCam. So in order to test the proposed tomography method, we applied it to the data analyzed by us in \cite{seds}. 
%Fig. 9 of K12 shows the power at 3.6 and 4.5 \um\ along with the cross-power from the two datasets analyzed there. 
The numbers there were used to construct according to eq. \ref{eq:excess_cib_1} the excess power component that arises at redshifts where the Lyman-break populations are present at 4.5 \um, but not at 3.6\um\ ($30\lsim z\lsim 40$ assuming the Ly-break at these pre-reionization epochs due to Ly$\alpha$ absorption).

The data analyzed in K12 consist of two regions of $21'\times21'$ (UDS) and $8'\times 62'$ of similar integration depth. The regions have full overlap between 3.6 and 4.5\um\ so both the auto power and the cross power were measured as shown in Fig. 9 of K12. However, because for the EGS region there appeared a low-level large-scale artifact at 3.6 \um\ which artificially suppressed cross-power at $>1,000''$, we restrict the analysis here to scales below 1,000$''$.

\begin{figure}[h!]
\includegraphics[width=5in]{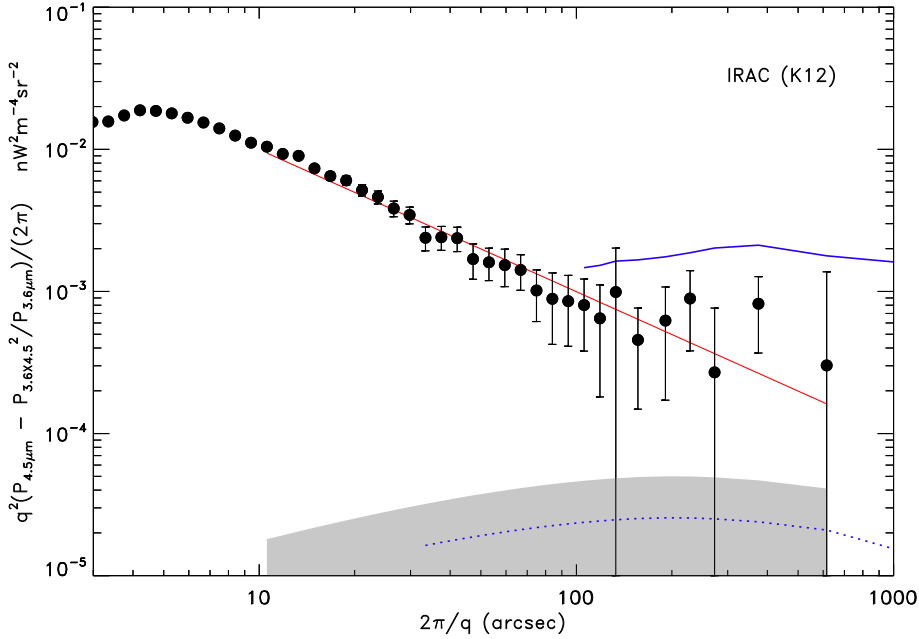}
\caption{\small 
The Lyman-break based tomography reconstructed from the current {\it Spitzer}/IRAC measurements of K12 at 3.6 and 4.5 \um\ are shown with filled circles. Red solid line shows the $P\propto q^{-1}$ template that fits the data and is consistent with the non-linear clustering of
known galaxies remaining after differential subtraction at the two bands.  The high-$z$ $\Lambda$CDM template  that fits the data at 4.5 \micron\ 
is shown with the blue solid line and the filled region shows the $1\sigma$ limit on the amount of power remaining then for populations at $z>z_{\rm Lyman-break}(4.5\mic) \gtrsim 30$. The amount of power left for these populations is at most $\sim2\%$ of that measured in K12. Blue dotted line shows the $1\sigma$ upper limit on the remaining power for the 1 deg$^2$ configuration discussed here.  Since coherence is defined to be strictly bounded by unity from above, the central points
lead to $\Delta P>0$.
}
\label{fig:seds_lybreak}
\end{figure}
Fig. \ref{fig:seds_lybreak} shows the resultant numbers with filled circles and $1\sigma$ errors, which are close to the estimate in eq. \ref{eq:sigma_tomography}. The slope of the fluctuations is very close to that of non-linear galaxy clustering produced by differentially removed
sources at the two IRAC bands.
We decompose the data shown in the figure into 1) shot-noise, 2) non-linear clustering from remaining differentially removed galaxies at the two
IRAC bands, assumed to follow $P\propto q^{-1}$, consistent with the 2MASS CIB measurements \citep{2mass}, and 3) high-$z$ LCDM and evaluate the amplitudes. 
%TBD: Kari - add discussion here of the HRK12 reconstructions in this context.

The red solid line in the figure shows the resultant fit from the non-linear clustering component. The fit uncertainty on its amplitude is about $\simeq 10\%$ and is not shown in this logarithmic plot.
In the presence of the empirically determined remaining galaxy component, the amplitude of clustering component with template of the concordance $\Lambda$CDM power spectrum at $z\simeq 30$ (see Fig. \ref{fig:lbreak}) is shown at its $1\sigma$ upper limit, which is in good agreement with estimating the error according to eq. \ref{eq:sigma_tomography}. 
The result shows that the method is robust and we can already constrain the contribution to the power measured at 4.5 \um\ by \cite{seds} to be at most 2\% from $z\gtrsim 30$, setting the best upper limits available to date for emissions
from these epochs. 
%TBD: Kari, can you quantify ``non-linear'' component of this via HRK12?
%Use curves in Fig. \ref{fig:cib_byz} to evaluate the redshift and uncertainty from fitting the templates to the power for the chosen field 
%configuration.

%\begin{figure}[ht!]
%\plotone{coh_og.eps}
%\caption{
%Coherence of remaining known populations between the adjacent NIRCam bands, starting with F090W/F115W pair.
%}
%\label{fig:coherence_og}
%\end{figure}

%\subsection{NIRCam band filter configuration}

Because the JWST proposed configuration will cover a larger area {\it and} remove known galaxies to much fainter levels, we would be able to isolate the CIB contribution to much lower levels by doing the Lyman-tomography at 3.6 and 4.5 \um. Assuming that the shot noise level from known galaxies is decreased on par with Fig. \ref{fig:shotnoise}, we assume that the clustering component from the new population will then dominate already at scales$\sim 30''$, leaving the net of $N_q\gtrsim 20,000$ there for the broad band power estimation when fitting the CIB fluctuation amplitude with the assumed $\Lambda$CDM template. This would enable probing the CIB emissions from $z\gtrsim 30$ down to the 1$\sigma$ levels below 1\% of the power measured with {\it Spitzer}/IRAC at 4.5 \um\ as shown with the blue dotted line in Fig. \ref{fig:seds_lybreak}.
%\section{Diffuse light contributions from noise, systematics and foregrounds}

\subsection{Application of Lyman-break tomography to {\it CIBER} data}
\label{sec:ciber_tomography}
Given the CIB results  of CIBER \citep{ciber} we also applied the proposed tomography to that data at the adjacent wavelengths of 1.1 and 1.6 \micron. %Fig. \ref{fig:ciber_lybreak} shows 
The results of this application further demonstrate the apparent problems with that data. The value of $P_{\Delta F}$
must always be positive yet here, as a consequence of the coherence exceeding unity (Fig. \ref{fig:coherence_ciber}),  the value of the
derived $P_{\Delta F}$ is negative at both the small and large angular scales, and where it is positive, no clear signature can be discerned and the signal
appears not consistent with the power spectrum from either the shot-noise or non-linear clustering from the remaining sources. 

%\begin{figure}[h!]
%\includegraphics[width=5in]{fig22.eps}
%\caption{\small Same as shown in Fig. \ref{fig:seds_lybreak}, except for CIBER data from \cite{ciber} at 1.1 and 1.6 \micron. The value of $P_{\Delta F}$
%must always be positive yet here it is significantly negative, as a consequence of coherence exceeding unity, is shown (in absolute value) with open circles. Filled circles show positive values of $P_{\Delta F}$, which is not consistent with either the shot-noise or non-linear
%clustering. %This further illustrates implicit problems with the CIBER data.
%}
%\label{fig:ciber_lybreak}
%\end{figure}

%begin Mather editing

%\section{Observing Approach}
\section{Non-extragalactic Signals}
\label{sec:foregrounds}

We have outlined above the science and methodology to gain, via CIB fluctuations with the suitably configured {\it JWST} NIRCam instrument, the important knowledge on the emergence of the Universe from the ``Dark Ages''. Here we describe our approach to estimating, modeling, and removing the effects of various astronomical foregrounds and potential faults of the instrumentation. We begin with instrumental and telescope dependent items, and then consider solar system and Galactic foregrounds.

\subsection{Calibration}
Photometric calibration will start with the standard calibration produced by the {\it JWST} pipeline processing, but we do not yet know how accurate this will be, nor how it will be accomplished. Self-calibration can be more efficient and more accurate than separated calibration steps \citep{holmes}, and has been developed and applied for the {\it Spitzer} (our work), SDSS \citep{padman}, and PanSTARRS \citep{schlaf}.  We anticipate that our calibration needs are more stringent than for observers interested in point sources, and that we will need to apply the self-calibration developed by \cite{fixsen}.  (See also \cite{anders}, and other online documentation from the {\it HST} and {\it Spitzer} projects.) Self-calibration measures the calibration parameters directly from observations of the sky, using least-squares fits or other optimizations such as likelihood functions, and can be very efficient and accurate if the model of the observatory and instrument includes all of the significant errors and features.

Flat fields (response to a uniform illumination)%, or to zero illumination) 
can be determined in many ways that effectively ignore bright sources, and average over the smooth backgrounds in many images.  However, this is not sufficiently accurate when measuring low-surface-brightness large-scale features.  In addition, smooth backgrounds are sometimes used to determine detector gains as well, but this may not be accurate enough if the backgrounds have spatial gradients for reasons described below.  In order to separate detector gain, offset, and smooth background light levels, observations must be taken that expose each detector pixel to a large dynamic range. For instance, observations of a star cluster where most pixels have significant brightness can be used; dithered observations change the brightness on each pixel and enable simultaneous solution for the detector offset, gain, and star cluster map. This method differs from using changes of the zodiacal light background, in that the star cluster has fine spatial structure tied to discrete objects (most are presumed to be stable), while the smooth zodiacal light is subject to contamination from stray light paths, as described below.

Note that photometric calibration for surface brightness of extended features is not identical to photometric calibration for point sources. In ideal cases the connection is simple if the point spread function is known and constant, but for {\it JWST} the point spread function varies with position on the detector, and might be significantly time-dependent, depending on the wavelength, the mirror figure stability and the occasional adjustments made to optimize image quality. In addition, we will have the opportunity to confirm from observation that the geometric mapping from pixels to celestial coordinates preserves surface brightness calibrations.

\subsection{Noise}
The NIRCam sensitivity is primarily determined by photon noise and not by the detectors. In our application, where we search for relatively large-scale features, this noise source is important for detecting and masking out the foreground objects (stars and galaxies) as it controls how many can be recognized. It is  the basis of the exposure time calculations given above.  While the random effects of photon noise and instrument noise cannot be removed from the images, we can use the difference between independent images of the same location (e.g. 
two images made from separate subsets of the data) to measure and subtract the contribution of the noise to the derived power spectrum.
After this step, ideally the dominant random noise term should be remaining shot noise from the unresolved galaxies. %However, systematic errors of many types are important, and we outline our approach below.

%\subsection{Stray light, gradients, dithering etc}

%\subsection{Additional Foregrounds: 1. Zodiacal light}

%\subsection{Foregrounds: 2. Galactic cirrus}

%\section{Unresolved sources}
%
%Evaluate source-subtracted CIB fluctuations from each component given a set of NIRCam observation parameters.
%
%\subsection{First stars and quasars}
%
%\subsection{Faint luminosity end of known galaxies}
%
%\subsection{Probing the Lyman break}
%

% Per John's e-mail of 1/26/14, updated by JCM Nov. 2014

%The JWST was optimized for sensitivity and angular resolution, and not for the measurement of low-surface-brightness features such as CIB fluctuations over scales larger than the field of view, so our observing strategy is designed to measure the possible error sources and enable their removal in analysis.  The measured CIB fluctuations are small residuals after compensating for detector and electronic features, and after removing (by masking and modeling) all point sources down to specified magnitude limits.%, and it is always difficult to be sure of such a process.
%
%The general strategy is to observe each field with a wide variety of instrument pointings and orientations, possibly with different detector parameters, and at many different times during the mission. In this way, either the systematic errors will be randomized to that they can average out, or they will be recognizable through internal comparisons of data that ought to give the same answers but do not. In that case, the instrument models would be augmented to include the newly discovered features.

%Specific issues to be addressed include:

\subsection{After-images}
As the measured CIB fluctuations are faint compared with bright foreground objects, we will  observe in a way that can reveal and eventually remove the effects of faint residual images. Each new pointing will need at least two image frames so that we can compare them and detect the recovery from bright objects, whether they were intentionally observed in the previous frames, or were only transient during slews to new positions. The requirement for recovery conflicts with the requirement for rapid rastering to construct mosaics, so optimization will be based on a model for the recovery behavior. %The Spitzer MIPS data show both gain and offset changes after exposure to bright sources (see the MIPS instrument handbook reference),  so both might also be needed in a model for the JWST photometry on extended sources.

\subsection{Stray Light and Ghosts}
The {\it JWST} was designed with attention to stray light and ghost images, the diffraction pattern due to the segmented hexagonal primary and known optical element errors is calculable, and the point source response function will be measured in flight by scans of bright objects during the on-orbit checkout period. However, due to its open design, unlike the {\it HST} and {\it Spitzer} observatories, it is vulnerable to illumination from angles far from the line of sight. Hence, there is a possibility that there are additional sources of low surface brightness  large-scale features that might escape standard measurement processes, so we suggest some observations that could detect them. There are several mechanisms that have to be checked.

\subsubsection{Multiple Internal Reflections and Glints}
In this case, a ray bundle from a celestial source comes through the telescope and instrument, but instead of making it all the way to a detector, or to absorption, it can bounce from a detector surface, from a filter or lens surface, and then again from another similar surface, returning to a possibly different detector as a ghost image, usually not in focus, so a point source would appear to be a circle or hexagon or an image of the primary mirror,  possibly partially obscured or vignetted or distorted. Sometimes, a chance alignment of a star image on a reflective edge (say the edge of a field stop or detector frame) can function as a new source of light that varies with pointing. The original source need not be in the instrument field of view for this to happen. One type of target that could reveal such ghosts would be a selected relatively dark (perhaps dust-obscured) field near a star cluster or galaxy, even the SMC or the LMC,  observed multiple times with different observatory orientations. Then, differences between the observations could be traced to stray light from the star cluster, and since the cluster has significant brightness contrast, patterns could be recognized.  The ideal field near a cluster might be at high ecliptic latitude so that the widest range of observatory roll angles would be available throughout the year. 

\subsubsection{Truant and Rogue Paths}
The {\it JWST} has  two known sources of out-of-focus stray light that might still produce spatial structure or gradients on the images. First, unfocussed sky light can pass directly into the aft-optics and instrument chamber through the small aperture that also permits the main beam to enter. Some of this light will fail to be absorbed by internal baffles, and could then pass through to the instruments. Second, light from the sky passing near the edge of the primary mirror can reflect from the secondary mirror or its support structures, and into the same instrument volume. In this case the light will be imaged by the tertiary mirror and will be focussed near the edge of the fine steering mirror, where a stop can block it. However, this stop was not optimized to control stray light, but rather to maximize observatory sensitivity, so some of the stray light can still reach the detectors.  In the instruments this stray light appears to originate near the edge of the primary mirror, and can be partially vignetted by other stops in the system, so there is a way for the stray light to produce a kind of shadow image on the detectors and hence a large-scale spatial gradient that is fixed in observatory coordinates. Such spatial gradients might be different according to where the offending bright object is located. Quantitative models of the JWST show that these effects should be negligible for all purposes except finding low surface brightness features.

Observations that could reveal the effects of these particular paths would put known bright objects (e.g. star clusters, or the Galactic plane or Galactic center) in the expected vulnerable areas on the sky. We would choose particular target areas that could be observed with multiple orientations of the bright celestial illuminators, and compare the observations. For example, assuming that the vulnerable area is 38\deg\ from the line of sight, we would find a target on the circle of radius 38\deg\ around the Galactic center, and arrange to take data when the Galactic center was filling one of the vulnerable areas. The same target would then be observed at another time when the Galactic center is hidden, say when the Sun is near the Galactic center so that the sunshield offers good protection. Similarly, for the paths going near the edge of the primary and bouncing from the secondary into the telescope, there is a circle of radius about 45\deg\ centered on the Galactic anticenter, where targets can be observed with maximum interference possibility, and again when the interference is well blocked by the sunshield.

Observations for the CIB will need to be carried out in a way to reveal and characterize or limit any such effects. The general strategy of observing with many observatory orientations should be sufficient once the general degree of the issue is known from specialized tests.

\subsubsection{Mirror Scatter}
The diffraction calculation for the {\it JWST} point spread function can not fully include ``wide'' angle scattering by dust on the mirrors or the details of the turned edges of each primary mirror segment. Also, the dust population is likely to change during launch, and the continuing bombardment by micrometeoroids will gradually build up a population of small pits, of the order of 30 $\mu$m in size.  The pre-launch model for these pits shows that about 0.1\% of the primary mirror will be covered with pits after 10 years of operation. However, such scattering could be significant to CIB fluctuation measurements if it can produce a spatial gradient at the detectors.  The necessary tests would be like those for multiple internal reflections and ghosts, using a dark target near a bright cluster. The test for scattering from the primary mirror turned edges would be best if the bright cluster were located in the directions perpendicular to a mirror edge, i.e. in the direction of a diffraction spike.

%\subsubsection{Mosaics}
%Measuring a large-scale map feature by building a mosaic from small patches requires multiple cross-checks. Our strategy is to build up mosaics from observations connected in as many different ways as possible, rastering in at least two directions as fast as possible to deal with time-dependent errors and brightness changes (such as zodiacal light), with significant overlaps of individual frames as well. In addition, mosaics taken at different times of year, when the telescope orientation is different, must give the same answers.  This problem is mathematically similar to building up a cosmic microwave background map from differential observations, as was done with the COBE and WMAP, and somewhat differently with the Planck mission. A full simulation of the map construction process is required to optimize the original data taking procedures. \cite{holmes} simulated four different surveys and found that the selection of overlapping regions is extremely important, so much that one method failed altogether in the self-calibration process, while another did very well at recovering the input parameters in a simulation.

\subsection{Cosmic Rays}
Like {\it Spitzer}, {\it JWST} will be in deep space and exposed directly to galactic and solar protons and heavier particles, at a rate of the order of 4 particles/cm$^2$/sec, depending on the solar cycle because the solar wind carries galactic cosmic rays outwards and protects the {\it JWST} from them.  The {\it JWST} detectors will be ``sampled up the ramp,'' meaning that the charge on each pixel will be read many times and reported before the pixel is reset.  The comparison of these multiple samples offers the possibility of detecting and compensating for the effects of individual cosmic rays.  We will not have detailed information about long-lasting after-effects of cosmic ray hits until the {\it JWST} is launched and checked out in orbit. But most cosmic rays deposit charges small compared with the full well depth of each pixel, most cosmic rays hit more than one pixel, and charge can leak from one pixel to another.  All of these are potentially important for accurate photometry. Depending on detector performance, we may be able to measure and compensate for each individual cosmic ray, or we may have to reject data taken by the affected pixels for some recovery time.

\subsection{Zodiacal Light}
The zodiacal light is bright relative to the CIB, and could in principle have structure on scales comparable to those of interest for the CIB fluctuations. 
Studies intended to measure random, small-scale structure in the zodiacal light 
have set limits on the structure at $\sigma_{I_\nu}/I_\nu < 0.2\%$ \citep{abraham}, and more restrictively, $\sigma_{I_\nu}/I_\nu < 0.03\%$ at scales of $200''$ \citep{pyo}.
More specific structures are generated by comets in the form of comet dust trails \citep{sykes}. These dust trails are typically observed at mid-IR
wavelengths and in close proximity to the parent comet \citep[e.g.][]{reach}.
Recently, in the 12 and 25 \micron\ {\it COBE}/DIRBE data, \cite{arendt14} has found that a few comet dust trails can be 
detected on larger angular scales, at higher ecliptic latitudes, and further from the parent body than expected. The peak brightness 
of the trails can be $\sim1\%$ of the zodiacal light intensity. However the 
trails exhibit high proper motion, and would be a transient perturbation 
($\sim1$ day duration) for any CIB observations.
In addition, the zodiacal light has large scale gradients from the ecliptic plane to the ecliptic poles, and from the Sun outwards, and near the ecliptic plane has several bright bands due to orbital debris from collisions of certain asteroid families. The standard zodiacal light model in use was developed as a parametric fit to the {\it COBE}/DIRBE data by \cite{kelsall}.  This model uses a main cloud distribution of dust that is a power law in distance from the Sun, combined with a function of (heliocentric) ecliptic latitude. It also includes components for the asteroidal dust bands, and for dust that is held in orbital resonance with the Earth including a trailing blob of dust along the Earth's orbit. The residuals from this zodiacal light model are typically in the percent range. \cite{kelsall} reported a residual periodic variations at the 1-2\% level, which is correlated the solar \ion{Mg}{2} index, although the mechanism for the correlation remains unknown.

The observing strategy for zodiacal light is to take measurements at many different times, so that features in the zodiacal light can be detected as residuals to a best fit, and so that the large scale gradients and time variations can be recognized and modeled.

\subsection{Galactic Stars}
At near-IR wavelengths, the luminosity of the Galaxy is dominated by starlight. 
This starlight is a significant impediment to CIB measurements with very
low angular resolution, such as DIRBE \citep[see][]{arendt}. However, with large telescopes and deep 
imaging in high latitude fields, essentially all foreground Galactic stars can 
be individually resolved and detected. Therefore they have no impact on
the measurement of source subtracted background fluctuations. 
At a depth of $m_{AB} = 27$, even faint M dwarfs can be detected out to 
distances of $\sim 20$ kpc, and at high Galactic latitude ($|b| > 30$, 
and away from the bulge), the density of stars is $<10$ amin$^{-2}$.

\subsection{Galactic ISM}
\label{sec:cirrus}

Galactic ISM emission (``cirrus'') and scattered light (diffuse Galactic light, DGL) represent a potentially large foreground which needs to be assessed when measuring CIB fluctuations. The cirrus emission at longer wavelengths ($\lambda \gtrsim 3$ \micron) is thermal emission from 
radiatively heated dust. At shorter wavelengths cirrus is visible as scattered light of the interstellar radiation field.
The cirrus emission is highly variable across the sky, with a general decrease in intensity with Galactic latitude, as any Galactic component. Many locations for deep extragalactic surveys are selected, in part, to lie in regions that are local minima in the cirrus emission. 
The Extended Groth Strip (EGS), Chanda Deep Field South (CDFS) and the Lockman Hole (LH) are some survey fields where the cirrus emission is near the absolute minimum.

The cirrus emission adds another term, $P_{\lambda}^{ISM}(q)$, 
to the measured power spectrum of the 
IR background. This contribution can be measured (or estimated) and subtracted,
but this subtraction does include associated uncertainties which 
add in quadrature to the total uncertainties of $P_{\lambda}^X(q)$. It is 
very difficult to make direct measurement of the intensity or power spectrum
of ISM emission at short wavelengths ($\lambda \lesssim 10$ $\mu$m) in the fields
typically used for CIB studies, because the ISM is intrinsically faint at these 
wavelengths and because these fields are specifically chosen
to be on lines of sight with minimal extinction, \ion{H}{1} column density, and
ISM emission. Therefore at near-IR wavelengths, the ISM emission and power 
spectra are generally extrapolated from the intensity or power spectrum measured
at other wavelengths and/or locations where the ISM is brighter. This measurement 
can then be rescaled appropriately:
\begin{equation}
P_{\lambda}^{ISM}(q, [l,b]) = 
P_{\lambda_0}^{ISM}(q, [l_0,b_0])\ 
\left(\frac{I^{ISM}_{\lambda}}{I^{ISM}_{\lambda_0}}\right)^2\
\left(\frac{I^{ISM}([l,b])}{I^{ISM}([l_0,b_0])}\right)^2\ =
P^{ISM}_0\ C(\lambda,\lambda_0)^2\ R(l,b;l_0,b_0)^2 
\label{eq:ism1}
\end{equation}
where the second factor on the right is the color of the ISM emission
between the reference and the desired wavelengths, and the last factor on the
right is the scaling between the mean ISM intensity at the location where the power 
spectrum was measured and the desired location. The color term will not 
be dimensionless if the power spectrum is measured using an alternate tracer 
of the ISM, such as the \ion{H}{1} column density. 
%For Eq. \ref{eq:ism1} the fractional variance of 
%$P_{\lambda}^{ISM}$ is given by 
%\begin{equation}
%\left(\sigma_{P_{\lambda}^{ISM}}\over{P_{\lambda}^{ISM}}\right)^2 = 
%\left(\sigma_{P_{0}^{ISM}}\over{P_{0}^{ISM}}\right)^2 +
%4\ \left(\sigma_C\over{C}\right)^2 +
%4\ \left(\sigma_R\over{R}\right)^2.
%\label{eq:ism2}
%\end{equation}

Measurements of the power spectra are rather diverse. For example, \cite{wright} 
analyzes DIRBE data to find that $P_{\lambda}^{ISM}(q) \propto q^{-3}$ ($60 \leq \lambda \leq 240$ $\mu$m), while 
\cite{ingalls2004} examine {\it Spitzer} and {\it IRAS} data ($8 \leq \lambda \leq 70$ $\mu$m) to find 
$P_{\lambda}^{ISM}(q) \propto q^{-2.6}$ at $2\pi/q > 250''$ and 
$P_{\lambda}^{ISM}(q) \propto q^{-3.5}$ at smaller scales.
There is also diversity in the colors measured for ISM emission, much of this dispersion 
may reflect intrinsic variations in the local properties of the dust and the radiation
field that heats the dust \citep{Flagey2006}. This makes the estimation of the cirrus
intensity and the amplitude and shape of the power spectrum rather uncertain.

Estimates of cirrus emission must be bounded from above by the measured diffuse flux fluctuation at 8 \micron\ and it has been argued in KAMM1, KAMM2, AKMM and K12, that this upper limit, when extrapolated to IRAC shorter bands leads to power significantly below that measured in the {\it Spitzer} data at 3.6 and 4.5 \micron. However, the estimates 
made by \cite{ciber} indicate that diffuse Galactic light (DGL, scattered light from 
the ISM) can be a significant contributor to the 1-2 $\mu$m power spectrum at large 
angular scales, assuming the $P_\lambda^{ISM}(q)\propto q^{-3}$ cirrus template.

\section{Discussion}

%1. Sum up JWST part

In this paper we have developed a {\it JWST}/NIRCam-based experiment and methodology to identify the origin of the source-subtracted CIB, discovered with current instruments, 
using the measurements to gain unique understanding of the emergence of the Universe out of the ``Dark Ages'' over $10\lsim z \lsim 40$. As discussed, the current 
CIB fluctuation measurements provide a consistent picture over 2--5\micron, but not at shorter IR bands. The measured fluctuations exceed the levels that
can be produced by remaining galaxies and were proposed to originate from first stars and black holes with the energy requirements discussed here, or in new diffuse sources at more recent epochs.  We have shown that, very generally, the bulk of the sources during first stars era will likely be within the confusion noise of the NIRCam beam requiring CIB to study that era. Observing for 400 hrs with NIRCam at {\it JWST} a 1 deg$^2$ low cirrus region would enable, with the methodology proposed and developed here, to reconstruct CIB power spectrum with sufficient fidelity to constrain its origin and probe the dependence
of the large-scale clustering component on the remaining shot-noise providing a unique probe of the  flux distribution of the sources producing the fluctuations. We have identified quantitatively the floor from remaining galaxies to show that it provides a serious obstacle to probing the epochs of the
CIB sources with experiments involving current instruments, while demonstrating the feasibility of this probe with the  proposed {\it JWST} observation.  Then we have pointed out that the NIRCam
wavelength coverage is uniquely suitable for doing the Lyman-based tomography method proposed and developed here, which can identify or
constrain the emission history of the Universe from $z\sim 10$ to $z\gsim30$. We have applied the method to the current {\it Spitzer}-based CIB 
measurements to show that it already leads to interesting upper limits at $z\gsim 30$. Applying this to the recent CIBER results, however, led to 
unphysical situation if the data are taken at face value. We then discussed the foreground and systematic components which must be accounted for 
in the course of this measurement.

%2. Analysis:

With the {\it JWST}/NIRCam data, our analysis will fit a model of the sky simultaneously with a model for the instrumentation, including possible stray light from the telescope.  The model will include the following terms for each detector pixel:  1)  the detector gain (bits per photon), 2) the detector offset (response if the sky were dark and the exposure were short), 3) the detector dark current (rate of increase of detector signal if the sky were dark), 4) linearity correction (possibly already available from the standard JWST pipeline calibration), and 5) rate of decay of after-images (possibly not a simple exponential). There might be differences in calibrated photometry based on detector bias conditions, exposure times, or readout methods. Ideally we would search for these effects by  observing in multiple conditions, rather than trying to control them by standardizing observations to a single condition.
For each observation, we would compute a model of the zodiacal light as a function of time and direction. Initially such a model would be the standard COBE-based \cite{kelsall} model maintained by the STScI for JWST, but we would search for model residuals and hunt for patterns.  A principal component analysis is likely to reveal spatial variations that are not apparent to the eye due to the presence of bright objects. We would also compute a model of the stray light due to the processes described above.  Then, at last the data will be ready for correlation analysis and comparison with theoretical predictions.

This measurement will also supply additional important data for cross-correlating  with the CIB to be measured by this team from the NASA/ESA-authorized project LIBRAE\footnote{\url{http://librae.ssaihq.com}} (Looking at Infrared Background Anisotropies with {\it Euclid}), but {\it Euclid} will
not reach the proposed depth and wavelength coverage of {\it JWST}; the former is critical in isolating CIB from progressively fainter sources and the latter makes the proposed tomography feasible over the full  range of expected $z$.
Cross-correlating CIB fluctuations with other wavelengths to measure coherence of the CIB with other backgrounds \citep{c13,ak14} would provide further useful information although the S/N of this would be limited by the much smaller area than will be covered by {\it Euclid}. 
Measurements of $\gamma$-ray absorption from high $z$ sources can supply additional information on the mean CIB excess levels \citep[][b]{k05-grb} which would be
important in fully interpreting the fluctuation measurements. The proposed techniques offer to probe the reionization process in a manner alternative to HI 21 cm tomographic studies  \citep[see review by][]{furlanetto}; CIB and the 21cm tomography methods being complementary, but subject to different foregrounds and systematics.

\appendix
\section{Source-subtracted CIB fluctuations: general and definitions}
\label{sec:cib_defs}
%\setcounter{section}{0}
%\setcounter{equation}{0}
%\setcounter{figure}{0}
%\setcounter{table}{0}

%\subsection{Theoretical candidates at high $z$}
%
%\begin{equation}
%\Gamma_{\rm Edd} = \frac{4\pi G m_p c}{\sigma_T} \simeq \frac{1}{3.1\times 10^4} \Gamma_\odot
%\label{eq:m2l_edd}
%\end{equation}
%
%\begin{equation}
%f_* = TBD
%\label{eq:fraction}
%\end{equation}
%
%\subsubsection{Pop III main sequence stars and SNe}
%
%\subsubsection{Black Holes}
%
%DCBHs, Begelman et al mechanism, stellar remnants
%
%\subsubsection{Dense stellar systems}
%

%\renewcommand{\thesection}{A\arabic{section}}
\subsection{Definitions}

In CIB fluctuation studies, resolved sources are removed down to some shot-noise level and the source-subtracted diffuse light fluctuations are then
evaluated. We assume that the shot-noise threshold is approximately equivalent to removing sources brighter some equivalent magnitude limit $m_0$ as discussed in KAMM3, \cite{kari}. We define CIB flux as $F\equiv \nu I_\nu=\lambda I_\lambda$. The mean flux from the remaining cosmological populations is then given by:
\begin{equation}
F(>m_0)= \int_{m_0}^\infty S(m) \frac{dN}{dm} dm
\label{eq:cib}
\end{equation}

\begin{equation}
\frac{dF_\lambda}{dz} = \frac{c}{4\pi} \int \nu^\prime \epsilon_{\nu^\prime} \frac{dt}{dz} \frac{dz}{1+z}
\label{eq:dfdz}
\end{equation}
where $^\prime$ denotes rest-frame quantities and $\epsilon_\nu=\int L_\nu \Phi(L_\nu)dL_\nu$ is the comoving volume emissivity related to the
luminosity function of the emitters, $\Phi$.

The maps of observed surface
brightness are clipped and masked of the resolved sources,
yielding the fluctuation field, $\delta F(${\mbox{\boldmath$x$}).
Its Fourier transform, $f($\mbox{\boldmath$q$}$)= \int \delta
F($\mbox{\boldmath$x$}$)
\exp(-i$\mbox{\boldmath$x$}$\cdot$\mbox{\boldmath$q$}$) d^2x$ is
calculated using the FFT. The power spectrum is $P_2(q)=\langle |
f($\mbox{\boldmath$q$}$)|^2\rangle$, with the average taken over
all the independent Fourier elements $N_q$ corresponding to the given $q$. A
typical rms flux fluctuation is $\sqrt{q^2P_2(q)/2\pi}$ on the
angular scale of wavelength $2 \pi/q$. If the fraction of masked
pixels in the maps is too high (e.g. $>$40\%), one cannot reliably
compute large-scale map properties using the Fourier transform and
instead the maps must be analyzed via the correlation function,
$C(\theta)$, which is immune to mask effects (e.g. Kashlinsky 2007). $C(\theta)$ and
$P_2(q)$ are uniquely related to each other via Fourier
transformation. The cross-power describing the correlations between fluctuations at different wavelengths (1,2) is
$P_{\rm 1\times2} (q)$=$\langle f_{1}(q) f^*_{2}(q)\rangle$=${\cal R}_{1}(q) {\cal R}_{2}(q) + {\cal I}_{1}(q) {\cal I}_{2}(q)$ with ${\cal R, I}$
standing for the real, imaginary parts of $f(\mbox{\boldmath$q$})$.  The 
cross-power spectrum is a real quantity which 
can assume positive or negative values.

The residual CIB fluctuations in IRAC data have two components: 1)
shot/white noise from the remaining (unresolved) sources dominates on
small angular scales, 2) fluctuation component due to clustering
of sources is found on scales $> 0.5^\prime$. These are discussed below.
\subsection{Shot-noise}

The shot noise,
produced by the variance in the number of sources within the beam,
contains contributions from all sources fainter than the limiting
magnitude $m_0$. Its power is characterized by number counts
$\frac{dN}{dm}$ as 
 \begin{equation}
P_{\rm SN} = \int_{m_0}^\infty S^2(m) \frac{dN}{dm} dm
\label{eq:shotnoise}
\end{equation}
(e.g., Kashlinsky 2005). Here $S_\nu(m)$ is the
flux corresponding to magnitude $m$ with these sources having $dN$
counts per unit solid angle and in any experiment this expression gives the white-noise amplitude which then should be convolved with the beam.  
In addition in some models there is also a 1-halo term, which is white noise convolved with the the typical halo-size (and the beam).

Combining eq. \ref{eq:cib} with eq. \ref{eq:shotnoise} leads to $P_{\rm SN} \sim S F(>m_0)$ if both the shot-noise and the bulk of the remaining CIB
arises in the same populations.

\subsection{Clustering component}

The clustering component is given by the Limber equation, which depends on the rate of flux production squared and the underlying 3-D power spectrum, $P_3$, of clustering of the sources integrated over the epochs spanned by them. Specifically the mean squared fluctuation due to cross-power between two bands, $\lambda_1$ and $\lambda_2$, at angular scale $2\pi/q$ is:
\begin{equation}
P_{12}(q) =\int
 \frac{dF_{\lambda_1^\prime}}{dz^\prime} \frac{dF_{\lambda_2^\prime}}{dz^\prime} Q(qd_A^{-1}; z^\prime) dz^\prime
\label{eq:cross-power-app}
\label{eq:limber}
\end{equation}
where $\lambda^\prime\equiv \lambda/(1+z)$ is the rest-frame wavelength, 
$Q(k, z) \equiv \frac{P_3(k, z)}{c(1+z)dt/dz d_A^2(z)}$ and $d_A$ is the comoving angular distance to $z$. If $\lambda_1=\lambda_2$ this gives the auto power, which we denote with just one subscript, $P_1$ or $P_2$. Multiplying both sides of eq. \ref{eq:limber} by $\frac{q^2}{2\pi}$ leads to a simple order-of-magnitude estimate of the mean squared flux fluctuation being $\sim$ (mean flux)$^2$ $\times$ (mean squared fluctuation in the source counts over a cylinder of diameter $qd_a^{-1}$ and length $\sim c\int(1+z)dt$) (Kashlinsky 2005a).
If the sources populated a brief epoch with a characteristic redshift $z$ at the co-moving angular diameter distance $d_a(z)$ emitting the net flux $F$, the typical fluctuation at angular scale $2\pi/q$ due to clustering is $\delta F\propto F \sqrt{q^2 P_3(qd_a(z)})$.

If both the large-scale CIB fluctuation, $\delta F$, from clustering and the shot noise arise in the same populations with typical flux $S$, the two components become coupled $P_{\rm SN} \propto S\cdot \delta F$. 

%\section{Diffuse light contributions from noise, systematics and foregrounds}
%
%
%\subsection{Noise}
%detector noise negligible for large scale features which are averaged over zillions of pixels. it's all cosmic variance, foreground variance, and amplificaiotn of detector/stray light structure in the mosaic process
%
%\subsection{Stray light, gradients, dithering etc}
%
%\subsection{Foregrounds: 1. Zodiacal light}
%
%\subsection{Foregrounds: 2. Galactic cirrus}
%
%\subsection{Foregrounds: 3. Galaxy clustering}
%methods we developed for Spitzer analysis, masking, testing clustered component of the foregrounds, comparison with the galaxy correlation function observed with NIRCam as point sources with angle and redshift,possibility of halos, 

\end{document}